\documentclass[11pt]{article}
\usepackage{color,amsfonts,amsmath,amssymb,mathrsfs,verbatim,epsf}

\def\inn{\in}

\def\pqg{\mbox{$ \!\! \mbox{\Large \_} \! $}}
\def\pqo{\mbox{$ \!\!\!\!\; \mbox{\textbf {\Large \_}} \! $}}
\def\sp{\; \!}

\def\kl{k \pqo l}
\def\jl{j \pqo l}

\def\il{i \mbox{$ \!\!\!\!\; \:\! \mbox{\textbf{\Large \_}} \! \!\!\; $} l}
\def\scirc{\mbox{\raisebox{0.2ex}{\scriptsize {$\circ$}}}}

\def\TM{\mbox{\it TM}}

\def\rrr{{\mathbb{R}}}
\def\ccc{{\mathbb{C}}}
\def\hhh{{\mathbb{H}}}
\def\ooo{{\mathbb{O}}}
\def\kkk{{\mathbb{K}}}

\def\kkk{{\mathbb{K}}}

\def\b1{\mbox{\boldmath $1$}}

\def\v{\mbox{\boldmath $v$}}
\def\btt{\mbox{\boldmath $t$}}

\def\bh{\mbox{\boldmath $h$}}
\def\bk{\mbox{\boldmath $k$}}
\def\bl{\mbox{\boldmath $l$}}

\def\bsl{\boldsymbol}

\def\br{\mbox{\boldmath $r$}}
\def\bv{\mbox{\boldmath $v$}}
\def\bvh{\hat{\bv}}
\def\hG{\hat{G}}

\def\lvh{L(\hat{\v})=1}

\def\lvf{L(\v_4)=1}

\def\lvni{L(\v_9)=1}

\def\lvt{L(\v_{27})=1}
\def\lvfs{L(\v_{56})=1}
\def\lvtfe{L(\v_{248})=1}
\def\lvn{L(\v_n)=1}

\def\htwc{\mbox{h}_2\ccc}
\def\hthc{\mbox{h}_3\ccc}

\def\htwo{\mbox{h}_2\ooo}
\def\htho{\mbox{h}_3\ooo}

\def\sltc{\mbox{SL}(2,\ccc)}
\def\sltca{\mbox{sl}(2,\ccc)}

\def\slthc{\mbox{SL}(3,\ccc)}

\def\sltho{\mbox{SL}(3,\ooo)}

\def\sltwoo{\mbox{SL}(2,\ooo)}

\def\uo{\mbox{U}(1)}

\def\sutw{\mbox{SU}(2)}
\def\sutwa{\mbox{su}(2)}
\def\suth{\mbox{SU}(3)}

\def\sufo{\mbox{SU}(4)}
\def\sufoa{\mbox{su}(4)}

\def\soot{\mbox{SO}^+(1,3)}
\def\soota{\mbox{so}^+(1,3)}

\def\sootn{\mbox{SO}^+(1,9)}

\def\gt{\mbox{G}_2}
\def\ff{\mbox{F}_4}
\def\ee{\mbox{E}_8}
\def\eeg{\mbox{E}_{8(-24)}}
\def\ese{\mbox{E}_7}
\def\eseg{\mbox{E}_{7(-25)}}
\def\esi{\mbox{E}_6}
\def\esig{\mbox{E}_{6(-26)}}

\def\SM{\mbox{SU}(3) \times \mbox{SU}(2) \times \mbox{U}(1)}
\def\SML{\mbox{SU}(3)_c \times \mbox{SU}(2)_L \times \mbox{U}(1)_Y}

\def\tthX{{\theta_{\! \mbox{\tiny{$X$}}}}}
\def\tthXd{{\theta_{\! \mbox{\tiny{$X$}}}^{\dag}}}

\def\ph{\phantom}

\def\lag{{\mathcal L}}

\def\mcT{{\mathcal T}}
\def\mcX{{\mathcal X}}
\def\mcY{{\mathcal Y}}
\def\mcW{{\mathcal W}}
\def\mcZ{{\mathcal Z}}
\def\mcI{{\mathcal I}}

\def\fh{\frac{1}{2}}

\def\gpath{ }

\def\setb{\setlength{\baselineskip}{0.625\baselineskip}}

\begin{document} 

{\setlength{\baselineskip}{0.625\baselineskip}

\begin{center}

 {\LARGE{\bf Time, $\ee$, and the Standard Model}} \\
  
\bigskip

 \mbox {{\Large David J. Jackson}}\footnote{email: david.jackson.th@gmail.com}  \\
  
  \vspace{10pt}
 
 { \large September 6, 2017 }

 \vspace{12pt}

{\bf  Abstract}

\vspace{-2pt} 
 
\end{center}

      Based upon the unique and simple starting point of the continuous flow of time a physical theory is 
derived through an analysis of the elementary arithmetic composition and symmetries of this one-dimensional 
progression. We describe how the explicit development of the theory leads to a prediction of the unique and 
largest exceptional Lie group $\ee$ as the full `symmetry of time', and hence as the unification group for 
the physical theory. This proposal results from the identification of a series of esoteric properties of 
the Standard Model of particle physics from a series of intermediate augmentations in the 
`multi-dimensional form of time'. These physical properties derive from the breaking of the full symmetry 
of time through the necessary interposition of an external 4-dimensional spacetime arena, itself 
constructed from a 4-dimensional form of time, as the background to all observations. The basic conceptual 
picture is presented together with reviews of a number of references regarding $\ee$ structures which may 
provide a significant guide in pursuing the goal of converging upon a complete unified theory.     

{

\vspace{-5pt}
\tableofcontents

\vspace{15pt}
}


\section{Introduction}
\label{ee1}

  The conception of a physical theory presented here rests upon the general 
presumption that all empirical phenomena are infused within the passage of time. On 
parametrising this temporal continuum with a single real variable, while utilising the 
elementary arithmetic structure of the real number system, infinitesimal intervals of 
time can be expressed in terms of a composition of a multi-dimensional set of real 
variables within well-defined constraints. 
In this manner the original linear flow of time can be simultaneously manifested in 
terms of quadratic structures, with a direct interpretation as underpinning a 
Euclidean spatial framework  -- as associated for example with a local inertial frame 
in general relativity. In turn
 higher-dimensional structures based on cubic or higher-order forms of time, with 
residual parameters and symmetries over and above those required to construct a 
4-dimensional Lorentzian manifold, give rise to matter fields in spacetime, as will be 
described in detail in this paper.

    A principle aim of this project is to see how far this very simple idea can be 
developed in terms of making contact with the empirical world, and with existing 
successful physical theories, and the degree of explanatory power that can be achieved 
in this way. Progress for the theory over a range of topics from connections with the 
Standard Model of particle physics to the standard model of cosmology, together with 
the relation to non-Abelian Kaluza-Klein theory and quantum field theory, has been 
collectively presented in the lengthy tome \cite{Unifi}. The means of obtaining the 
Standard Model connections are summarised in \cite{Novel}, while the comparisons and 
contrasts between this theory originating from the one dimension of time and 
Kaluza-Klein theories constructed with extra dimensions of space are further 
emphasised in \cite{KKone}. 

     Building upon the above references, and in particular (\cite{Unifi} chapters 
6--9, \cite{Novel}), we again set out to motivate the basic conception of the theory 
and here describe in detail how its development leads to consideration of $\ee$ as the 
ultimate unification group and the manner in which this proposal may meet criteria of 
testability. To this end a significant degree of contact is made with existing work in 
the mathematical physics literature regarding the structure and properties of $\ee$. 
Correspondingly we first review the relevant literature in the following section, with 
emphasis upon the larger exceptional Lie groups more generally and their relation with 
the four division algebras. In particular, in subsection~\ref{ee23} we describe how 
some of these structures can be associated with a notion of `generalised spacetime'.
 
   In section~\ref{ee3}, by contrast, we set out the motivation and argument for 
founding a theory upon one dimension of time only, with references to historical 
developments in physics and mathematics. The straightforward technical means by which 
a physical theory can be constructed from this single dimension is described via a 
minimal example in subsection~\ref{ee41}. The development of the theory through 
subsequent natural extensions of the multi-dimensional form of time leads 
\textit{directly} to significant contact with the Standard Model, as will be presented 
in subsections~\ref{ee42} and \ref{ee43}.

  The progression of the theory towards larger symmetries of time culminates in 
section~\ref{ee5} in a prediction of  $\ee$ as the full unification group. This 
observation is justified in terms of the need to converge upon a complete description 
of the Standard Model symmetry and particle multiplet structure and through 
connections made with various studies regarding $\ee$ as reviewed in 
section~\ref{ee2}. This discussion is continued in section~\ref{ee6} where we also 
summarise the other areas in which the theory has been developed, with the aim of 
converging upon a complete unified theory more generally. The overall status of the 
theory will be further assessed in the concluding section. While successes that have 
been achieved mark a proof of principle for the basic idea of the theory, we also 
consider its potential predictive power.

  In recent decades there has been much interest in theories based on extra spatial 
dimensions and also in unification schemes employing the exceptional Lie algebras. 
These two areas are drawn together in the framework presented here through a change of 
emphasis in constructing the theory from the one dimension of time only.  
 While some empirical success has already been achieved, and directions for further 
progress can be identified, one of the main strengths of the theory lies in the unique 
simplicity of founding the theory on this single dimension of time, as will be a 
central theme running through this paper.  In summary the main aims of this paper are 
to:
\begin{itemize}

\item Motivate the conceptual basis for the theory in describing how it is possible to 
build a full physical theory from the one dimension of time,

\item Demonstrate the explanatory power of the theory in uncovering a series of 
properties of the Standard Model with minimal redundancy, and
 
\item Justify the proposed culmination of this progression in $\ee$ as the full 
symmetry of time and consider the possible form this might take. 
  
\end{itemize}
    
	 The final stage is ultimately presented in the manner of a puzzle that remains to 
be solved. With this goal in mind in the following section we begin by considering 
some of the mathematical structures involving the exceptional Lie group $\ee$ that may 
provide essential input.


\section{Review of Selected Studies in $\ee$}
\label{ee2}

\subsection{Unification in Physics -- Unique in Mathematics}
\label{ee21}

   Several applications of $\ee$ in physical theories and the status of $\ee$ as a 
mathematical structure itself will be reviewed in this subsection. In the latter case, 
as the largest exceptional Lie algebra with rich symmetry properties, $\ee$ occupies a 
unique position in mathematics. First though we consider a well known proposal for a 
fundamental role for $\ee$ in physics in a branch of superstring theory -- itself 
conceived as a framework incorporating a quantum theory of gravity (see for 
example~\cite{Polch}). Heterotic string theory combines 10-dimensional superstring 
theory with the original 26-dimensional bosonic theory, with the additional 16 
dimensions compactified on a torus. This torus can be defined by the root lattice of  
the rank-16 Lie algebra of either SO(32) or $\ee \times \ee$ in order to obtain a 
consistent theory free from anomalies.

    (In this paper we generally employ upper-case letters for classical Lie groups 
such as SO(32), with lower-case denoting the corresponding Lie algebra such as so(32); 
while upper-case is used for both the exceptional Lie groups and their algebras, such 
as E$_8$, with the distinction taken from the context).

  With the further six extra dimensions of the 10-dimensional superstring also being 
compactified over an external 4-dimensional spacetime, the case with the local gauge 
group $\ee \times \ee$ emerging at low energies describes a gauge field theory 
comfortably able to accommodate the internal gauge symmetry $\SM$ of the Standard 
Model. That is,  the 
 $\ee \times \ee$ symmetry can be considered  as  an elaborate example of a `Grand 
Unified Theory' (GUT) within this component of the string theory. Within this setting 
it is possible to obtain string vacua containing three generations of quarks and 
leptons of the Standard Model, with additional matter multiplets in `hidden sectors' 
of the theory which are not empirically observed (see for example~\cite{Braun}). 
However, 
one of the challenges for string theory, which is further exacerbated when subsumed 
into M-theory, is in handling the vast number of different solutions admitted by the 
equations. The corresponding lack of a unique vacuum solution is referred to as the 
`landscape problem', while the difficulty in addressing it can be used to motivate a 
`multiverse' interpretation of the theory.

   In fact the Lie group $\ee$ itself is comfortably large enough to contain as a 
subgroup the Standard Model gauge group $\SM$ together with the external Lorentz 
symmetry $\soot$, and hence on its own has the potential to be utilised by a theory 
seeking, beyond the ambition of a GUT, to unify the internal gauge forces together 
with gravity through a single symmetry group. At the level of the Lie algebra 
structure alone, and motivated in part by a notion of `mathematical beauty', this is 
the approach adopted in \cite{Lisi}.
 However, while components associated with the external gravitational field and 
internal gauge fields as well as three generations of `quarks' and `leptons' are 
identified in the $\ee$ root lattice the second and third generations of these 
`fermion' states lack the appropriate external and internal symmetry properties other 
than through a `graviweak' $\mbox{SO(8)} \subset \ee$ `triality' transformation with 
respect to the first (\cite{Lisi} subsections~2.2.3 and 2.4.2).

  The impossibility of amending this discrepancy with the Standard Model, while 
keeping strictly within the goal of embedding these structures within the $\ee$ Lie 
algebra, owing to the insufficient number of non-compact generators for any real form 
of $\ee$ is described in \cite{DiGa}. Nevertheless, the fact that structures 
resembling the Standard Model can be identified for the exceptional Lie algebras 
(including $\esi$ and $\ese$ GUT models dating from the 1970s \cite{Gur1,Gur7}), 
together with the observation that $\ese$ and $\ee$ are large enough to incorporate 
the external Lorentz group alongside the Standard Model gauge group, is suggestive.
  A further exploration of some of these mathematical structures and tentative 
connections with physics is seen for example in \cite{MaTr}, in which 4-dimensional 
spacetime itself is proposed to emerge through fundamental interactions which in turn 
can be defined in terms of the structure of the $\ee$ Lie algebra.

   A review of the more purely mathematical properties of $\ee$, dating from the 
inception of this 248-dimensional Lie algebra by Wilhelm Killing in 1887 through to 
much more recent developments, can be found in \cite{Gari}. Here we briefly consider 
some of these algebraic structures. 

   For the rank-8 Lie algebra $\ee$ there are eight independent Casimir operators, 
that is elements that are defined by the centre of the universal enveloping algebra of 
$\ee$ and hence commute with all elements of the Lie algebra. These operators are of 
order 2, 8, 12, 14, 18, 20, 24 and 30 in the Lie algebra generators, where the first 
of these is the quadratic invariant defined by the Killing metric. The second of these 
is an eighth order invariant tensor as explicitly constructed for the compact real 
form E$_{8(-248)}$, via the adjoint and spinor representations of the maximal 
subalgebra $\mbox{so(16)} \subset \mbox{E}_{8(-248)}$, in (\cite{CedP} equation 2.3) 
and also described in \cite{Tala,GarG,BeRu}. These references imply such an octic 
invariant can be constructed for any of the three real forms of $\ee$, namely 
E$_{8(-248)}$ as well as the two non-compact forms E$_{8(-24)}$ and E$_{8(8)}$. For 
each case, in being composed of the 248 elements of a real $\ee$ Lie algebra, the 
tensor invariant further implies the existence of an eighth-order real-valued 
polynomial function of 248 real variables that is invariant under the adjoint action 
of $\ee$. In fact invariant homogeneous polynomials over $\rrr^{248}$ could in 
principle be identified corresponding to each of the eight Casimir invariants and for 
each of the three real forms of $\ee$.

  One of the challenges in studying the Lie group $\ee$ is the unique feature of 
lacking a non-trivial representation smaller than the adjoint representation in 248 
dimensions, which describes $\ee$ in terms of a set of symmetries acting upon its own 
Lie algebra. However, as for other exceptional Lie algebras, a description of $\ee$ is 
possible in terms of the octonions $\ooo$, the largest division algebra, as presented 
towards the end of \cite{Baez1}. 
 One such construction is in terms of the final entry in the $4 \times 4$ 
Freudenthal-Tits `magic square' of Lie algebras. Each of the 16 entries in the magic 
square can be denoted M$(\kkk,\kkk')$, where $\kkk$ and $\kkk'$ each denote a division 
algebra $\rrr$, $\ccc$, $\hhh$ or $\ooo$, and involves a Jordan algebra 
$\mbox{h}_3\kkk'$ of $3 \times 3$ Hermitian matrices over $\kkk'$ (see for example 
\cite{Baez1} section~4.3, \cite{BaSu}). 
 
   The largest Lie algebra constructed in this way is $\ee$ which occupies a unique 
position as the fourth row and fourth column entry M$(\ooo,\ooo)$ of the magic square. 
Different real forms of $\ee$ can be obtained by employing the `split' octonions 
$\ooo_s$; for example with $\kkk = \ooo_s$ the real Lie algebra corresponding to 
M$(\ooo_s,\ooo)$ is E$_{8(-24)}$ while for the doubly-split magic square the entry  
M$(\ooo_s,\ooo_s)$ yields E$_{8(8)}$. This formulation is not explicitly utilised in 
this paper. However we shall make significant reference to the realisation of 
E$_{8(-24)}$ described below in subsection~\ref{ee23}, with potential connections to 
physics, which also employs in a central role the octonions and the exceptional Jordan 
algebra $\htho$, the basic properties of which we hence review in the following 
subsection.

\subsection{The Octonions and the Exceptional Jordan Algebra}
\label{ee22}

    For any two elements $a,b \in \kkk$ of any one of the four normed division 
algebras $\kkk = \rrr, \ccc, \hhh$ or $\ooo$ (of real dimension $1,2,4$ and 8 
respectively) the relation:
\begin{equation} 
\label{abnorm}
  \vert ab \vert = \vert a \vert \vert b \vert
\end{equation}
 holds. Here $\vert a \vert = (a\bar{a})^{\fh}$ is the norm of $a\in\kkk$, where the 
conjugate $\bar{a}$ is obtained by negating each of the imaginary components of $a$. 
Each element $a\neq 0$ also has a unique multiplicative inverse $a^{-1}=\bar{a}/\vert 
a \vert^2$ (see for example~\cite{Baez1}, \cite{Unifi} section 6.2). These algebras 
hence naturally describe symmetry operations since  any $a \in \kkk$ with $\vert a 
\vert =1$ acting by multiplication on any $b\in \kkk$  leaves the norm of the latter 
invariant, that is    
$\vert b \vert \to \vert ab \vert = \vert b \vert$ by equation~\ref{abnorm}, and has 
an inverse operation.

  Unlike the real $\rrr$ and complex $\ccc$ numbers the quaternions $\hhh$ form a 
non-commutative algebra and the octonions $\ooo$ are also non-associative, hence 
multiplication in the latter case can not directly \textit{represent} the actions of a 
symmetry \textit{group}. This non-associativity does not however prohibit the 
octonions from playing an important role in describing symmetries for a physical 
theory (see for example \cite{Pais,Gamb,Gunay}, the references in \cite{SoLo} for 
further early studies, and \cite{Dix1,Borst,Man4}).   

   Indeed Lie algebras, with elements being the infinitesimal generators of the finite 
symmetry transformations of a Lie group,  themselves provide a very familiar example 
of non-associative algebras that are of central importance in physics. For the case of 
employing the octonion algebra as considered here the corresponding non-associativity 
is apparent at the level of compositions of finite symmetry transformations.
 In fact the algebraic properties of the octonions allow a high degree of symmetry to 
be described, as noted for the relation between octonion composition and the Lie group 
SO(7) in (\cite{Unifi} section 6.2).
 However owing to the non-associativity care is  needed in the algebraic manipulation 
of these operations, and in particular many of the general results of group theory, 
which rely on the group axiom of associativity, cannot be directly applied.

    As described for (\cite{Unifi} equation~6.6, \cite{Novel} equation~21) here we 
adopt the notation of~\cite{Man4,Man5,Wang,Wang2} for many of the conventions 
regarding the octonions and the exceptional Jordan algebra, with for example an 
octonion element $a\in\ooo$ written as:
\begin{equation}
\label{octa}
  a \; = \; a_1 \; + \; a_2\,i \; + \; a_3\,j + \; a_4\,k \; + \; a_5\,{\kl}
        \; + \; a_6\,{\jl}   \; + \; a_7\,{\il} \; + \; a_8\,l
\end{equation}  
   Here $a_1, \ldots, a_8 \inn \rrr$ are  eight real components and    
    $\{i,j,k,\kl,\jl,\il,l\}$ are the seven imaginary octonion units with 
$i^2=j^2=\ldots=\il^2=l^2 = -1$. These seven units are mutually anticommuting, with 
the notation motivated by products such as $i \sp l = \il = -l\sp i$. The full 
multiplication table is given for example in (\cite{Man4} figure 2, \cite{Unifi} 
figure 6.1) with the non-associativity exhibited by products such as $(i\sp j) l = 
-i(j \sp l) = +\kl$.

  The exceptional Jordan algebra $\htho$ of $3 \times 3$ Hermitian matrices over the 
octonions is of particular significance for this paper, and here we review some of the 
structures relating to this algebra (see also for example \cite{Ohwa,Krute}). For
 any two elements  $\mcX,\mcY \in \htho$ the Jordan product is defined by:
\begin{equation}
\label{joralg}
 \mcX \circ \mcY \; = \; \frac{1}{2}(\mcX\mcY +\mcY\mcX)  \; \inn \; \htho  
\end{equation}
   where $\mcX\mcY$ denotes the usual $3\times 3$ matrix multiplication. As for any 
Jordan algebra the product $\mcX \circ \mcY$ is commutative but not in general 
associative. In turn the Freudenthal product for elements of $\htho$ can be defined by 
the commutative composition:
\begin{equation}
\label{fjprod}
  \mcX \times \mcY  \, = \,  \mcX \circ \mcY - \fh 
    \big( \mbox{tr}(\mcX)\mcY + \mbox{tr}(\mcY)\mcX  \big) + \fh
	\big( \mbox{tr}(\mcX) \mbox{tr}(\mcY) - \mbox{tr}(\mcX \circ \mcY) \big)\b1_3
	 \, \inn \, \htho  \;\:
\end{equation}
  where $\mbox{tr}(\mcX)$ is the trace of the matrix $\mcX \in \htho$ and $\b1_3$ is 
the unit $3 \times 3$ matrix. In the final term an inner product on the space $\htho$ 
is employed, that is the bilinear map denoted and defined by:
\begin{equation}
 \label{trabimap}
 (\mcX, \mcY) \; = \; \mbox{tr}(\mcX \circ \mcY) \; \in \; \rrr
\end{equation}

   This map is invariant under $\ff$ transformations, that is under the automorphism 
group of the exceptional Jordan algebra. We adopt the notation for the components of  
$\mcX,\mcY \in \htho$ from (\cite{Unifi} equation 9.25), which in turn for $\mcX \in 
\htho$ was adopted from \cite{Wang}, and write:
\begin{equation}
 \label{hthoxy}
   \mcX = 
   \left( \begin{array}{ccc}
       p & \bar{a} & c  \\
       a &   m     & \bar{b}        \\
 \bar{c} &   b     & n 
          \end{array}  \right), \qquad \qquad
   \mcY = 
   \left( \begin{array}{ccc}
       P &  \mathbf{\bar{\mathnormal{A}}} & C  \\
       A &   M     & \mathbf{\bar{\mathnormal{B}}}       \\
 \mathbf{\bar{\mathnormal{C}}} &   B     & N 
          \end{array}  \right)
\end{equation}
   with  $p,m,n\in \rrr$ and $a,b,c \in \ooo$ for $\mcX$ and similarly for their 
upper-case counterparts in $\mcY$.  The inner product of equation~\ref{trabimap} can 
then be written out explicitly as:
\begin{equation}
 \label{incomp}
 (\mcX, \mcY) \; = \; pP + mM + nN + 2\langle a,A \rangle 
           + 2\langle b,B \rangle + 2\langle c,C \rangle 
\end{equation}  
\begin{equation}
\label{octinner}
  \mbox{with} \qquad \langle a,A \rangle \; = \;  
    \frac{1}{2}(a\mathbf{\bar{\mathnormal{A}}} + A\bar{a})
    \; = \; \mbox{Re}(a\mathbf{\bar{\mathnormal{A}}} )  \; = \; \sum_{h=1}^8 a_h A_h.
\end{equation} 
  Here equation~\ref{octinner} defines an inner product on the octonion algebra 
itself, where `Re' denotes the real part of an octonion and the $a_h \inn \rrr$ are 
the eight components of $a\in \ooo$ of equation~\ref{octa}, with $A_h \inn \rrr$ 
similarly the components of $A\in \ooo$.
 
  The inner product on the Jordan algebra of equation~\ref{trabimap} can be used in 
conjunction with the Freudenthal product of equation~\ref{fjprod} to define a cubic 
form for any three elements $\mcX,\mcY,\mcZ \in \htho$ as:
\begin{equation}
\label{cubicj}
  (\mcX,\mcY,\mcZ) \; := \; (\mcX,(\mcY \times \mcZ)) 
      \; = \; \mbox{tr}(\mcX \circ (\mcY \times \mcZ))  \;  \in  \; \rrr
\end{equation} 
  which is totally symmetric in the three arguments. This trilinear form is invariant 
under transformations of the Lie group $\esi$ upon the space $\htho$. 
 The quadratic adjoint map $\mcX \to \mcX^{\sharp}$ can also be defined via the 
Freudenthal product, with:
\begin{eqnarray}
  \mcX^{\sharp} & = & \mcX \times \mcX  \label{quadmap}  \\
\mbox{or explicitly:} \quad \mcX^{\sharp}  & = & 
   \left( \begin{array}{ccc}
       mn - \vert b \vert^2 \, & \, cb - n\bar{a} \, & \, \bar{a}\bar{b} - mc  \\
       \bar{b}\bar{c} - na \, & \, pn - \vert c \vert^2 \, & \, ac - p\bar{b} 
	             \\
       ba - m\bar{c} \, & \, \bar{c}\bar{a} - pb \, & \, pm - \vert a \vert^2 
          \end{array}  \right) \; \inn \htho		\label{hthoshp}							  			 
\end{eqnarray}
   from the components of $\mcX$ in equation~\ref{hthoxy}.  
   On setting $\mcX = \mcY = \mcZ$ in equation~\ref{cubicj}
    a cubic norm or determinant can be defined for $\mcX \in \htho$ as:
\begin{eqnarray}
  \det(\mcX) & = &  \frac{1}{3}(\mcX,\mcX,\mcX)
                 \;  = \; \frac{1}{3}\mbox{tr}(\mcX \circ (\mcX \times \mcX)) 
                 \;  = \; \frac{1}{3}\mbox{tr}(\mcX \circ \mcX^{\sharp})
				          \label{detxxx}   \\
      & = &   pmn - p\vert b \vert^2 - m\vert c \vert^2 - n\vert a \vert^2  
                + 2 \mbox{Re}(\bar{a}\bar{b}\bar{c})     \label{detpmn}		  			 
\end{eqnarray}
  which is of course also invariant under $\esi$. This $\esi$ symmetry of $\det(\mcX)$ 
is explicitly constructed in \cite{Man4,Man5,Wang,Wang2}.    
   
   Alternatively the above structures can be motivated by \textit{beginning} with
    the definition of $\det(\mcX)$ from equation~\ref{detpmn} which exhibits the 
$\esi$ symmetry.
	 Equation~\ref{detpmn} has a very similar form to the usual definition of the 
determinant for a $3 \times 3$ matrix, with the final term adapted for the 
non-associativity of the octonions.
 This determinant $\det(\mcX)$ can be used to \textit{define} the quadratic adjoint of 
$\mcX$ as the `cross product'
   $\mcX^{\sharp} = \mcX \times \mcX$ on $\htho$ \textit{through} 
    equation~\ref{detxxx} (\cite{Baez1} section 3.4), with the adjoint identity 
	 $(\mcX^{\sharp})^{\sharp} = \det(\mcX)\mcX$ also satisfied (\cite{Borst} 
equation~9.8). Equation~\ref{detpmn} can also be written as:
\begin{eqnarray}
  \det(\mcX) & = &  \frac{1}{3}\mbox{tr}(\mcX^3) \, - \,
             \frac{1}{2}\mbox{tr}(\mcX^2)\mbox{tr}(\mcX) \, + \,
			 \frac{1}{6}\mbox{tr}(\mcX)^3 
				          \label{dettra}   \\
  \mbox{while} \quad  						  
      \mcX^{\sharp} & = &  \mcX^2 \, - \, \mbox{tr}(\mcX)\mcX \, + \,
	  \frac{1}{2} \big( \mbox{tr}(\mcX)^2 - \mbox{tr}(\mcX^2) \big)\b1_3   
	      \label{shptra}		  			 
\end{eqnarray} 
 in terms of trace functions, 
	 consistent with equation~\ref{detxxx}.
	 The quadratic adjoint $\mcX^{\sharp}$ in equation~\ref{detxxx},
	 with $\det(\mcX) = \frac{1}{3}(\mcX,\mcX^{\sharp})$,
	  plays a similar role to the classical adjoint of a matrix as used in the 
standard method of constructing the determinant of the matrix,
  as can be seen from equations~\ref{incomp} and \ref{hthoshp}.  
 The `cross product' defined in this way can be linearised to define:
\begin{equation}
\label{linxy}
   \mcX \times \mcY \; = \;
    \fh \lbrack (\mcX+\mcY)^{\sharp} - \mcX^{\sharp} - \mcY^{\sharp} \rbrack
\end{equation}
  which is equivalent to the Freudenthal product of equation~\ref{fjprod}, as can be 
seen via equation~\ref{shptra}. Similarly the determinant or cubic norm 
   $\det(\mcX) = \frac{1}{3}(\mcX,\mcX,\mcX)$ can be linearised to define a cubic 
form, that is:
\begin{eqnarray}
   (\mcX,\mcY,\mcZ) \; = \; 3 \times \frac{1}{6}
   \big\lbrack \det(\mcX+\mcY+\mcZ) \!\!\! & - & \!\!\! \det(\mcX+\mcY) 
    \,  -  \, \det(\mcY+\mcZ)
    \,  - \, \det(\mcZ+\mcX)   \nonumber  \\
  &   &	 + \, \det(\mcX) \, + \, \det(\mcY) \, + \, \det(\mcZ) \big\rbrack
    \label{cubiclin}
\end{eqnarray}
   which is manifestly symmetric in $\mcX,\mcY,\mcZ \in \htho$  and equivalent to the 
cubic form defined in equation~\ref{cubicj}. 
 Hence the role of the Jordan product and Freudenthal product in equation~\ref{cubicj} 
can be justified in this way, considering the determinant $\det(\mcX)$  defined in 
equation~\ref{detpmn} to be of primary significance.

   In general matrix determinants exhibit the property that for any elements $a,b \inn 
M$ belonging to a particular $m \times m$ matrix algebra $M$ the product satisfies:
\begin{equation} 
\label{abdet}
    \det(ab)   = \det(a)\det(b)
\end{equation}
   This property of matrix algebras is analogous to that of  the normed division 
algebras described for equation~\ref{abnorm} in the opening of this subsection, and 
hence both types of algebra can naturally be applied to describe symmetries. In the 
case of the $3\times 3$ matrices of $\htho$ over the octonions \textit{both} kinds of 
algebraic structure are used together in equation~\ref{detpmn}. The actions of the 
real Lie group $\esig \equiv \sltho$ preserve $\det(\mcX)$ for any $\mcX \in \htho$, 
as described explicitly in \cite{Man4,Man5,Wang,Wang2} and extensively used in 
\cite{Unifi,Novel}, and as will be of central importance in this paper.

\subsection{Generalised Spacetime and $\ee$ Symmetry}
\label{ee23}

   The space $\htho$ can be considered a `generalised spacetime' as a natural 
extension from 4-dimensional Minkowski spacetime, as described for example in 
\cite{Gunay3,Gunay2} in which such `spacetimes' are defined as having coordinates 
parametrised by the elements of a Jordan algebra. We first note that Lorentz 
transformations on the 4-vector 
  $\bv_4 = (v^0,v^1,v^2,v^3)\in \rrr^4$ can be represented by the double cover group 
$\sltc$ acting on the $2\times 2$ matrices:
\begin{equation}
  \label{hinvfour}
      \bh \, = \, v^0\sigma^0 +v^1\sigma^1 + v^2\sigma^2 + v^3\sigma^3 
	 \,  = \, \left(   
	   \begin{array}{cc} v^0 + v^3 & v^1 - v^2i \\
		   v^1 + v^2i  & v^0 - v^3  \end{array} \right) 	\, \inn  \, \htwc  
\end{equation}
   with $\sigma^0 = \binom{1 \;\, 0}{0 \;\, 1}$ together with the three Pauli matrices 
$\sigma^1 = \binom{0 \;\, 1}{1 \;\, 0}$,  $\sigma^2 = \binom{0 \, -i}{i \;\;\, 0}$, 
$\sigma^3 = \binom{1 \;\;\, 0}{0 \,\, - \! 1}$ (see for example \cite{Novel} equations 
17 and 18). The $S \inn \sltc$ matrix actions $\bh \to S \bh S^{\dag}$ leave invariant 
$\det(\bh) = \bv_4 \! \cdot \! \bv_4 = \vert \bv_4 \vert^2$, where $\bv_4 \! \cdot \! 
\bv_4 = \eta_{ab}v^av^b$ is the Lorentz inner product (with the conventional summation 
over repeated indices used throughout this paper, here for $a,b = 0,1,2,3$ and with 
$\eta = \mbox{diag}(+1,-1,-1,-1)$ being the Minkowski metric).

    On generalising from $2 \times 2$ Hermitian matrices over $\ccc$ the space $\htwo$ 
of $2 \times 2$ Hermitian matrices over $\ooo$, with ten real components, can 
represent 10-dimensional spacetime. The quadratic norm $\det(X)$, with $X \in \htwo$, 
is preserved by actions of the group $\sltwoo$ as the double cover of the 
10-dimensional Lorentz group $\sootn$ \cite{Man2}. This space can be further extended 
to $3 \times 3$ Hermitian matrices over $\ooo$, with the group $\esig \equiv \sltho$ 
acting on the space $\htho$ preserving the cubic norm $\det(\mcX)$, for any $\mcX \in 
\htho$. This latter symmetry was described in the previous subsection for 
equation~\ref{detpmn} and can be constructed in terms of a composition of three 
interlocking $\sltwoo$ subgroup actions \cite{Man4,Man5,Wang,Wang2}. This natural 
progression justifies consideration of the space $\htho$ as a generalised spacetime.

  As for $\htho$ the set of elements of $\htwc$ is  closed under the symmetric 
anticommutator  product of equation~\ref{joralg} and in fact also forms a Jordan 
algebra. As described in \cite{Gunay3} the `conformal group' SU$(2,2)$ (the double 
cover of SO$(2,4)$) associated with the space $\htwc$ is defined as leaving invariant 
the light cone:
\begin{equation}
   \label{htwclc}
   \det(\bh \! - \! \bk) = 0
\end{equation}  
     for the separation between $\bh, \bk \in \htwc$. The corresponding Lie algebra 
$\mathfrak{g}= \mbox{su}(2,2) \equiv \mbox{so}(2,4)$ possesses a 3-graded structure:
\begin{eqnarray}
     \mathfrak{g} & = & \mathfrak{g}^{-1} \quad \oplus \quad\;
	        \mathfrak{g}^{0} \quad\; \oplus \quad \mathfrak{g}^{+1} \nonumber \\
 \mbox{of dimension:} \quad 15 & = &
                 4 \quad + \quad (6+1) \quad + \quad 4     \nonumber \\ 
 \mbox{with elements:} \qquad  & &  
            U_{\bsl{k}}, \qquad\quad S_{\bsl{k}\:\!\bsl{l}},
			    \qquad\quad \tilde{U}_{\bsl{k}} \label{grada}
\end{eqnarray}
  labelled by $\bk,\bl \in \htwc$. The generators $U_{\bsl{k}}$ correspond to 
translations and the $\tilde{U}_{\bsl{k}}$ to conformal transformations, while the set 
$S_{\bsl{k}\:\!\bsl{l}}$ is composed of the six generators of the Lorentz group 
together with a dilation generator. These act upon an element $\bh \inn \htwc$ as 
(\cite{Gunay3} equation 8):
\begin{equation}
\label{usuact}
   U_{\bsl{k}}(\bsl{h}) = \bk, \quad S_{\bsl{k}\:\!\bsl{l}}(\bh) =
                            \{\bsl{k},\bsl{l},\bsl{h}\}, 
     \quad \tilde{U}_{\bsl{k}}(\bh) = -\fh\{\bh, \bk, \bh\}
\end{equation}
\begin{equation}
  \label{jtphtwc}
  \mbox{where:} \quad  \{\bh,\bk,\bl\} \; = \;
       (\bh\!\cdot\! \bk)\,\bl + (\bl\!\cdot\! \bk)\,
	    \bh - (\bh\!\cdot\! \bl)\,\bk 
\end{equation}  
  is the Jordan triple product on $\htwc$. 
  Here $\bh\!\cdot\! \bk =  
  \frac{1}{2}\big(\mbox{tr}(\bh)\mbox{tr}(\bk) - \mbox{tr}(\bh\circ \bk)\big)$ 
   is the Lorentz inner product between the 4-vectors associated with these elements 
of $\htwc$ via equation~\ref{hinvfour}. This Jordan triple product can be defined in 
terms of the Jordan product (\cite{Gunay3} equation~3) in a form similar to 
equation~\ref{jtphtho} below.
  The triple product is also utilised in the $\mbox{su}(2,2)\equiv \mbox{so}(2,4)$ Lie 
bracket (\cite{Gunay3} equation 9):
\begin{eqnarray}
  \lbrack U_{\bsl{k}}, \tilde{U}_{\bsl{l}} \rbrack 
                  =  S_{\bsl{k}\:\!\bsl{l}} \quad\; & \quad &
  \lbrack S_{\bsl{k}\:\!\bsl{l}}, S_{\bsl{h}\bsl{m}} \rbrack  = 
     S_{\{\bsl{k},\bsl{l},\bsl{h}\}\bsl{m}}
	          - S_{\{\bsl{l},\bsl{k},\bsl{m}\}\bsl{h}} \nonumber \\
  \lbrack S_{\bsl{k}\:\!\bsl{l}}, U_{\bsl{h}} \rbrack  = 
                 U_{\{\bsl{k},\bsl{l},\bsl{h}\}}  & \quad &
  \lbrack S_{\bsl{k}\:\!\bsl{l}}, \tilde{U}_{\bsl{h}} \rbrack  = 
               \tilde{U}_{\{\bsl{l},\bsl{k},\bsl{h}\}}  
    \label{commhtwc}  
\end{eqnarray}

   Similarly the conformal group for 10-dimensional spacetime $\htwo$ can be 
identified as $\mbox{SO}(2,10)$ (see for example \cite{DrayMW}). However here we 
consider the  Lorentz symmetry $\sootn$ acting, via the double cover $\sltwoo$, on the 
10-dimensional vector space  $\htwo$  and augment  this space by an $\sootn$ scalar 
$n\inn \rrr$. The product $n\det(X)$, for any $X \in \htwo$, is then invariant under 
both $\sootn$ and a dilation symmetry defined with reciprocal scaling actions on $n$ 
and $\det(X)$. Upon further extension by the components of an $\sootn$ spinor $\theta 
\in \rrr^{16}$ a cubic form can be constructed (\cite{Gunay2} equation 63):
\begin{equation}
 \label{vcubic}
  \mathcal{V}(n,X,\theta) \; = \; n\det(X) \; - \; 2X \!\cdot\! (\theta\theta^{\dag})
\end{equation} 
 where here the inner product is again defined by 
  $X\!\cdot\! Y = \frac{1}{2}\big(\mbox{tr}(X)\mbox{tr}(Y) - \mbox{tr}(X\circ 
Y)\big)$, now for $X,Y \inn \htwo$, (as adopted from (\cite{Man5} equations~38 and 39, 
\cite{Wang} section~3.3) with the sign adapted for Lorentz metric signature 
 $(+1,-1, \ldots, -1)$ when interpreted as a 10-dimensional spacetime inner product).
  The dilation, which also acts on the $\theta$ component such that the cubic form of 
equation~\ref{vcubic} is invariant, is a component of the full invariance group which  
 is found to be the Lie group $\esig$. In fact the components $(n,X,\theta)$ contain 
the 27 real parameters of an element of $\mcX \in \htho$, that is such an element 
$\mcX$ of equation~\ref{hthoxy} can be written as (\cite{Unifi} equation~6.26, 
\cite{Wang} section~3.3):
\begin{equation}
  \label{xoct3}
  {\mathcal X} \; = \;     
	\left( \!\!\!\!\!\!\!\!\;\! \begin{array}{cc} 
               \begin{array}{cc}   p \,\, & \; \bar{a}  \\ a \,\, & \; m \end{array} 
\!\!\!\!\!\!   &
          \!     \begin{array}{c}    c  \\  \bar{b}   \end{array} \!\! 
				                         \\  
        \;\;\,\,\,\; \bar{c} \;\;\;\;\;\:\, b \!\!\!\!\!\!\!  \begin{array}{cc}        
&   \end{array}  &	  
		  \:  n      \end{array}  \right) \; \equiv \;
	\left( \begin{array}{c|c} 
        \,\,\,\, X                \begin{array}{cc} &  \\  &  \end{array} \!\!\!   &
        \,  \theta  \begin{array}{cc} &  \\  &  \end{array} \!\!\!\!\!\!\!\!\!\! 
				                         \\  \hline
        \,\,\,\,\,\, \theta^{\dagger} \!\! \begin{array}{cc}        &   \end{array}   
&	  
		\,  n      \end{array}  \right)   
	\;	 \inn \htho  
\end{equation}
    with $p,m,n \inn \rrr$, $a,b,c \inn \ooo$, $X \in \htwo$ and $\theta \in \ooo^2$.  
Further, we note that the cubic form
 of equation~\ref{vcubic} is identical to that of equation~\ref{detpmn}
  (\cite{Unifi} equations~6.27 and 6.28 respectively), 
  that is $\mathcal{V}(n,X,\theta) = \det(\mcX)$
  through the correspondence of equation~\ref{xoct3},
 constructed here explicitly as an extension of the quadratic form $\det(X)$ on 
10-dimensional spacetime.

  The construction of the conformal group of the generalised spacetime $\htho$ is 
analogous to that for $\htwc$ in equation~\ref{grada}. It can also be described by a 
Lie algebra $\mathfrak{g}$ with a 3-graded structure, here with:
\begin{eqnarray}
     \mathfrak{g} & = & \mathfrak{g}^{-1} \quad\; \oplus \quad\;\;
	        \mathfrak{g}^{0} \quad\;\; \oplus \quad\; \mathfrak{g}^{+1} \nonumber \\
 \mbox{of dimension:} \quad 133 & = &
                 27 \quad + \quad (78+1) \quad + \quad 27     \nonumber \\ 
 \mbox{with elements:} \qquad  & &  
            U_{\mcY}, \qquad\quad\; S_{\mcY\mcZ}, 
			\qquad\quad\; \tilde{U}_{\mcY} \label{gradx} 
\end{eqnarray}
  now labelled by $\mcY,\mcZ \inn \htho$. The generators $U_{\mcY}$, $S_{\mcY\mcZ}$ 
and $\tilde{U}_{\mcY}$ have similar interpretations as those described for the 
corresponding actions of equation~\ref{grada}, here with the $S_{\mcY\mcZ}$ 
collectively describing the $\det(\mcX)$ preserving group $\esig$ together with a 
further dilation. Explicitly, the actions of these 133 generators upon $\mcX\in \htho$ 
are given by equation~\ref{usuact} with $\bh,\bk,\bl \to \mcX, \mcY, \mcZ$, where here 
the Jordan triple product for elements $\mcX, \mcY, \mcZ \in \htho$ is of the form:
\begin{equation}
  \label{jtphtho}
  \{\mcX,\mcY,\mcZ\} \; = \;
       (\mcX \circ \bar{\mcY})\circ \mcZ 
   + (\mcZ \circ \bar{\mcY}) \circ \mcX
   - (\mcX \circ \mcZ) \circ \bar{\mcY} 
\end{equation} 
in place of equation~\ref{jtphtwc}, where $\bar{\mcY}$ denotes a conjugation
  on $\htho$ (\cite{Gunay3} appendix A). Similarly the Lie bracket relations for 
$U_{\mcY}$, $S_{\mcY\mcZ}$ and $\tilde{U}_{\mcY}$  are closely analogous to those of 
equation~\ref{commhtwc}. Collectively the 133 elements of equation~\ref{gradx} 
generate a symmetry group identified as the exceptional Lie group $\eseg$ which, as 
the conformal group for $\htho$, leaves invariant the generalised cubic `light cone' 
separation:
\begin{equation}
  \label{htholc}
    \det(\mcX \! - \! \mcY) = 0
\end{equation}  
  for $\mcX,\mcY \inn \htho$.
  This is a generalisation from equation~\ref{htwclc} with $\eseg$ acting via a 
non-linear realisation on the components of the 27-dimensional space $\htho$.
 The $\eseg$ conformal group for $\htho$ is also contrasted with SO$(2,10)$
  as the conformal group for $\htwo$ in (\cite{DrayMW} section~6).

   Combining the $\esig$ cubic invariant $\det(\mcX)$ with an appropriate  scalar 
singlet $\alpha \inn \rrr$ the quartic product  $\alpha \det(\mcX)$ is invariant under 
$\esig$ and the additional dilation action. Extending via further components $\mcY \in 
\htho$ and $\beta \inn \rrr$ the quartic norm:
\begin{equation}
   q(x) \; = \; -2\lbrack \alpha\beta - (\mcX,\mcY)\rbrack^2 \, - \,
       8\lbrack\alpha \det(\mcX) + \beta\det(\mcY) - (\mcX^{\sharp},
	                                          \mcY^{\sharp})\rbrack \label{fquartic}									  			 
\end{equation}
 can be constructed using the inner product of equations~\ref{trabimap} and 
\ref{incomp} and the quadratic adjoint of equations~\ref{quadmap} and \ref{hthoshp} 
(\cite{Unifi} section~9.2, with the sign convention for $q(x)$ adopted for example 
from \cite{Borst,Krute,Rios}).
The argument $x$ of equation~\ref{fquartic} belongs to the `Freudenthal triple 
system',  denoted $F(\htho)$, an element of which can be written
  in  `$2 \times 2$ matrix' form:
\begin{equation}
  \label{ftscomp}
   x =  \left(\begin{array}{cc} \alpha & \mcX \\
                          \mcY & \beta \end{array} \right)
\end{equation}
   with $\mcX,\mcY \inn \htho$ of equation~\ref{hthoxy} and $\alpha,\beta \inn \rrr$.
With 56 real components  $F(\htho)$ forms a natural  space for the smallest 
non-trivial representation of the Lie group $\ese$. In fact
the quartic norm $q(x)$ of equation~\ref{fquartic} is invariant under a full set of 
$\eseg$ linear transformations (\cite{Unifi} equations~9.29--9.32) on the 
56-dimensional space $F(\htho)$.

     A non-degenerate bilinear antisymmetric form can also be defined on the space 
$F(\htho)$, on introducing a second element 
 $y = \binom{\gamma \;\: \mcW}{\;\!\!\!\mcZ \;\; \delta} \inn F(\htho)$ with
   $\mcW,\mcZ \inn \htho$ and $\gamma,\delta \inn \rrr$, by:
\begin{equation}
 \{x,y\} \; = \; \alpha\delta \, - \, \beta\gamma \, + \, 
    (\mcX, \mcZ) \, - \, (\mcY, \mcW) \; \in \; \rrr  \label{biasqf}
\end{equation}
   that is also invariant under the $\eseg$ transformations. 
 Further, a symmetric four-linear form can be defined by the linearisation of the 
quartic norm of equation~\ref{fquartic}, that is for $x,y,z,w \in F(\htho)$:
\begin{eqnarray}
 q(x,y,z,w) \!\! & := & \!\!  \frac{1}{24}  \Big[ \, q(x+y+z+w)  \nonumber  \\
     & &  - q(x+y+z) - q(x+y+w) -q(x+z+w) -q(y+z+w) \nonumber  \\
	 & &  + q(x+y) + q(x+z) + q(x+w) + q(y+z) + q(y+w) + q(z+w) \nonumber  \\
	 & &  - q(x) - q(y) - q(z) - q(w) \, \Big]  \label{qxyzw} 
\end{eqnarray}
   such that $q(x,x,x,x) = q(x)$. In turn a symmetric trilinear product $T$ can be 
defined uniquely via equations~\ref{biasqf} and \ref{qxyzw} such that (see for 
example~\cite{Krute} equation~35, following~\cite{Brown}):
\begin{equation}
 \label{tqdef}
    \{T(x,y,z),w\} \, = \, q(x,y,z,w)
\end{equation}
  for any $x,y,z,w \inn F(\htho)$.
  This is the triple product $T(x,y,z) \inn F(\htho)$  from which the Freudenthal 
triple system takes its name. This ternary product is presented explicitly in 
(\cite{Borst2} section 3.3) as:
\begin{equation}
  \label{texpl}
   T(x,y,z) \; = \; \frac{2}{9} \Big[ (x \wedge y)\:\!z
     \,  + \,  (y\wedge z)\:\!x \,  + \, (z \wedge x)\:\!y \Big]
\end{equation}
  with the Freudenthal product $x \wedge y$ between elements of $F(\htho)$ also 
defined by (\cite{Borst2} equations~3.9 and 3.10). 

  An alternative definition for the triple product $T(x,y,z)$ for any elements $x,y,z 
\in F(\htho)$ is given for example in (\cite{Gunay2} section 3.2), following the 
convention of \cite{Faulk}, which is \textit{not} symmetric in the three arguments. In 
turn a four-linear form $q(x,y,z,w)$ can be \textit{defined} in equation~\ref{tqdef} 
via this axiomatically introduced ternary product $T$ and the bilinear form of 
equation~\ref{biasqf}.  This latter definition of $q(x,y,z,w)$ is not symmetric in the 
four arguments, but can be symmetrised as described for (\cite{Gunay4} 
equations~4.1.19 and 4.1.20).
       However here we adopt the former definition of $q(x,y,z,w)$ from 
equation~\ref{qxyzw} and interpret equation~\ref{tqdef} as the definition of 
$T(x,y,z)$, since here the quartic norm $q(x)$ of equation~\ref{fquartic} is of 
central importance.

  While the conformal action of $\ese$, with the generator composition of 
equation~\ref{gradx}, extends the $\esi$ action on the space $\htho$, a 
`quasiconformal' extension of the $\ese$ action on the space $F(\htho)$ can also be 
constructed \cite{Gunay3, Gunay2}. A quasiconformal realisation is generated by a 
5-graded structure for which the spaces $\mathfrak{g}^{\pm 2}$ are each 
one-dimensional, here with (see also \cite{Gunay4,Gunay5}): 
\begin{eqnarray}
     \mathfrak{g} & = & \mathfrak{g}^{-2} \quad \oplus \quad
	             \mathfrak{g}^{-1} \quad \oplus \quad\;
	        \mathfrak{g}^{0} \quad\; \oplus \quad \mathfrak{g}^{+1}
			\quad \oplus \quad \mathfrak{g}^{+2}   \nonumber \\
 \mbox{of dimension:} \quad 248 & = &  1  \quad\, + \quad\,
                 56 \quad\! + \quad\! (133+1) \quad\! + \quad\! 56 
				   \quad\, + \quad\, 1    \nonumber \\ 
 \mbox{with elements:} \qquad  & & K_{\rho}, \qquad\quad 
            U_{y}, \qquad\quad S_{yz}, 
			\qquad\quad \tilde{U}_{y},
           		\qquad\quad  \tilde{K}_{\rho}	 \label{gradxf}
\end{eqnarray}
  labelled by $y,z \in F(\htho)$ and $\rho \in \rrr$. The elements $S_{yz}$ include 
the 133 generators of $\ese$ together with a dilation $\Delta$; $U_y$ and 
$\tilde{U}_y$ are analogous to the corresponding generators described for 
equations~\ref{grada} and \ref{gradx}; while $K_{\rho}$ and $\tilde{K}_{\rho}$ 
 are each one-dimensional and  together with $\Delta$ form a distinguished 3-element 
closed $\mbox{sl}(2,\rrr)$ subalgebra (\cite{Gunay5} section~3).

 Collectively the 248 elements of equation~\ref{gradxf} form an $\ee$ Lie algebra and 
generate the real form $\eeg$ of this largest exceptional Lie group acting via a 
non-linear realisation on the `extended Freudenthal triple system', denoted 
$eF(\htho)$. This is a 57-dimensional space with elements such as 
  $e=(x,\tau)\in eF(\htho)$, with $x\in F(\htho)$ and $\tau \in \rrr$
 associated respectively with the grade $+1$ and grade $+2$ subspaces in 
equation~\ref{gradxf}
   \cite{Gunay3,Gunay2,Rios,Gunay4}.
    Both the quartic norm $q(x)$ of equation~\ref{fquartic} and the extra real 
parameter $\tau \in \rrr$  are invariant under the $\eseg \subset \eeg$ subgroup. The 
quartic symplectic distance between any two elements $e = (x,\tau), f = (y, \kappa) 
\in eF(\htho)$ is defined by (\cite{Rios} equation~61):
\begin{equation}
  \label{qusydi}
   d(e,f) \; = \; q(x-y) \, - \, (\tau - \kappa \, + \, \{x,y\})^2
\end{equation} 
  with the full set of $\eeg$ actions on the space $eF(\htho)$ leaving invariant the 
generalised light cone:
\begin{equation}
 \label{defcone}
   d(e,f) \; = \; 0
\end{equation} 

  This `quartic light cone' is hence a further generalisation from the cubic light 
cone of equation~\ref{htholc} and the quadratic light cone of equation~\ref{htwclc}.
  While a non-zero symplectic distance $d(e,f)$ may be transformed up to an overall 
factor by the  generators of $\eeg$, the invariance of the 57-dimensional light cone 
in equation~\ref{defcone} under the full group $\eeg$ justifies the term 
`quasiconformal' realisation. For example the dilation $\Delta$ scales the components 
in terms such as $q(x-y)$, $\tau-\kappa$ and $\{x,y\}$ in equation~\ref{qusydi} in the 
appropriate proportions such that equation~\ref{defcone} is invariant.

  The transformations of the 248 generators of equation~\ref{gradxf}, generalising 
from equation~\ref{usuact} and here acting on any element $e=(x,\tau) \in eF(\htho)$, 
are presented explicitly in (\cite{Rios} equation 63, following \cite{Gunay3} equation 
29) and include for example $\tilde{K}_{\rho}(e)$ with the non-linear actions:  
\begin{equation}
\label{kkact}
   \tilde{K}_{\rho}(x) = -\frac{1}{6}\rho \;\! T(x,x,x) + \rho\:\! x \tau,  \qquad 
    \tilde{K}_{\rho}(\tau) = \frac{1}{6}\rho\:\! \{T(x,x,x),x\} + 2\rho \:\!\tau^2
\end{equation}
  Care is needed for differences in notation and the consistency of the definitions 
involving the asymmetric quadratic form and ternary product as discussed following 
equation~\ref{tqdef} in the construction of this non-linear realisation of $\ee$. 
The full $\ee$ Lie algebra bracket itself is listed for example in \cite{Gunay3, 
Gunay5}.

   Different real forms of $\ee$ can also be described in this way.
  Employing the above analysis for the Freudenthal triple system $F(\htho_s)$, defined 
over the split octonions $\ooo_s$, leads to the real form E$_{8(8)}$ as the 
corresponding quasiconformal group \cite{Gunay3,Gunay4}. This group contains 
  $\mbox{E}_{7(7)} \subset \mbox{E}_{8(8)}$ as the conformal subgroup and in turn
  E$_{6(6)}$ as the reduced structure group of $\htho_s$.
  On the other hand the real form of interest here $\eeg$, with the subgroup chain 
  $\esig \subset \eseg \subset \eeg$, is obtained on employing the non-split octonion 
algebra $\ooo$ and the Freudenthal triple system $F(\htho)$. This difference is 
analogous to that discussed for the `magic square' towards the end of 
subsection~\ref{ee21} for which the real form $\eeg$ or E$_{8(8)}$ obtained depends on 
the choice of $\kkk'=\ooo$ or $\kkk'=\ooo_s$ respectively.
 On the physics side, applications of the real forms of the exceptional Lie algebras 
discussed above in supergravity theories are described in
  \cite{Gunay3,Gunay2,Gunay4,Gunay5}.

   The conclusion of most relevance for the physical theory to be considered in this 
paper is that in progressing from the Lorentz symmetry of 4-dimensional spacetime, 
represented by the double cover $\sltc$ acting on the space $\htwc$ as described in 
the opening of this subsection, via a sequence of generalised spacetimes with norm or 
generalised light cone preserving symmetries, we are led ultimately to the largest 
exceptional Lie group $\ee$, which leaves equation~\ref{defcone} invariant. This 
distinguished role for $\ee$ adds to the unique properties of this group reviewed in 
subsection~\ref{ee21}, here with a tentative connection to physics through the notion 
of a `generalised spacetime'. By contrast, in the following section we motivate the 
conception of a `general form of time', and then in section~\ref{ee4} we shall propose 
that certain structures described in this subsection might rather be interpreted as 
multi-dimensional temporal forms, leading in section~\ref{ee5} to the proposal that 
this progression may lead to $\eeg$ as the full symmetry of time.


\section{The General Form of Time}
\label{ee3}

    The notion of an observer drifting through space in a spacecraft with the engines 
turned off, or located within a freely falling lift near the surface of the Earth, 
following a \mbox{world line} parametrised by a real proper time variable $s$ 
recording the progression along a trajectory in 4-dimensional spacetime, with an 
apparent local absence of any force of gravity, is central to Einstein's theory of 
general relativity. This idea is encapsulated in the `equivalence principle' (see for 
example~\cite{Unifi} section~3.4), the strong form of which states that at any 
location in spacetime, in the limit of arbitrarily small spacetime volumes, local 
inertial coordinates $(x^0,x^1,x^2,x^3)$ can be constructed within which special 
relativity holds for all laws of physics other than gravity.
  The proper time interval $\delta s$, for $\delta s \to 0$ in such a local inertial 
frame, can be expressed with a local Minkowski metric as:
\begin{equation} 
 \label{propint}
  \delta s^2 = (\delta x^0)^2-(\delta x^1)^2-(\delta x^2)^2-(\delta x^3)^2
\end{equation}       
   with the coordinates transforming under a local Lorentz symmetry that leaves 
$\delta s$ invariant. With the local gravitational force vanishing, as seen from any 
other location nearby objects within the same inertial frame (which in practice 
extends to a very good approximation over finite spacetime volumes such as the 
vicinity of the spacecraft or interior of the lift in the above examples) will appear 
to `fall' together (such as a person and a ball in the freely falling spacecraft or 
lift), hence following a preferred extended path in spacetime independent of the 
constitution of the falling objects. This implies that the global properties of the 
gravitational field can be ascribed to the global structure of spacetime itself, as 
formulated via the curvature tensor and Einstein's field equation in the theory of 
general relativity. 

  For the theory presented in this paper we begin with an even simpler structure than 
a local inertial frame in spacetime, as expressed by the coordinate intervals 
  $(\delta x^0, \delta x^1, \delta x^2, \delta x^3)$ on the right-hand side of 
equation~\ref{propint}, and take the irreducible element of the theory to be simply 
the interval of proper time $\delta s$ on the left-hand side of that equation.
 Having stripped this structure down to one dimension of time only a theory can be 
constructed which in some sense generalises general relativity for the notion of an 
observer `drifting through' a higher-dimensional parameter space, over and above 
4-dimensional spacetime, as an expression of and deriving directly from the 
one-dimensional temporal progression of the observer, as we describe in the following.

  While in Newtonion physics time $(x^0)$ and space $(x^1,x^2,x^3)$ are independent, 
in Einstein's relativity they are drawn together in a 4-dimensional spacetime element 
through the Lorentz invariant proper time interval $\delta s$ in 
equation~\ref{propint}. Here, in placing the emphasis upon this one-dimensional flow 
of time, rather than a specific spacetime structure, we aim to combine 4-dimensional 
spacetime together with `extra dimensions' collectively in a higher-dimensional 
structure with a full symmetry (augmenting the Lorentz group) acting upon a 
\textit{general form of time}.
 A simple generalisation of equation~\ref{propint} can be written, balancing the order 
of the infinitesimal elements on each side, as the homogeneous $p^{\mathrm{th}}$-order 
polynomial:
\begin{equation} 
 \label{propgen}
  \delta s^p = \alpha_{abc\ldots}\delta x^a \delta x^b \delta x^c \ldots
\end{equation}  
   for any integer power $p\ge 1$ and each $\alpha_{abc\ldots} = -1,0$ or 1 with 
 indices  $a,b,c = 1, \ldots ,n$ for an $n$-parameter space. This general expression 
naturally contains equation~\ref{propint} as a particular quadratic case for $p=2$, 
that is:
\begin{equation} 
 \label{propeta}
  \delta s^2 = \eta_{ab }\delta x^a \delta x^b
\end{equation} 
  with $4 \times 4$ Minkowski metric $\eta = \mbox{diag}(+1,-1,-1,-1)$ and here with 
the index convention $a,b = 0,1,2,3$. This 4-dimensional spacetime form can be 
embedded in higher-dimensional homogeneous polynomials in the form of 
equation~\ref{propgen} which are \textit{not} restricted to the quadratic structure of 
extra \textit{spatial} dimensions. In fact, since we do not `see' the extra dimensions 
there is no compelling reason for them to be artificially limited to quadratic  
extensions of equation~\ref{propeta} with a higher-dimensional Minkowski metric and 
corresponding local Euclidean properties for the additional `spatial' components. 
Cubic and higher-order polynomial forms are equally permitted given the emphasis upon 
generalising the form of the proper time interval on the left-hand side of these 
equations. 

     This approach is hence distinct from, and more general than, the class of models 
based purely on extra spatial dimensions as initiated by Kaluza and Klein 
\cite{Kaluza,Klein}, as described in \cite{KKone}.  As for Kaluza-Klein models matter 
fields and physical structures in 4-dimensional spacetime will be associated with the 
extra-dimensional components, however here the properties directly deriving from these 
components will differ from Kaluza-Klein theory  owing to the more general form of 
equation~\ref{propgen}. The ultimate goal will then be to assess the degree to which 
derived properties of the resulting matter fields match empirically observed phenomena 
for the theory presented here based upon the general form of time.

   We note that given we have expressed an interval of time $\delta s$ as a 
homogeneous polynomial in equation~\ref{propgen} in principle a similar decomposition 
could also be applied to any of the parameter intervals $\delta x^a$. However if we 
were to substitute for example $(\delta x^1)^q = 
\beta_{ijk\ldots}\delta y^i \delta y^j \delta y^k \ldots$ 
as a $q^{\mathrm{th}}$-order homogeneous polynomial, having a structure analogous to 
equation~\ref{propgen}, 
into equation~\ref{propgen} itself we could express the latter equation in the form 
$\delta s^{(pq)} = \alpha_{abc\ldots}(\delta x^a)^q (\delta x^b)^q (\delta x^c)^q 
\ldots$, 
 again balancing the order of the infinitesimal elements. However this new 
$(pq)^{\mathrm{th}}$-order homogeneous polynomial is of the same form as 
equation~\ref{propgen} with $p \to pq$, except with a more restrictive structure,
 and hence such particular cases are already implicitly incorporated. We shall 
primarily be interested in forms of time with a high degree of symmetry, avoiding 
preferred components such as $\delta x^1$ above that might be distinguished in this 
way, and hence equation~\ref{propgen} can be consistently adopted as the most general 
form of time.

  The conceptual basis of the theory is also described in \cite{Unifi,Novel,KKone} 
where it is also noted that in order to avoid dealing directly with infinitesimal 
quantities such as $\delta s$ and $\delta x^a$ we define the differentials 
  $v^a = \frac{dx^a}{ds} = \frac{\delta x^a}{\delta s} 
   {\big{\vert}}_{\mbox {\tiny $\delta s \! \to \! 0$}}$. Hence upon dividing both 
sides of equation~\ref{propgen} by $\delta s^p$ and taking the limit $\delta s \to 0$, 
we have:
\begin{eqnarray}
   \delta s^p & = & \alpha_{abc\ldots}\delta x^a \delta x^b \delta x^c \ldots  
      \nonumber \\
   \Rightarrow \qquad 1 & = & 
    \alpha_{abc\ldots}\:\, v^a\;\: v^b\;\: v^c \ldots    \nonumber \\
   \mbox{written as:} \quad   L(\bv_n) \!\! & \,\,  = \!\! & \,\, 1   \label{lvo}
\end{eqnarray}
  where $L(\bv_n) = \alpha_{abc\ldots}v^a v^b v^c \ldots $ is a homogeneous polynomial 
in the generally finite components $(v^1, v^2, \ldots ,v^n)$ of the $n$-dimensional 
vector  $\bv_n \in \rrr^n$.
 Technically, the $\delta x^a$ and $\delta s$ are considered to be `infinitesimals of 
the same order' in defining the components $v^a =  \frac{\delta x^a}{\delta s} 
   {\big{\vert}}_{\mbox {\tiny $\delta s \! \to \! 0$}}$, which is reasonable since 
the $x^a$ express the flow of time $s$ itself, with $s=x^1$ for the trivial 
one-dimensional case. The simple general expression of equation~\ref{lvo} for the 
multi-dimensional form of time, and its symmetries,  will provide the basis for a full 
unified field theory. 

   Such a physical theory is obtained initially on applying a local translation 
symmetry of equation~\ref{lvo} to a substructure of four components $\delta x^a$ 
exhibiting the quadratic form on the right-hand side of equation~\ref{propeta} in 
order to construct a local inertial frame in 4-dimensional spacetime, as will be 
described in subsection~\ref{ee41}. This preferential treatment of four components is 
necessary to identify an external spacetime within which all observations and 
experiments are framed and to provide the necessary background for any physical 
structures to be `seen' at all. In providing the mathematical framework through which 
physical structures can be observed in space as well as time, the full symmetry of the 
general form of time of equation~\ref{lvo} is necessarily broken, as described for 
example in (\cite{KKone} subsection 2.3) and in the follow section of this paper, 
completing the basic conceptual picture upon which a full physical theory can be 
developed.

  Historically the most successful physical theories, including Newtonian mechanics, 
Maxwell's equations, the Dirac equation, quantum theory and general relativity, have 
in common a notion of a  continuous flow of  time that can be parametrised by a real 
number $s \in \rrr$. It is this feature alone that we have taken as our starting 
point. Founding the theory purely on the notion of a continuous progression in time, 
through which all observations of the physical world are made, marks a minimal and 
conservative basis for a physical theory. On taking the infinitesimal limit the 
general expression for the multi-dimensional form of time of equation~\ref{propgen} or 
\ref{lvo} follows directly from the basic arithmetic composition of the real line. It 
is \textit{through} these defining characteristics of the real numbers, as 
representing the one-dimensional continuum of time, that arithmetic forms together 
with their associated symmetries can be identified which in principle describe both 
the properties of external spacetime together with the matter fields it contains, all 
carried \textit{simultaneously} within the flow of time itself.    

  The mathematical structure of the theory hence originates from the real number 
parametrisation of the continuous flow of time that permeates all of our experiments 
in, and observations of, the world around us. With the structure of spacetime itself 
deriving from the symmetries of forms of time, and the properties of matter deriving 
from multi-dimensional temporal forms over and above that needed to describe 
4-dimensional spacetime, in principle the conceptual basis of the theory presents an 
opportunity to account for the apparently `unreasonable effectiveness of mathematics 
in the natural sciences' \cite{Wign}. That is, the mathematical origins of the theory 
are \textit{anchored} in the simple structure of the one-dimensional progression of 
time, through which we necessarily observe the world, that can be identically 
expressed in the multi-dimensional form of equation~\ref{lvo}, through which the 
physical properties of matter are proposed to derive directly.

  An historical precedent for founding a general mathematical structure on the notion 
of the temporal continuum can be found in the work of William Rowan Hamilton. These 
ideas were influenced by the relation between time and continuous progression that had 
been so successfully employed in Newton's method of fluxions in underpinning Newtonian 
mechanics, and were being considered by Hamilton around the same time that he was 
developing his own more general formulation of mechanics. In the introduction to his 
lengthy paper of 1837 \cite{Ham1} Hamilton contrasts three general approaches to the 
study of algebra; namely practical, symbolic and theoretical. In seeking theoretical 
clarification for the `science of algebra' Hamilton proposed that the notion of time 
might provide such a basis, complementing the relation between the `science of 
geometry' and the concept of space. Seemingly influenced also by the ideas of the 
philosopher Immanuel Kant half a century earlier, regarding the \textit{a priori} 
necessity of the forms of space and time as innate structures through which the world 
is perceived \cite{Kant}, Hamilton comments (\cite{Ham1} in the `General Introductory 
Remarks', with the upper-case of the original):

\begin{quotation}
    The notion or intuition of ORDER IN TIME is not less but more deep-seated in the 
human mind, than the notion or intuition of ORDER IN SPACE; and a mathematical Science 
may be founded on the former, as pure and as demonstrative as the science founded on 
the latter. There is something mysterious and transcendent involved in the idea of 
Time; but there is also something definite and clear: and while Metaphysicians 
meditate on the one, Mathematicians may reason from the other.
\end{quotation}

  With the properties of the real line $\rrr^1$ considered to represent directly the 
continuum of time, in developing this idea
   Hamilton introduced `number-couples' $(a_1,a_2) \in \rrr^2$ associated with two 
\textit{independent} steps in time, termed primary and secondary but \textit{not} 
mutually related by succession in the same one-dimensional progression.
 Basic arithmetic operations $(+,-,\times, \div )$ were then defined between the 
number-couples in the spirit of `an algebra of pure time' by analogy with such 
operations for $\rrr^1$ alone, while also being guided by considerations of 
simplicity. He found the resulting properties to be equivalent to the arithmetic of 
the complex numbers, elements of which can be written as $a_1 + a_2 i \inn \ccc$.
 Noting that the \textit{imaginary} unit $i=\sqrt{-1}$, denoting an `impossible 
extraction', can be represented by the \textit{real} number-couple $(a_1,a_2) = (0,1) 
= \sqrt{(-1,0)}$ provided some of the 
 justification for this approach. 
 The significant observation for the present paper is that Hamilton's number-couples 
$(a_1,a_2)$ effectively express \textit{time} progressing as a  \textit{two}-parameter 
entity.

  Having been unable to construct a system of `number-triplets' in a similar spirit, 
Hamilton conceived a 4-dimensional algebra, known as the quaternions $\hhh$, in 1843 
while speculating upon `an additional illustration of his view respecting the Science 
of Pure Time'~\cite{Ham3}. However, with the multiplication of the three quaternion 
imaginary units $i,j,k$ being non-commutative and having a natural geometric 
interpretation, the new algebra immediately became associated more with the properties 
of space. Hamilton subsequently focussed upon promoting the quaternions for their 
practical applications in this geometrical sense, rather than with regard to his 
earlier more theoretical view as an algebra of time. 

  In the mid-1840s, shortly after Hamilton's first presentation of the quaternions, 
Graves and Cayley independently discovered the 8-dimensional octonions $\ooo$ as a 
further generalisation. This latter development was made by considering the symbolic 
manipulation of imaginary units, of which there are seven for the octonions as 
described for equation~\ref{octa} in subsection~\ref{ee22}. This completed the set of 
four normed division algebras $\rrr$, $\ccc$, $\hhh$ and $\ooo$, which were proven to 
be uniquely the only such algebras in 1898 by Hurwitz (see for example \cite{Baez1}). 
In the case of the discovery of the octonions it took over one hundred years of 
developments in physics until this largest division algebra began to be taken 
seriously for possible applications in a scientific theory, as noted in 
subsection~\ref{ee22} and illustrated by the examples of 
\cite{Pais,Gamb,Gunay,SoLo,Dix1,Borst,Man4}.

   The point of view adopted in this paper is not to regard the continuum of time as a 
foundation of a \textit{mathematical algebra} in isolation but rather as the starting 
point for the mathematical description of a full \textit{physical theory}, while 
maintaining some of the philosophical influences as alluded to for Hamilton above. 
Such a physical theory can be developed from the basic arithmetic composition of the 
real numbers $\rrr$ as embodying the structure of the continuous flow of time. In 
contrast to Hamilton's invention of algebraic rules for composing finite
 number-couples $(a_1,a_2)$ by analogy with the algebraic properties of 
one-dimensional time, here we simply write down a \textit{direct} and \textit{exact} 
identity for the general multi-dimensional form of time expressed for the limit of 
infinitesimal intervals in equation~\ref{propgen}. This expression can then be written 
in terms of the finite components $v^a = \frac{dx^a}{ds}$ as described for 
equation~\ref{lvo}.

  Any of the four division algebras $\kkk = \rrr$, $\ccc$, $\hhh$ or $\ooo$ can be of 
significance for forms of time owing to the norm compatibility of the composition of 
their elements, as described for equation~\ref{abnorm}. For example with $a,b \in 
\kkk$ and $\bv_n = b$ for $n=1,2,4$ or 8, equation~\ref{lvo} could take the quadratic 
form $L(\bv_n) = \vert b \vert^2 = b\bar{b} = 1$ and the mapping $b \to ab$ for $\vert 
a \vert  =1$ then represents a symmetry leaving the form $\lvn$ invariant. As noted in 
equation~\ref{abdet} at the end of subsection~\ref{ee22} this norm-preserving property 
is shared by the determinant for matrix algebras, and hence we are naturally drawn to 
consider matrices over the division algebras to identify possible forms for $\lvn$ and 
the corresponding symmetry. In particular, while being non-associative, compositions 
involving octonions can incorporate a high degree of symmetry in a compact algebraic 
form, as noted in subsection~\ref{ee22}. Hence it is a curious observation that the 
octonion algebra, which will feature heavily in the `symmetry of time' for the present 
theory, was discovered historically via a short sequence of developments initiated by 
Hamilton's ambitions concerning the `algebra of time'. 

   Considered as a form of time the
   4-dimensional quadratic `spacetime' form in equation~\ref{propeta}, with Minkowski 
metric, can be written via equation~\ref{lvo} as $\lvf$.  As noted for 
equation~\ref{hinvfour}  the Lorentz symmetry $\soot$ of 
  $L(\bv_4) = \vert \bv_4 \vert^2 = 1$ can be represented by its double cover $\sltc$ 
acting on the space of $2\times 2$ matrices $\htwc$, via:
\begin{equation}
 \label{lvfr}
  L(\bv_4) \; = \; \eta_{ab} v^av^b  \; = \;  \det (\bh)  \;= \; 1 
   \qquad \mbox{with} \qquad  \bv_4 \equiv \bh \in \htwc
\end{equation}
  While the structure of such a metric or the norm of a division algebra is limited to 
a quadratic form, on employing determinants of matrices higher-order polynomial forms 
for equation~\ref{lvo} may be introduced. For example the spaces h$_m\kkk$ of
 $m\times m$ Hermitian matrices over $\kkk$ can be considered and in particular, via 
the space $\hthc$ or $\htwo$ as described for (\cite{KKone} equation~95), we can 
construct the homogeneous cubic form:
\begin{equation}
 \label{lvts}
  L(\bv_{27}) \; = \;  \det (\mcX)  \;= \; 1 
   \qquad \mbox{with} \qquad  \bv_{27} \equiv \mcX \in \htho
\end{equation}
  As noted for equations~\ref{detpmn} and \ref{vcubic} this 27-dimensional form has an 
\mbox{$\esig \equiv \sltho$} symmetry, which can be constructed in terms of the 
actions of octonion-valued matrices. In turn this cubic form further embeds in the 
homogeneous quartic form:
 \begin{equation}
  \label{lvfs}
  L(\bv_{56}) \; = \;  q(x)  \;= \; 1 
   \qquad \mbox{with} \qquad  \bv_{56} \equiv x \in F(\htho)
\end{equation} 
  which has an $\eseg$ symmetry, as described for equation~\ref{fquartic}.
  Given the minus signs in equation~\ref{fquartic} and the fact we are looking for 
expressions of the form $L(\bv_n) = +1$ a change of sign $q(x) \to -q(x)$, in place of 
the convention for equation~\ref{fquartic}, might be more appropriate, but would not 
change the results to be presented in this paper. We also note that the appearance of 
integer coefficients differing from $\pm 1$ or 0 in equation~\ref{fquartic}, even 
after a possible scaling of the components, is compatible with the general form of 
equation~\ref{lvo} with coefficients $\alpha_{abc\ldots} = -1,0$ or +1 owing to the 
multiple opportunities for double counting in the index summations over the $n$ real 
components.

   Hence the
 spaces $\htho$ and $F(\htho)$ of the above homogeneous polynomial forms of 
equations~\ref{lvts} and \ref{lvfs}, rather than representing structures considered as 
`generalised spacetimes' as reviewed in the previous subsection, can be interpreted as 
spaces underlying higher-dimensional \textit{forms of time}, that is 
equation~\ref{lvo} with $n=27$ and $n=56$ respectively.

 Constructed from the original one dimension of time the form of 4-dimensional 
spacetime in equation~\ref{lvfr} is itself embedded as an intermediate form of time of 
particular significance. This significance arises from the need to identify an arena 
of space as well as time, with the local geometrical properties described by this 
4-dimensional form and its symmetries, within which observations are made.
 That is, from the philosophical perspective, equation~\ref{lvfr} underlies the 
\textit{a priori} forms of space and time through which the world is perceived.
  The higher-dimensional forms of time of equations~\ref{lvts} and \ref{lvfs} 
incorporate this 4-dimensional spacetime \textit{together with} a structure of `extra 
dimensions'.  
 The necessary imposition of an extended 4-dimensional spacetime arena is then central 
to the breaking of the 
 symmetry of the higher-dimensional forms of time, with the properties of the residual 
extra dimensions identified as matter fields in spacetime. This symmetry breaking, and 
the physical structures of matter over the extended spacetime manifold deriving from 
it, will be explicitly described in the following section.


\section{Symmetry Breaking and the Standard Model}
\label{ee4}

   In this section we aim to demonstrate how particular features of the Standard Model 
of particle physics emerge in the present theory for the 27-dimensional form of time 
of equation~\ref{lvts} and accumulate further for the 56-dimensional form of time of 
equation~\ref{lvfs}. These two cases will be presented in subsections~\ref{ee42} and 
\ref{ee43} respectively, summarising and further analysing the progress made in
 (\cite{Unifi} chapters 6--9, \cite{Novel}). In the first subsection below we first
 consider how an extended spacetime itself can be identified from the symmetries of 
the 4-dimensional form of time of equation~\ref{lvfr}. We then 
  review a  minimal extension from this 4-dimensional spacetime form in order to 
describe the mechanism of symmetry breaking for this simpler case (following 
\cite{KKone} section~2.3), for which, nevertheless, a non-trivial physical structure 
will be identified.

\subsection{Extended Spacetime and $\slthc$ Symmetry}
\label{ee41}

  The structure of a local inertial frame, central to general relativity as described 
in the opening of section~\ref{ee3}, can be constructed in a straightforward manner 
from the arithmetic properties of an interval of time alone. The form of time $\lvf$ 
of equation~\ref{lvfr} not only possesses an $\soot$ Lorentz symmetry but also an 
$\rrr^4$ translation symmetry with:
\begin{eqnarray}
    L(\bv_4) \!\! & = & \;\;\;\, 
	       \left( v^0 \right)^2 \:\;\quad - \:\;\quad \left( v^1 \right)^2 
		   \:\;\quad  - \:\;\quad \left( v^2 \right)^2
		    \:\;\quad - \:\;\quad \left( v^3 \right)^2  \nonumber \\
& = &  \;\,  \left( \frac{dx^0}{ds} \right)^{\!\! 2} \;\!\!\quad - \;\!\!\quad
	     \left( \frac{dx^1}{ds} \right)^{\!\! 2} \;\!\!\quad - \;\!\!\quad
	     \left( \frac{dx^2}{ds} \right)^{\!\! 2} \;\!\!\quad - \;\!\!\quad  
    	 \left( \frac{dx^3}{ds} \right)^{\!\! 2}     \nonumber   \\
& = & \!\!\!\! \left(\! \frac{d(x^0\!+\!r^0)}{ds} \!\right)^{\!\! 2} \!-\!
	       \left(\! \frac{d(x^1\!+\!r^1)}{ds} \!\right)^{\!\! 2} \!-\!
	       \left(\! \frac{d(x^2\!+\!r^2)}{ds} \!\right)^{\!\! 2} \!-\!
		   \left(\! \frac{d(x^3\!+\!r^3)}{ds} \!\right)^{\!\! 2}  \, = \: 1  
		   \qquad   \label{fourtran}
\end{eqnarray}
  for any constant $\br_{\!\!\: 4} = (r^0,r^1,r^2,r^3) \inn \rrr^4$. These symmetries 
are pictured in figure~\ref{fourd}, which exhibits the basic geometric properties of a 
local inertial frame of an empty 4-dimensional spacetime. This demonstrates how the 
structure of an extended spacetime can itself be derived from the one dimension of 
time alone, given that the time interval $\delta s$ on the left-hand side of 
equation~\ref{propint} and figure~\ref{fourd} is interpreted as the fundamental 
element of the theory.
\begin{figure}[htbp]  
\centering
\epsfxsize=14.3cm
\leavevmode
\epsffile[0 0 2341 873]{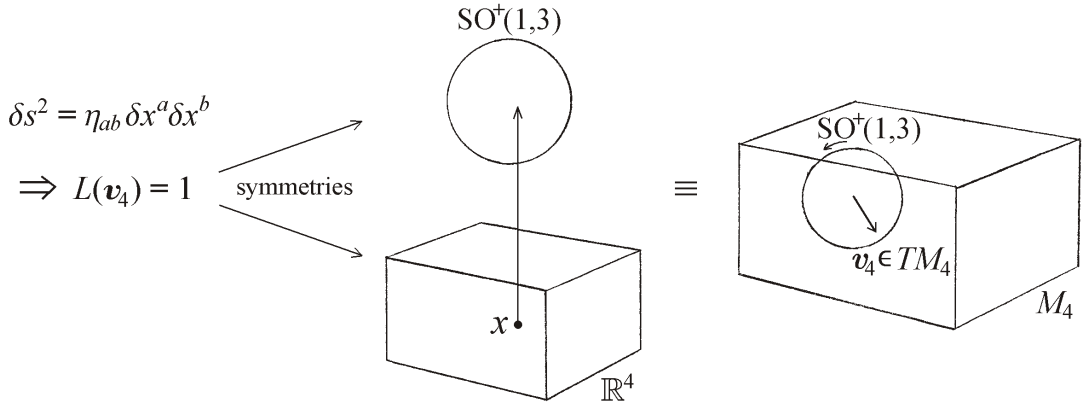}
\caption{\setb A one-dimensional time interval $\delta s$ expressed in the form of 
equation~\ref{propeta} or \ref{lvfr} possesses the symmetries described for 
equation~\ref{fourtran}
 which parametrise a 4-dimensional spacetime volume element
  $M_4 \equiv \rrr^4$ of arbitrary extension that can be interpreted as describing a 
matterless vacuum state.}
\label{fourd}
\end{figure}  

  The manner in which 4-dimensional spacetime can `pop-up' out of the one-dimensional 
flow of time is analogous to the construction of the 3-dimensional creations of 
origami from the folding of a 2-dimensional sheet of paper. One significant difference 
is that here spacetime is \textit{itself} created from the original one-dimensional 
element, rather than being constructed within a pre-existing space as for the origami 
analogy. The differing mathematical structure for the two cases is however similarly 
simple to describe, as expressed through equation~\ref{fourtran} and 
figure~\ref{fourd} here.

  The essential point is that the 4-dimensional spacetime element of 
figure~\ref{fourd} expresses directly the symmetries of $\lvf$ which in turn derives 
directly from the \textit{equality} for $\delta s$ in equations~\ref{propint} and 
\ref{propeta}  
 \textit{without adding anything to} the flow of time itself. By further exploring 
this idea we shall find that the structure of matter in spacetime can also be
 `enfolded' within higher-dimensional forms of time, as we describe in the following.   
  
   The cubic form of time to be considered here represents the simplest and most 
direct generalisation of the  4-dimensional quadratic form of time in 
equation~\ref{lvfr} to a higher-order homogeneous polynomial based on the determinant 
of a matrix.  
  This is achieved by first expressing the Lorentz 4-vector 
   $\bv_4 = (v^0,v^1,v^2,v^3)$ as an element $\bh \in \htwc$ via the Pauli matrices, 
as described for equation~\ref{hinvfour} in the opening of subsection~\ref{ee23} and 
utilised in equation~\ref{lvfr},
 and then simply  embedding this $2 \times 2$ Hermitian complex matrix  inside a $3 
\times 3$ Hermitian complex matrix as:
\begin{equation}
  \label{hinvnine}
     \begin{array}{cc} 
     \htwc \ni  \bh  = \left(   
	   \begin{array}{cc} v^0 + v^3 & v^1 - v^2i \\
		   v^1 + v^2i  & v^0 - v^3  \end{array} \right) \!\!\!   &
        \,   \mbox{{\Large $\to$}}  \begin{array}{cc} &  \\  &  \end{array} 
\!\!\!\!\!\!\!\!\!\! 
				                         \\  
         \begin{array}{cc}        &   \end{array}   	        
		   \end{array}  
	\left( \begin{array}{c|c} 
        \,\,\,\, \bh \!     \begin{array}{cc} &  \\  &  \end{array} \!\!\!   &
        \,  \psi  \begin{array}{cc} &  \\  &  \end{array} \!\!\!\!\!\!\!\!\!\! 
				                      \\ \hline
        \,\,\,\,\,\;\! \psi^{\dagger} \!\!\! \begin{array}{cc}  
		        &   \end{array}   &	  
		\,  n      \end{array}  \right)   
	\; \equiv \bv_9	 \inn \hthc  
\end{equation}
  Here there are five real parameter `extra dimensions', which could be labelled
   for example by
    $(v^4,v^5,v^6,v^7,v^8)$ in the form of $\psi = \binom{v^4 + v^5i}{v^6+v^7i}
	 \in \ccc^2$ and $n=v^8 \in \rrr$ (with the notation $n \inn \rrr$ consistent with 
equation~\ref{xoct3} used here). This structure then represents a full 9-dimensional 
cubic form of time:
\begin{equation}
 \label{lvni}
  L(\bv_9) \; = \;  \det (\bv_9)  \;= \; 1 
   \qquad \mbox{with} \qquad  \bv_9  \in \hthc
\end{equation}
  which is invariant under an $\slthc$ symmetry, as a direct augmentation of the 
$\sltc$ symmetry of $\lvf$ in equation~\ref{lvfr}. (This is the `Lorentz group'
 $\slthc$ of (\cite{Gunay2} equation~5) with $\hthc$ there considered a `generalised 
spacetime', while here $\slthc$ is the symmetry of a `general form of time').

   In general $n$-dimensional translation symmetries, similar to 
equation~\ref{fourtran}, can also be identified for any form $\lvn$ since $x^a$ for 
each component $v^a = \frac{dx^a}{ds}$ implicitly represents any value $x^a \in \rrr$ 
of the real line, as described for example in (\cite{KKone} figure~1 and equation~5). 
The \textit{necessary} identification of an external 4-dimensional spacetime manifold 
$M_4$ can be realised through a preferred choice of four components for which this 
translation symmetry is explicitly employed. That is, for the case of the full form 
$\lvni$ of equation~\ref{lvni} we note that for the subspace vector
  $\bv_4 \equiv \bh \in \htwc \subset \hthc$ we can take, explicitly: 
\begin{eqnarray}
 \label{lvnitr}           
  L(\bv_9)  & = & \det 
	\left( \begin{array}{cc|c} 
	 \frac{dx^0}{ds} + \frac{dx^3}{ds} 
   & \frac{dx^1}{ds} - \frac{dx^2}{ds}i  
   & \frac{dx^4}{ds} + \frac{dx^5}{ds}i    \\
	 \frac{dx^1}{ds} + \frac{dx^2}{ds}i 
   & \frac{dx^0}{ds} - \frac{dx^3}{ds}
   & \frac{dx^6}{ds} + \frac{dx^7}{ds}i   \\  \hline
     \frac{dx^4}{ds} - \frac{dx^5}{ds}i 
   & \frac{dx^6}{ds} - \frac{dx^7}{ds}i  
   & \frac{dx^8}{ds}    \end{array}  \right) 
				\nonumber	 \\		
		& &		\nonumber \\
       = \;\; \det \!\!\!\!\!\!  & \!\!\!\!\! & \!\!\!\!\!\!\!\!
	  \left( \begin{array}{cc|c} 
	 \frac{d(x^0+r^0)}{ds} + \frac{d(x^3+r^3)}{ds} 
   & \frac{d(x^1+r^1)}{ds} - \frac{d(x^2+r^2)}{ds}i  
   & \frac{dx^4}{ds} + \frac{dx^5}{ds}i    \\
	 \frac{d(x^1+r^1)}{ds} + \frac{d(x^2+r^2)}{ds}i 
   & \frac{d(x^0+r^0)}{ds} - \frac{d(x^3+r^3)}{ds}
   & \frac{dx^6}{ds} + \frac{dx^7}{ds}i   \\  \hline
     \frac{dx^4}{ds} - \frac{dx^5}{ds}i 
   & \frac{dx^6}{ds} - \frac{dx^7}{ds}i  
   & \frac{dx^8}{ds}    \end{array}  \right) 
	   = 1 						  
\end{eqnarray}     
   for any constant $\br_{\!\!\: 4} = (r^0,r^1,r^2,r^3) \in \rrr^4$, similarly
  as for equation~\ref{fourtran}. The identification of $M_4 \equiv \rrr^4$, with 
$\bv_4 \in \TM_4$ on the local tangent space, breaks the full $\slthc$ symmetry of 
equation~\ref{lvni}, as described for (\cite{KKone} figure 3) and reproduced here in 
figure~\ref{mtogmaph}.
\begin{figure}[htbp]  
\centering
\epsfxsize=12.5cm
\leavevmode
\epsffile[0 0 1765 918]{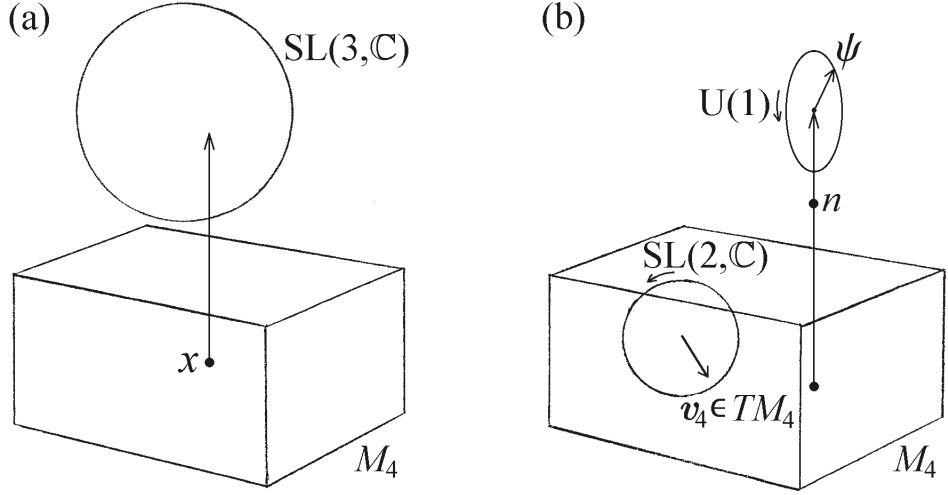}
\caption{\setb  (a) The full symmetry $\slthc$ of  $\lvni$ over the translational 
parameter subspace
 $x \inn M_4$ (b) is broken to the subgroup  $\sltc \times \uo$ by the necessary 
identification of the external 4-dimensional spacetime manifold $M_4$ itself through 
the translation symmetry of
 equation~\ref{lvnitr}.}
\label{mtogmaph}
\end{figure}  

 The choice of a preferred \textit{external} symmetry $\sltc \subset \slthc$ acting on 
the external tangent space $\TM_4$ reduces the original element $\bv_9 \in \hthc$ 
 to the Lorentz vector $\bv_4 \in \TM_4$ ($\equiv \bh \in \htwc$), a spinor $\psi \in 
\ccc^2$ and a scalar $n \in \rrr$,
  as anticipated by the structure of equation~\ref{hinvnine}.
 A residual \textit{internal} $\uo$ is also identified and we have the symmetry 
breaking pattern:
\begin{eqnarray}
        \slthc & \to \quad\;\;  \sltc \quad\!\! \times \quad\!\! \uo &   \nonumber \\
	    \bv_9  \quad\! \to  &  \!\!\!   
    \left\{ \!\!\! \begin{array}{lcl}
     \bv_4  \quad\!  &  \! \mbox{vector}\,  \qquad\;  &   0   \\
	 \psi   \quad\!  &  \! \mbox{spinor}\,  \qquad\;  &   1   \\
	 n      \quad\!  &  \! \mbox{scalar}\,  \qquad\;  &   0   
	  \end{array}   \right. &    \label{slthbr}
\end{eqnarray}  
  where the final column entries denote the $\uo$ charges with unit normalisation and
 with $\psi$ hence identified as a `charged' spinor field over $M_4$.

   The internal $\uo$ symmetry over the base space $M_4$ in figure~\ref{mtogmaph}(b) 
implies a `principle bundle' structure $P \equiv M_4 \times \uo$. This leads to a 
relationship between the curvature of the external spacetime and the internal 
curvature of a gauge field $A(x)$ associated with the $\uo$ gauge symmetry in a manner 
analogous to and guided by Kaluza-Klein theory (as formulated for example in 
\cite{Cho, Katan}), which is the main topic of~\cite{KKone} (see also 
equation~\ref{gchift} here). While both the external and internal symmetry derive from 
the same unification group, namely $\slthc$ in figure~\ref{mtogmaph}(a), compatibility 
with the Coleman-Mandula theorem~\cite{ColMan} for the quantised theory follows from 
the \textit{absolute} symmetry breaking structure in figure~\ref{mtogmaph}(b) and 
equation~\ref{slthbr} as described in (\cite{KKone} subsection~5.3) and as will be 
reviewed in section~\ref{ee6} of this paper.

   Here we note that through both the Kaluza-Klein relation between the external and 
internal curvature and the subsequent quantisation of the theory the field values over 
$M_4$ in general vary and hence the translation symmetry of equation~\ref{lvnitr} is   
only strictly valid in a vanishingly small spacetime volume.
  As will also be described in section~\ref{ee6} here the dynamic degrees of freedom 
of the continuous spacetime geometry are not themselves `quantised', and the resulting 
physical notion of a local inertial frame in the extended spacetime is essentially 
equivalent to that reviewed in the opening of section~\ref{ee3} as central to general 
relativity.

   The gauge field $A(x)$ associated with the internal symmetry $\uo$ together with 
the residual temporal components $\psi(x)$ and $n(x)$ of $\bv_9 \in \hthc$ over $\bv_4 
\in \TM_4$, as pictured in figure~\ref{mtogmaph}(b), are  interpreted as `matter 
fields' over $M_4$. These fields, and their symmetry properties summarised in 
equation~\ref{slthbr}, derive directly from the elementary structure of the theory 
based on the symmetries and possible arithmetic composition of the continuum of time 
alone.

  For a standard Kaluza-Klein theory, which already begins with the non-trivial 
structure of a principle fibre bundle $P \equiv M_4 \times G$ with an internal gauge 
symmetry group $G$ over an extended spacetime base manifold $M_4$, a further 
complication is required in order to introduce spinor fields, for example via a 
supersymmetric extension of the theory (see for example~\cite{DuNiPo} and the 
references therein). A suitable Lagrangian may also be proposed to introduce 
interactions between the various fields by hand. 

 For the theory developed here from the simple starting point of one dimension of 
time, even for this minimal model based on $\slthc$ as the symmetry of a cubic form of 
time, we have identified both an internal $\uo$ gauge field $A(x)$ and a spinor field
 $\psi(x)$ together with their mutual interaction, as represented in 
figure~\ref{mtogmaph}(b) and determined by the constraints of the theory. These 
constraints, such as equation~\ref{lvni} itself, are implied since the theory derives 
from the very simple structure of one dimension only, and in turn the need to add 
further restrictions by hand through postulated Lagrangian terms can in principle be 
avoided. The question is then whether these constraints lead to mathematical 
structures that are recognisable in comparison with the physical structures deduced 
from empirical observations.

   For the $\slthc$ model described in this subsection, on interpreting the internal 
$\uo$ as an electromagnetic gauge symmetry, we obtain a primitive `electrodynamics' 
with the gauge field $A(x)$ coupled to the spinor field $\psi(x)$.
  This is in addition to 
 the Kaluza-Klein relationship deduced between the external gravitational curvature 
and the electromagnetic field curvature as alluded to above, while the neutral scalar 
field $n(x)$ of equation~\ref{slthbr} can in principle be identified as a `dark 
matter' candidate. We can then ask how these possible physical structures might be 
augmented as we progress towards natural higher-dimensional extensions for the general 
form of time in equation~\ref{lvo}, as will be described in the following subsections.

\subsection{$\esi \equiv \sltho$ Symmetry}
\label{ee42}

  The natural generalisation of the complex numbers through the division algebras 
$\ccc \to \hhh \to \ooo$ to the octonions and the existence of the cubic norm
 or `determinant', as described for equation~\ref{detpmn}, suggests that
 the form $\lvni$ of equation~\ref{lvni}, with $\bv_9 \inn \hthc$  and an $\slthc$ 
symmetry,   can naturally be extended to the 27-dimensional form of time of 
 equation~\ref{lvts}, that is:
\begin{equation}
 \label{lvtspmn}
  L(\bv_{27}) \; = \;  \det (\mcX)  \;= \;
  pmn - p\vert b \vert^2 - m\vert c \vert^2 - n\vert a \vert^2  
                + 2 \mbox{Re}(\bar{a}\bar{b}\bar{c})  \; = \; 1    	
\end{equation} 
  with $\bv_{27} \equiv \mcX \in \htho$ of equation~\ref{hthoxy}, and with an
 $\esig \equiv \sltho$ symmetry.

  As noted in subsection~\ref{ee22} this real form of the $\esi$ Lie group symmetry is 
explicitly constructed in \cite{Man4,Man5,Wang, Wang2}  as summarised in (\cite{Unifi} 
chapter 6) and employed extensively in (\cite{Unifi} chapter 8, \cite{Novel}). 
Following those references a basis for the 78-dimensional Lie algebra of $\esig$ can 
be represented by a linearly independent set of vector fields in the tangent space 
$T\htho$ of the form:
\begin{equation}
 \label{ththo}
  \dot{R} \; = \; 
   \left( \begin{array}{ccc}
       \dot{p} & \dot{\bar{a}} & \dot{c}  \\
       \dot{a} &   \dot{m}     & \dot{\bar{b}}        \\
 \dot{\bar{c}} &   \dot{b}     & \dot{n} 
          \end{array}  \right) \;  \inn \; T\htho 
\end{equation}
  Here the notation convention for $\mcX \in \htho$ in equation~\ref{hthoxy} is 
employed with a `dot', as for $\dot{p}$, denoting a tangent vector component. 
 A complete preferred basis for this Lie algebra is listed 
 in (\cite{Wang} table A.1) and also described for (\cite{Unifi} table 6.3, 
\cite{Novel} table 1), consisting of `boosts' and `rotations' when considered as 
generators of a generalised Lorentz group.

 Each of the 78 tangent vector fields in the form of equation~\ref{ththo} is 
determined via the derivative of a corresponding finite $\esig$ symmetry 
transformation preserving $\det(\mcX)$ and is written out explicitly in (\cite{Unifi} 
tables~6.6 and 6.7, \cite{Novel} tables~6 and 7). The Lie bracket between any pair of 
these $\esig$ generators was calculated for the entries of the full $78 \times 78$ 
algebra commutation table determined for \cite{Wang}, with the various sign and other 
conventions adopted in \cite{Unifi,Novel} oriented through consistency with that
 full $\esig$ Lie algebra table.

 In the context of the theory presented in this paper the breaking of the full $\esig$ 
symmetry of the 27-dimensional form of time $L(\bv_{27}) = \det(\mcX) = 1$ follows 
from the same argument as described for the $\slthc$ symmetry of $\lvni$ in the 
previous subsection, through the necessary interposition of an intermediate external 
4-dimensional spacetime form. However, in place of equation~\ref{hinvnine}, here the 
spacetime components $\bv_4 = (v^0,v^1,v^2,v^3) \equiv \bh \in \htwc$ are embedded 
within an element of $\bv_{27} \equiv \mcX \in \htho$ as:
\begin{equation}
  \label{hinvts}
 \!\!\!\!\;    \begin{array}{cc} 
    \htwc \ni \bh \; \mbox{{\Large $\to$}}  \! 
	  \begin{array}{cc} &  \\  &  \end{array} \!\!\!\!\!\!\!\!\!\!\!\!\! 
				                         \\  
         \begin{array}{cc}        &   \end{array}   	        
		   \end{array}  
	\left( \!\! \begin{array}{cc|c} 
	 v^0 + v^3 
   & v^1 - v^2l + \bar{a}(6)  
   & c    \\
	 v^1 + v^2l + a(6)
   & v^0 - v^3
   & \bar{b}  \\  \hline
     \bar{c} 
   & b  
   & n    \end{array} \!\!  \right) 	
	\!	\equiv \!
	\left( \! \begin{array}{c|c} 
         \bh  + a(6) \!
		    \begin{array}{cc} &  \\  &  \end{array} \!\!\!\!\!\!\!\!\!   &
        \,  \theta  \begin{array}{cc} &  \\  &  \end{array} \!\!\!\!\!\!\!\!\!\! 
				                      \\ \hline
      \,\,\,\,\, \theta^{\dagger} \!\!\! \begin{array}{cc}        &   \end{array}   &	  
		\,  n      \end{array} \! \right)   
   \! \equiv \bv_{27}	 \in \htho 
\end{equation}

   With respect to the components of $\mcX \in \htho$ in equation~\ref{xoct3} we have 
substituted $p=v^0+v^3$ and $m=v^0-v^3$, while for the component $a\inn \ooo$ with the 
real subcomponents of equation~\ref{octa} we have $a_1 = v^1$ and $a_8 = v^2$, that is
with the octonion imaginary unit $l$ employed for the $v^2$ external spacetime 
component, in place of the complex unit $i$ in equation~\ref{hinvnine}, in line with 
the conventions of the preferred basis for equation~\ref{ththo} (as explained in 
\cite{Unifi} following equation~6.58).
 The six remaining imaginary components are denoted by:
\begin{equation}
\label{asix} 
  a(6) \; = \; a_7{\il}  + a_2 i \;\, + \;\, a_6{\jl} + a_3 j \;\, + \;\, a_5{\kl} + 
a_4 k
\end{equation}  
   ordered here for later reference.
   In the second $3 \times 3$ matrix in equation~\ref{hinvts} these components are 
embedded in the 
   upper-left-hand $2 \times 2$ part as 
 $a(6) \equiv a(6)\binom{0 \: -\!1}{1 \;\;\; 0}$, consistent with 
equation~\ref{xoct3}.
    There are now a total of 23 real parameter `extra dimensions' including $a(6)$ 
together with the sixteen real components of $\theta = \binom{c}{\bar{b}} \in \ooo^2$ 
and again with $n\in \rrr$. 

  On identifying a 4-dimensional spacetime manifold $M_4$ the
   components of \mbox{$\bv_4 \in \TM_4$} projected onto the tangent space transform 
as a Lorentz 4-vector under an external $\sltc \subset \esig$ symmetry, similarly as 
depicted in figure~\ref{mtogmaph}(b).
 The symmetry breaking pattern is now found to incorporate four Weyl spinors, denoted:
\begin{equation}  
\label{thelijk}
 \theta_l  = \binom{c_1 + c_8l}{b_1 - b_8l}, \quad
 \theta_i = \binom{c_7\il + c_2i}{- b_7\il - b_2i}, \quad   
 \theta_j = \binom{c_6\jl + c_3j}{- b_6\jl - b_3j}, \quad 
 \theta_k = \binom{c_5\kl + c_4k}{- b_5\kl - b_4k} 
\end{equation}
  from the components of $\theta = \binom{c}{\bar{b}} \in \ooo^2$,  and seven scalars 
corresponding to the components of $a(6)$ and $n$ under the external Lorentz symmetry, 
as described in detail in (\cite{Unifi} section~8.1) and summarised in 
 (\cite{Novel} section~5). Each of $\theta_l, \theta_i$, $\theta_j$ and $\theta_k$ 
transform in the same way as two-component Weyl spinors under $\sltc$ and they are 
taken to be `left-handed' by convention.

 An internal 
 $\suth \times \uo \subset \esi$ symmetry is also identified. The three components of 
the broken symmetry $\sltc \times \suth \times \uo \subset \esi$ are generated by the 
Lie algebra elements in the form of equation~\ref{ththo} as listed here:
\begin{eqnarray}
    \sltc:  &   &
	  \{\dot{B}_{t \pqg z}^{1}, \dot{R}_{x \pqg l}^{1}, \dot{B}_{t \pqg x}^{1},
     \dot{B}_{t \pqg l}^{1}, \dot{R}_{x \pqg z}^{1}, \dot{R}_{z \pqg l}^{1} \}
	    \nonumber \\
	\suth: & &	\{\dot{A}_q, \dot{G}_l\} \quad \mbox{for} \;\,  q= 
\{i,j,k,\kl,\jl,\il,l \}   \label{esibas}   \\
    \uo: &  &   \dot{S}_{l}^{1}   \nonumber
\end{eqnarray}	
  These elements form part of the preferred basis of (\cite{Wang} table A.1, as 
reproduced in \cite{Unifi} table 6.3, \cite{Novel} table 1) where the notation is 
explained in full. (These generators are listed here explicitly in 
equation~\ref{esibas} largely as a reference for equation~\ref{esisufo} later in this 
subsection).
  Any element of any one of the three sets of generators in equation~\ref{esibas} 
commutes with any element of the other two sets, by the Lie algebra table of
 \cite{Wang}. 	
 
  The internal $\suth \times \uo$ symmetry is
    associated with the internal colour $\suth_c$ and electromagnetic $\uo_Q$ 
	symmetry of the Standard Model, owing to the transformation properties of the 
components of $\bv_{27} \equiv \mcX \in \htho$  under the broken $\esi$ symmetry, as 
described in detail in  (\cite{Unifi} section~8.2) and summarised in (\cite{Novel} 
section~5), with the results listed here in table~\ref{esibr}.

\begin{table}[htbp]
\centering 
\begin{tabular}{|rlccccc||c|}
 \hline 
 $\esi \;\;$ & $\to $ & 
  $\sltc$ & $\times$ & $\suth_c$ & $\times$ & $\uo_Q \quad$
                     &  (B)SM \\
 \hline
    &  $\bv_4\quad$  &  vector  & & $\mathbf{1}$ 
	                  & &  $  0 \quad $   &  (Higgs)   \\ \cline{2-8}
	&  $\theta_l$ & $L$-spinor & & $\mathbf{1}$ & & $1 \quad$  & 
	   $e$-lepton \\  \cline{2-8}
 \raisebox{0pt}[0pt][0pt]{ {\raisebox{-2.0ex}{$\bv_{27} \to 
  \left\{ \begin{array}{c} \\ \\ \\ \\ \\ \\ \end{array}  \right. 
   \!\!\!\!\!\!\!\!\!\!\! $}} }
	&  $\theta_{i,j,k}$ &  $L$-spinor & & $\mathbf{3}$ 
	                                         & &  $\frac{1}{3} \quad$  & 
	    $d$-quarks \\  \cline{2-8}
	&  $\!\!\!\!{\big \lbrack} \: a_{1,8}$ &  X  & & $\mathbf{1}$
	                              &  & $0 \quad$  & 
					  $\;\;\nu$-lepton	${\big \rbrack}$    \\ \cline{2-8}
	&  $a(6)$ &  scalar & & $\mathbf{3}$
	                                    & &  $\frac{2}{3} \quad$  & 
                            $u$-quarks \\   \cline{2-8}
    &  $n$ &  scalar & & $\mathbf{1}$ & &  $ 0 \quad$  &  (DM)   \\
 \hline
\end{tabular} 
\caption{\setb Decomposition of the components of $\lvt$ as the $\esi$ symmetry is 
broken through the identification of an external spacetime with projected components 
$\bv_4 \in \TM_4$, augmenting the structure of figure~\ref{mtogmaph} and 
equation~\ref{slthbr}. The final column  lists the (Beyond the) Standard Model 
correlations that can be drawn from the symmetry breaking pattern.}
\label{esibr}
\end{table} 

    The provisional association of the 4-vector $\bv_4$ with a non-standard Higgs 
sector of the theory in table~\ref{esibr} will be explained in the following 
subsection. The interpretation of the scalar $n$ as a dark matter (DM) candidate was 
alluded to at the end of the previous subsection and will be discussed further in 
section~\ref{ee6}. Here we focus on the direct connections established with the 
Standard Model.

 The Weyl spinor $\theta_l$ of equation~\ref{thelijk}  transforms as a singlet $\b1$ 
under the internal $\suth_c$ but non-trivially under the internal $\uo_Q$. Hence 
$\theta_l$ (essentially the equivalent of the spinor $\psi$ of equation~\ref{slthbr}) 
is provisionally associated with the electron state, and motivates the charge 
normalisation of 1 for this case. The remaining $\uo_Q$ charges listed in 
table~\ref{esibr} are given by their magnitudes relative to the unit electron charge, 
with no interpretation of particle as compared with antiparticle state made here. 
These relative charges
  are \textit{fixed} by the structure of the $\esi$ Lie algebra and in particular the 
$\uo_Q$ generator in equation~\ref{esibas}.
 The three Weyl spinors $\{\theta_i,\theta_j,\theta_k \}$ in equation~\ref{thelijk} 
transform as an $\suth_c$ triplet $\mathbf{3}$  with relative $\uo_Q$ charges of 
magnitude $\frac{1}{3}$, and are hence associated with a generation of Standard Model 
$d$-quarks. 
 (As well as the $\uo_Q$ charge signs here we also do not distinguish between the 
 $\suth_c$ representations \mbox{$\mathbf{3}$ and $\mathbf{\overline{3}}$}).

  The six components of $a(6)$ in equation~\ref{asix} transform as $\sltc \subset 
\esi$ scalars rather than spinors, but they can be paired up in the set 
$\{a_{7,2},\:\!a_{6,3},\:\!a_{5,4} \}$  which also transforms as an $\suth_c$ triplet, 
and owing to their $\uo_Q$ charges of $\frac{2}{3}$ these components are provisionally 
correlated with $u$-quarks via their transformation properties under the internal 
symmetry. The two remaining components $a_{1,8}$ of the octonion element $a\in \ooo$ 
similarly provide a natural slot to be assigned to the neutrino state owing to the 
invariance of the $a_{1,8}$ components under the internal $\suth_c \times \uo_Q$ 
symmetry. The neutrino component $a_1 + a_8l$
 is then related to the electron component $\theta_l  = \binom{c_1 + c_8l}{b_1 - 
b_8l}$ similarly as the $u$-quark states in the remaining $a(6)$ components are 
 related to the $d$-quark states in the corresponding remaining imaginary components 
of  $\theta = \binom{c}{\bar{b}} \in \ooo^2$, on comparing equations~\ref{asix} and 
\ref{thelijk}, suggesting doublets of leptons and quarks. 
  However these $a_1 + a_8l$ components, as for those of $a(6)$, also do not transform 
via a spinor representation, and in fact the $a_{1,8}$ slot is already `occupied' by 
two of the four external spacetime components, explicitly with 
  $\bv_4 = (v^0,v^1,v^2,v^3) = (\fh(p+m),a_1,a_8,\fh(p-m)) \in \TM_4$ as described for 
equation~\ref{hinvts}; and hence this provisional `$\nu$-lepton state' is bracketed in 
table~\ref{esibr}.       

   Hence it remains to be explained how the neutrino can be accommodated within this 
theory and how a neutrino and $u$-quark components transforming as $\sltc$ spinors can 
be identified while retaining the internal symmetry properties of table~\ref{esibr}.
 Further, even the $e$-lepton and $d$-quark states in table~\ref{esibr} are only 
identified here as two-component left-handed Weyl spinors rather than the 
four-component  Dirac spinors of the Standard Model. In addition these patterns will 
need to be repeated for a full three generations of states. 
 The incompleteness of this picture is further demonstrated by the  
 well known fact that the Lie group $\esi$ is not large enough to contain the subgroup 
$\sltc \times \suth \times \sutw \times \uo$ representing the external Lorentz 
symmetry and the full Standard Model internal gauge symmetry together, and hence the 
full structure of electroweak theory cannot be incorporated here. 

  However, some of the properties of the Standard Model electroweak theory 
\textit{can} be identified based on $\sutw \times \uo \subset \esi$ subgroups for the 
$\esi$ symmetry of $\lvt$ presented in this subsection. These properties include an 
analogue of the standard symmetry breaking pattern $\sutw_L \times \uo_Y \to \uo_Q$ in 
the projection of $\bv_4 \in \TM_4$, as described in (\cite{Unifi} section~8.3) where 
a `mock electroweak theory' is constructed, in part motivating the link between the 
4-vector $\bv_4$ and the Higgs in table~\ref{esibr}, (a link that will be further
 justified after figure~\ref{mforese}).
 These observations hint that a full electroweak theory might be accommodated in a 
further augmentation of the theory towards a higher-dimensional form of time with a 
larger symmetry group.

 In the meantime simply by generalising from the complex numbers $\ccc$ in 
equations~\ref{hinvnine} and \ref{lvni} to the octonions $\ooo$
  in equations~\ref{lvtspmn} and \ref{hinvts},
 consistent with the form of equation~\ref{lvo}, the simple matter fields of 
equation~\ref{slthbr} have been augmented to those of table~\ref{esibr} and 
established a recognisable foothold in the structures of the Standard Model.
  That is, in place of the `primitive electrodynamics' described at the end of the 
previous subsection the matter fields identified here now resemble one generation of 
Standard Model leptons and quarks.

  The inability to fit an $\sltc \times \suth \times \sutw \times \uo$ subgroup inside 
$\esi$ can be seen by analysis of the Dynkin diagrams for the corresponding complex 
Lie algebras. In typical unification models the Lie group $\esi$ is generally 
considered from the point of view of internal symmetries alone, that is as a `Grand 
Unified Theory' with $\suth_c \times \sutw_L \times \uo_Y \subset \esi$ (see for 
example~\cite{Gur1,Slan}).
  Through the conceptual scheme presented here, as pictured for the $\slthc$ model in 
figure~\ref{mtogmaph} and now with the matter fields over $M_4$ described in 
table~\ref{esibr} for the $\esi$ case, the full symmetry necessarily includes 
\textit{both} the external symmetry and the internal symmetry, with the latter 
identified as far as $\suth_c \times \uo_Q$ for the $\esi$ symmetry described in this 
subsection. (For the overall theory this framework is consistent with the 
Coleman-Mandula theorem due to the absolute nature of the symmetry breaking, as 
discussed in the previous subsection for figure~\ref{mtogmaph} and further
 in section~\ref{ee6}).

  However, Dynkin analysis can also be employed to study the symmetry breaking pattern 
for $\sltc \times \suth \times \uo \subset \esig$ on the 
27-dimensional representation of $\esi$. Here for the complex Lie algebra we take the 
27 weights of the 
  $\mathbf{27}$  representation of $\esi$ described by Dynkin labels such as 
 $(1\;0\;0\;0\;0\;0)$, as listed in (\cite{Slan} table~11b) and (\cite{Geor} chapter 
27) for example, with the six Dynkin coefficients ordered in correspondence with the 
six simple roots $\alpha_i$ $(i=1,\ldots, 6)$ of the rank-6 $\esi$ Lie algebra, which 
in turn are matched with the six nodes of the $\esi$ Dynkin diagram as shown in 
 figure~\ref{dynesi}(a).
\begin{figure}[htbp]  
\centering
\epsfxsize=12.5cm
\leavevmode
\epsffile[0 0 1683 410]{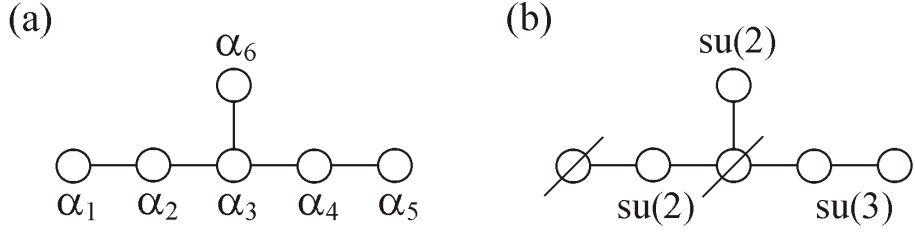}
\vspace{-3pt}
\caption{\setb  (a) Dynkin diagram for the complex Lie algebra $\esi$ with six simple 
root labels and (b) a possible breaking pattern that at the level of the corresponding 
real Lie groups describes $\sltc \times \suth \times \uo \subset \esig$.}
\label{dynesi}
\end{figure}  
  \indent
   Dynkin analysis deals with complex Lie algebras and with the `complexified' Lie 
algebra for the Lorentz algebra $\soota \equiv \sltca$ being $\sutwa \oplus \sutwa$ we 
are hence looking for a symmetry breaking structure of the semi-simple form:
\begin{equation}
 \label{sssuesi}  
  \sutw \times \sutw \times \suth_c \times \uo_Q \subset \esi
\end{equation}  
  in the analysis of the Dynkin diagrams (see also \cite{Unifi} section~7.3).
   This can be achieved in a number of ways, including that depicted in 
figure~\ref{dynesi}(b) with the corresponding weight projection matrices:
\begin{equation}
 \label{dynproj}
  P = \left( \begin{array}{cccccc}  \vspace{-7pt}
      0 & 1 & 0 & 0 & 0 & 0  \\ \vspace{-7pt}
	  0 & 0 & 0 & 0 & 0 & 1  \\ \vspace{-7pt}
	  0 & 0 & 0 & 1 & 0 & 0  \\
	  0 & 0 & 0 & 0 & 1 & 0  
    \end{array}  \right), \qquad \qquad
  Q = {\big\lbrack} \begin{array}{cccccc}
      0 & 1 & 2 & \frac{4}{3} & \frac{2}{3} & 1  
    \end{array}  {\big\rbrack}
\end{equation}
Here the $4\times 6$ matrix $P$ projects each weight of the $\mathbf{27}$ 
representation of $\esi$, such as $(1\;0\;0\;0\;0\;0)$, onto four Dynkin labels 
aligned with the rank-4 $\sutw \times \sutw \times \suth_c$ subgroup, while the $1 
\times 6$ matrix $Q$ projects out the $\uo_Q$ charges; such that the 27 weights of 
this $\esi$ representation collectively branch under the symmetry breaking as listed 
in table~\ref{dynkbr}.

\begin{table}[htbp]
\centering 
\begin{tabular}{|rlccccc||c|}
 \hline 
 $\esi$ & $\to $ & 
  Lorentz & $\times$ & $\suth_c$ & $\times$ & $\uo_Q \quad$
                     & $\bv_{27} \in \htho$ \\
 \hline
    &  $\mathbf{4}\quad$  &  vector  & & $\mathbf{1}$ 
	                  & &  $ \;\; 0 \quad $   &  $\bv_4$  \\ \cline{2-8}
   \vspace{-5pt}
	&  $\mathbf{2}$ & $L$-spinor & & $\mathbf{1}$ & & $+1 \quad$  & 
	   \raisebox{0pt}[0pt][0pt]{ {\raisebox{-1.5ex}{$\theta_l$}} } \\  
	&  $\mathbf{2}$ & $R$-spinor & & $\mathbf{1}$ & & $-1 \quad$  &   \\   \cline{2-8}
   \vspace{-5pt}
 \raisebox{0pt}[0pt][0pt]{ {\raisebox{-1.5ex}{$\mathbf{27} \to 
  \left\{ \begin{array}{c} \\ \\ \\ \\ \\ \\ \vspace{10pt}  \end{array}  \right. 
   \!\!\!\!\!\!\!\!\!\!\! $}} }
	&  $\mathbf{6}$ & $L$-spinor & & $\mathbf{\overline{3}}$ 
	                                         & &  $-\frac{1}{3} \quad$  & 
	   \raisebox{0pt}[0pt][0pt]{ {\raisebox{-1.5ex}{$\theta_{i,j,k}$}} } \\
	&  $\mathbf{6}$ & $R$-spinor & & $\mathbf{3}$
	                              &  & $+\frac{1}{3} \quad$  &  \\ \cline{2-8}
   \vspace{-5pt} 
	&  $\mathbf{3}$ &  scalar & & $\mathbf{\overline{3}}$
	                                    & &  $+\frac{2}{3} \quad$  & 
 \raisebox{0pt}[0pt][0pt]{ {\raisebox{-1.5ex}{$a(6)$}} } \\
	&  $\mathbf{3}$ &  scalar & & $\mathbf{3}$ 
	                              & & $-\frac{2}{3} \quad$  &  \\ \cline{2-8}
    &  $\mathbf{1}$ &  scalar & & $\mathbf{1}$ & &  $\;\; 0 \quad$  &  $n$  \\
 \hline
\end{tabular} 
\caption{\setb Branching of the $\mathbf{27}$ representation of $\esi$ under the 
subgroup of equation~\ref{sssuesi} via the Dynkin label projection matrices of 
equation~\ref{dynproj}.
The overall choice for each of $L \leftrightarrow R$, $\mathbf{3} \leftrightarrow 
\mathbf{\bar{3}}$, and $\pm Q$ is a matter of convention for the Lorentz, $\suth_c$ 
and $\uo_Q$ representations respectively. The final column lists the best-matching 
correspondence with the components of table~\ref{esibr}.}
\label{dynkbr}
\end{table} 
  
  In comparison with the matrix $P$ of equation~\ref{dynproj} a very different and 
more complicated projection for $\esi$ representations can be constructed from a 
combination of the weight projection matrices described in (\cite{Slan} section~7) 
via:
\begin{equation}
 \label{slanp}
  P_{20}(\sufo \subset \mbox{SU}(5)) \;\, \scirc \;\, 
  P_{17}(\mbox{SU}(5) \subset \mbox{SU}(6)) \;\, \scirc \;\,
  P_{16}(\mbox{SU}(2) \times \mbox{SU}(6) \subset \esi)
\end{equation}
  In this case the subgroup $\suth_c \times \uo_Q \subset \sufo \subset \esi$ is 
identified as a further projection after the above $P_{20}$ factor, with an 
appropriate $Q^{\mathrm{em}}$ projecting the electromagnetic charge, with the 
resulting branching structure for the 
$\mathbf{27}$ representation of $\esi$ listed in (\cite{Slan} table~21).
 While \cite{Slan} only deals with \textit{internal} gauge symmetries a further $\sutw 
\subset \esi$ projection can be identified along with the $\sutw$ factor from $P_{16}$ 
in equation~\ref{slanp} which combined together can be interpreted as an 
\textit{external} $\mbox{Lorentz} \equiv \sutw \times \sutw$ symmetry as described for 
equation~\ref{sssuesi}.
 The structure of the full projection for the corresponding complete subgroup of 
equation~\ref{sssuesi} applied to the $\mathbf{27}$ representation of $\esi$ is found 
to match that of table~\ref{dynkbr} above. A third calculation, projecting the same 
subgroup symmetry this time via the extended Dynkin diagram for $\esi$ (using the 
diagram and weights listed in 
 \cite{Geor} section~27.2), yields again the same result and conclusions.

  The Dynkin analysis for the broken symmetry  branching of the 27-dimensional 
representation of $\esi$ described in table~\ref{dynkbr} has a recognisable connection 
with that obtained through the explicit $\esig$ actions on the space $\htho$ for 
table~\ref{esibr}, as noted by the correspondence with subcomponents of $\bv_{27} \in 
\htho$ identified in the final column of table~\ref{dynkbr}. However there are also 
significant differences. Formulated in terms of complex Lie algebras the 27 weights 
for the Dynkin analysis are defined over a complex $\ccc$-valued space. On the other 
hand each of the 27 components of $\bv_{27} \in \htho$ can be considered as a real 
$\rrr$-valued parameter, while the explicit determinant preserving symmetry 
transformations of $\esig \equiv \sltho$ acting on the space $\htho$ involve both the 
quaternion $\hhh$ and octonion $\ooo$ algebras!

 This difference is seen for example with both left $L$ and right $R$ Weyl spinors 
being listed in table~\ref{dynkbr} while only left-handed $L$-spinors are identified 
in table~\ref{esibr}. This feature arises in the latter case since we have a 
non-standard representation of the Lorentz group $\soot$, via the Clifford algebra 
$C(1,3)$, as effectively embedded in $2\times 2$ quaternion matrices $\hhh(2)$ acting 
on the spinor space $\hhh^2 \subset \ooo^2$. As described for (\cite{Unifi} 
equations~8.10--8.13) this leads to the set of four Weyl spinors
 of equation~\ref{thelijk} of the
 \textit{same} handedness. 
  This observation is key for the left-right structure of the theory under the larger 
$\ese$ symmetry, in comparison with the Standard Model, as will be described in the 
following subsection.

 While the structure of complex Lie algebras and Dynkin analysis provides a useful 
guide the above differences highlight the need to study the breaking of the full 
symmetry of $\lvn$ of equation~\ref{lvo} explicitly if the octonion algebra is 
involved. As noted in subsection~\ref{ee22} non-associative octonion composition, 
while not directly forming a group structure, can describe a high degree of symmetry. 
This is the case with the construction of the $\esig$ symmetry of $\lvt$ of 
equation~\ref{lvtspmn}, and hence the explicit anatomy of this particular structure 
(following \cite{Man4,Man5,Wang, Wang2}) and symmetry breaking pattern has been 
analysed in detail in \cite{Unifi,Novel} in order to uncover the properties of the 
resulting matter fields listed in table~\ref{esibr}.

  Direct analysis of the extended Dynkin diagram for $\esi$ shows that in addition to 
the breaking pattern of figure~\ref{dynesi}(b) and equation~\ref{sssuesi} the Lie 
group $\esi$ also contains the larger subgroup:  
\begin{equation}
 \label{sssesi}  
  \sutw \times \sutw \times \sufo \subset \esi
\end{equation} 
  in turn with $\suth \times \uo \subset \sufo$ contained as a further possible 
decomposition.
 A similar subgroup structure for $\esi$ was implied above in the discussion of 
equation~\ref{slanp}. 
   Taking this as a guide for the  breaking of the explicit $\esig$ action on 
   \mbox{$\lvt$}, through the projection of the external $\bv_4 \in \TM_4$ components, 
this suggests that complementing the external Lorentz symmetry a full internal $\sufo$ 
symmetry might also be identified. This is indeed the case with the following set of 
15 elements of the $\esi$ Lie algebra in the form of equation~\ref{ththo} from the 
basis of (\cite{Wang} table A.1, \cite{Unifi} table 6.3 or \cite{Novel} table 1)
 generating such an $\sufo$: 
\begin{equation}
   \label{esisufo} 
   \left.  \begin{array}{rl}
	\suth: & 	\{\dot{A}_q, \dot{G}_l\} \quad \mbox{for} \;\,  q= 
\{i,j,k,\kl,\jl,\il,l \}     \\
    \uo: &    \dot{S}_{l}^{1}  \\
	  +:  &    \{ \dot{G}_{q} + 2\dot{S}_{q}^{1} \} \quad \mbox{for} \;\,  q= 
\{i,j,k,\kl,\jl,\il \}   \end{array}  \right\}  \;\; \sufo
\end{equation} 
 With 6 further generators added to the 8 of $\suth$ and the one for $\uo$ in 
equation~\ref{esibas}, this set of 15 forms a closed $\sufoa$ algebra under the Lie 
bracket commutator described in \cite{Wang}, while each of the 15 commutes with the 6 
external Lorentz generators on the top line of equation~\ref{esibas}.

 The action of this $\sufo \subset \esi$ subgroup on the components of 
 $\bv_{27} \in \htho$ can be determined by explicit examination of the 15 generators
  of equation~\ref{esisufo} as elements  $\dot{R} \in T\htho$, equation~\ref{ththo}, 
as listed in (\cite{Unifi} table~6.7 or \cite{Novel} table~7). Interpreted as an 
internal symmetry this $\sufo$ acts by the fundamental 4-dimensional representation 
$\mathbf{4}$ upon the set of four left-handed Weyl spinors $\{\theta_l, \theta_i, 
\theta_j, \theta_k \}$ of equation~\ref{thelijk}, 
 which would be to mix the $e$-lepton and $d$-quark components according to the 
Standard Model assignments proposed in table~\ref{esibr}. The same 
 $\sufoa \equiv \mbox{so}(6)$ internal algebra acts via the fundamental 6-dimensional  
representation $\mathbf{6}$ of the rotation group SO(6) on the set of six scalar 
components $a(6)$ in equation~\ref{asix}. The remaining $5=27-16-6$ real subcomponents 
listed in table~\ref{esibr}, namely of $\bv_4$ (including the $a_{1,8}$ part) and $n$, 
are invariant under this $\sufo$ subgroup.
 
  In fact a simple Dynkin analysis shows that the  rank-5 subgroup in 
equation~\ref{sssesi} can be also be obtained as  $\sutw \times \sutw \times \sufo 
\subset \mbox{D}_5 \equiv \mbox{SO}(10)$. A real form of the rank-5 Lie group D$_5$ 
can be constructed as $\sltwoo$, the double cover of $\sootn$, acting as the 
	 symmetry of $L(\bv_{10}) = \det(X) = 1$ for $\bv_{10} \equiv X\in \htwo$ as a 
quadratic form of time (that could be interpreted as representing 10-dimensional 
spacetime, as described in leading to equation~\ref{vcubic}). In this case $\sufo$ is 
the double cover of an internal SO(6) which acts on the six Lorentz scalar extra 
dimensions over 4-dimensional spacetime as pictured in (\cite{Novel} figure~1). The 
same six scalars $a(6)$ remain for the augmentation to the full $\esi$ symmetry of the 
cubic form of time $\lvt$, as embedded in equations~\ref{xoct3} and \ref{hinvts} and 
listed in table~\ref{esibr}, and again transform as a 6-dimensional vector under this 
internal SO(6).
  
  The proposal here is that under a further extension to a larger symmetry for a yet 
higher-dimensional form of time $\lvn$, with $n>27$, the components of 
 $a(6) = (a_7{\il}  + a_2 i)  + (a_6{\jl} + a_3 j) + (a_5{\kl} + a_4 k)$ from 
equation~\ref{asix}
  will effectively pair up 
  while being subsumed into a set of three Weyl spinors, essentially aligned with the  
existing spinors $\{\theta_i, \theta_j,\theta_k\}$ of equation~\ref{thelijk}.
   This will break
    the $\sufo$ symmetry to the $\suth_c \times \uo_Q$ of table~\ref{esibr} according 
to which the components underlying the three new spinors transform under $\suth_c$ as 
a triplet of $u$-quarks \mbox{with $\uo_Q$ charges of $\frac{2}{3}$}. In principle 
with $\binom{u}{d}$-quark doublets identified and a surviving $\uo_Q$ symmetry 
component this structure is hence expected to be closely related to the identification 
of a full electroweak symmetry with the breaking pattern $\sutw_L \times \uo_Y \to 
\uo_Q$ for the augmented theory. This extension of the theory will clearly also need 
to account for the left-right asymmetry of weak interactions. While keeping in mind 
these well known properties of the Standard Model, the general approach here is to 
seek natural mathematical augmentations to higher-dimensional homogeneous polynomial 
forms of time, one of which we describe in the following subsection.

\subsection{$\ese$ Symmetry}
\label{ee43}

Here we describe a further natural extension of the multi-dimensional form of time, 
consistent with the general homogeneous polynomial form $\lvn$ introduced in 
equations~\ref{propgen} and \ref{lvo}. The 27-dimensional cubic form 
 $L(\bv_{27}) = \det(\mcX)$ of equations~\ref{lvts} and \ref{lvtspmn} is invariant 
under $\esig$, providing an explicit description of the smallest non-trivial $\esi$ 
representation.
 With $\bv_{27} \equiv \mcX \in \htho$ this structure is embedded within the 
56-dimensional quartic form $q(x)$, with $x \in F(\htho)$, as can be seen from the 
explicit structure of equation~\ref{fquartic}, which suggests that we can consider:
\begin{equation}
 \label{lvfsq}
    L(\bv_{56}) \: = \:
   q(x) \: = \: -2\lbrack \alpha\beta - (\mcX,\mcY)\rbrack^2 \, - \,
       8\lbrack\alpha \det(\mcX) + \beta\det(\mcY) - (\mcX^{\sharp},
	                                          \mcY^{\sharp})\rbrack
    \: = \: 1 \;\;
\end{equation}
for $\bv_{56} \equiv x \in F(\htho)$ as a suitable candidate for a higher-dimensional 
form of time with an $\eseg$ symmetry, as noted for equation~\ref{lvfs}.
  
 In fact an element $x = \binom{\alpha \;\, \mcX}{\mcY \;\, \beta} \in F(\htho)$ of 
the Freudenthal triple system with 56 real components,
 as described for equation~\ref{ftscomp}, is composed of two elements $\mcX, \mcY \in 
\htho$ of the exceptional Jordan algebra together with two real variables
 $\alpha, \beta \in \rrr$. Collectively this object provides a description of the 
smallest non-trivial representation of $\ese$ which is known to branch under the $\esi 
\subset \ese$ subgroup as:
\begin{equation}
  \label{esetoesi}
    \mathbf{56}_{\mathrm{E}_7} \, \to \, 
	(\mathbf{27} + \mathbf{\overline{27}} + \mathbf{1} + \mathbf{1})_{\mathrm{E}_6}
\end{equation}	
  as described for (\cite{Unifi} equation~9.33, \cite{Novel} equation~62). Unlike the 
case for $\esi$ the Lie group $\ese$ does not have complex representations. Here the 
complex $\mathbf{27}$ representation of $\esi$, corresponding to $\mcX \in \htho$, is 
combined with the complex conjugate $\mathbf{\overline{27}}$ representation of $\esi$, 
corresponding to $\mcY \in \htho$, within the full $\ese$ symmetry action.   
 This implies a doubling up of the components listed in table~\ref{esibr} with the 
addition of a corresponding `complex conjugate' set. As noted in the previous 
subsection, since the octonion algebra is used to describe these symmetry 
transformations, an explicit analysis is required. Such an analysis is described in 
(\cite{Unifi} section~9.2, \cite{Novel} section 6) where each left-handed component 
under $\sltc \subset \esi$ in table~\ref{esibr} is found to be now partnered with a 
corresponding right-handed component for the augmented $\ese$ case, as might have been 
expected on appending the `complex conjugate' $\esi$ representation. 

  Hence this natural augmentation from a cubic $\esi$ to a quartic $\ese$ form of time 
leads from the original set of four 2-component left-handed Weyl spinors of 
equation~\ref{thelijk} and table~\ref{esibr} to the identification four 4-component 
Dirac spinors associated with the $e$-lepton and $d$-quark states. The internal 
$\suth_c \times \uo_Q$ transformation properties identified now also for the 
right-handed components are identical to those for the left-handed partner components 
as originally listed in  
 table~\ref{esibr}.
 
 A further significant observation can be made relating to this structure. The 
identification of the external 4-dimensional spacetime through the  component 
 $\bv_4 \in \TM_4$, described for the 4-, 9- and 27-dimensional forms of time via 
equations~\ref{fourtran}, \ref{hinvnine} and \ref{hinvts} respectively, now for the 
case of the 56-dimensional form of time $\lvfs$ is necessarily projected out from 
\textit{either} the $\mcX \in \htho \subset F(\htho)$ component, that is the 
`left-handed' part of $x\in F(\htho)$, or from the $\mcY \in \htho \subset F(\htho)$ 
component, that is the `right-handed' part of $x\in F(\htho)$,
  but \textit{not from both}. This  concrete difference between the left and 
right-handed sectors of the theory with respect to the identification of the external 
spacetime manifold $M_4$ itself, at this $\ese$ level, in principle underlies the 
manifestation of parity violating phenomena observed in the laboratory.

   This latter feature is introduced by hand in the Standard Model via Lagrangian 
terms, as is typically also the case for many unification schemes. For example
in \cite{PaSa} the Standard Model gauge group derives from a local
\begin{equation}
\label{sspasa}
 G \, = \, \sutw_L \times \sutw_R \times \sufo  
\end{equation}
symmetry for which the right-handed gauge bosons
 $W_R$, associated with the $\sutw_R$ component, are \textit{assumed} to have a large 
mass relative to their left-handed counterparts $W_L$, in order to be compatible with 
empirically observed parity violating phenomena. This is arranged by
  \textit{assigning} appropriate parameters to terms in the Lagrangian for the 
spontaneous symmetry breaking structure (\cite{PaSa} equations~17--19).  The `colour 
$\sufo$' component of the internal symmetry group $G$
 in equation~\ref{sspasa} is analogous to the internal $\sufo$ in 
equation~\ref{sssesi}, in both cases mixing leptons and quarks, although in the latter 
equation the $\sutw \times \sutw$ subgroup represents the external Lorentz symmetry 
rather than a further internal component. In the present theory a further augmentation 
in the symmetry of time will ultimately be required in order to incorporate a full
 $\sutw_L \times \uo_Y$ electroweak symmetry, but a natural origin for a left-right 
asymmetry is clear at the $\ese$ level as described above.   

 We also note that one puzzling feature for the symmetry breaking from the $\esi$ 
level, described in the previous subsection, was the observation that the natural slot 
to be assigned to neutrino state, that is the $\suth_c \times \uo_Q$ invariant 
components $a_{1,8}$ of table~\ref{esibr}, is already occupied by two of the four
 components of $\bv_4 \in \TM_4$ in the
 external spacetime tangent space. This slot in the left-handed sector $\mcX \in 
\htho$ is now \textit{freed up} on projecting the external $\bv_4 \in \TM_4$ from the 
corresponding components of $\mcY \in \htho$ instead, which in turn \textit{prohibits} 
the identification of a neutrino state in the right-handed sector of the theory (the 
choice of $\mcX$ or $\mcY$, as for the naming of `left' or `right', is a matter of 
convention that does not affect the conclusions). Hence the empirical asymmetry in the 
handedness of the physical neutrino particle state can in principle be naturally 
accounted for, while still identifying both left and right-handed components (in 
$\mcX$ and $\mcY$ respectively) for the $e$-lepton and $u$- and $d$-quark states, 
within the caveat for the `$\nu$'-lepton and `$u$'-quark components recalled below.

 In comparison with the simpler case of subsection~\ref{ee41} for the breaking of the 
$\slthc$ symmetry of $\lvni$, as pictured in figure~\ref{mtogmaph},
and the intermediate case of subsection~\ref{ee42} for the breaking of the $\esi$ 
symmetry of $\lvt$, as described for table~\ref{esibr},
 the matter fields resulting from the breaking of the $\ese$ symmetry of $\lvfs$ over 
the base manifold $M_4$ are summarised in figure~\ref{mforese} (see also \cite{Unifi} 
equation~9.46, \cite{Novel} equation~66).
 These primarily include gauge fields associated with the internal gauge symmetry 
$\suth_c \times \uo_Q$ in interaction with  states associated with the decomposition 
of $\bv_{56}$ with properties further resembling a generation of Standard Model 
leptons and quarks. While the $e$-lepton and $d$-quark components are identified as 
Dirac spinors, the `$\nu$'-lepton and `$u$'-quark components are listed in quote marks 
in figure~\ref{mforese} since they lack the appropriate spinor structure and are 
provisionally identified via the internal symmetry transformation properties alone.

\vspace{47pt}
\def\ngap{\vspace{-5pt}}
\begin{figure}[htb]  
\centering
\vspace{100pt}
\epsfxsize=12cm
\leavevmode
\epsffile[0 0 1765 918]{\gpath aEfig4e}
%
\setlength{\unitlength}{25pt}
\begin{picture}(5.0,0.0)(0.0,0.0)
	\put(-6.0,10.0){
\begin{tabular}{rllccccc}
 $L(\bv_{56}) = q(x) = 1$  symmetry  $\ese$ 
                          & \!\!\!\!$-\!\!\!-\!\!\!\longrightarrow \!\! $ & 
  $\sltc$ & $\!\!\times\!\!$ & $\suth_c$ & $\!\!\times\!\!$ & $\uo_Q \quad$
                     &  (B)SM \\ 
	&    & $\,$Dirac & & $\mathbf{1}$ & & $1 \quad$  & 
  $\biggl\{ \!\! \begin{array}{c} e_L \vspace{-7pt} \\ e_R \end{array} \!\!\biggr\} $ 
\\	   
	&    & $\,$Dirac & & $\mathbf{3}$ 
	                                         & &  $\frac{1}{3} \quad$  & 
  $\biggl\{\!\! \begin{array}{c} d_L \vspace{-7pt} \\ d_R \end{array} \!\!\biggr\} $ 
\\ 
  &    &  $\,$vector  & & $\mathbf{1}$
	                              &  & $0 \quad$  & 
					  `$\nu_L$'    \\ 
  &    &  $\,$scalar & & $\mathbf{3}$
	                                    & &  $\frac{2}{3} \quad$  & 
  $\biggl\{\!\! \begin{array}{c}
     \mbox{`$u_L$'} \vspace{-7pt} \\ \mbox{`$u_R$'} \end{array} \!\!\biggr\} $ \\  
     &    &  $\,$scalar & & $\mathbf{1}$ & &  $ 0 \quad$  &  (DM)   \\  					 
    &     &  $\,$vector  & & $\mathbf{1}$ 
	                  & &  $  0 \quad $   &  (Higgs)   \\ 
\end{tabular} 
   }
    \put(-1.8,11.2){
	  $\begin{array}{r} \theta_l \ngap \\ \theta_{i,j,k} \ngap \\ 
	   n \ngap \\ a(6) \ngap  \\ p,\!m,\!a_1,\!a_8 \end{array}$
	}
	\put(-1.0,7.5){
	  $\begin{array}{r} \theta_{\lag} \ngap \\ \theta_{I,J,K} \ngap \\
	   N \ngap \\ A(6) \ngap  \\ \bv_4 \end{array}$
	}
%
%
  \put(-2.3,10.1){$X \left\{ \begin{array}{c} \vspace{7.5pt} \\ \end{array} \right.$}
  \put(-2.3,6.5){$Y \left\{ \begin{array}{c} \vspace{7.5pt} \\ \end{array} \right.$}
  \put(-3.2,11.0){$\mcX \left\{ \begin{array}{c} 
                    \\ \ngap \\ \\ \\ \end{array} \right.$}  
  \put(-3.2,7.4){$\mcY \left\{ \begin{array}{c} 
                    \\ \ngap \\ \\ \\ \end{array} \right.$} 
  \put(-5.4,8.75){$\bv_{56} \;\; \mbox{{\LARGE $\to$}} \; \left\{ \begin{array}{c} 
                 \\ \vspace{-3pt} \\ \\ \\ \\ \\ \\ \\ \\ \end{array} \right.$} 	
  \put(-3.2,5.25){$\alpha,\beta \qquad \qquad \;\;\; \alpha,\beta$}	
  \put(2.2,5.8){$\underbrace{\qquad \qquad \qquad \qquad \qquad
                             \qquad \qquad \qquad \qquad \quad}_{ }$}  	
  \put(-4.06,4.0){$\bv_{56} \equiv x
     = \binom{\alpha \;\;\! \mcX}{\!\mcY \;\;\! \beta} \inn F(\htho)$}	
  \put(-2.79,3.1){with $\quad\mcX,\mcY \inn \htho$}
  \put(-4.65,2.2){$X = \binom{p \;\: \bar{a}}{a \; m},
                  Y = \binom{P \;\:\! \mathbf{\bar{\mathnormal{A}}} }{A \; M}  \inn 
\htwo$}
  \put(-3.07,1.3){and $\quad\bv_4 \equiv \bh \inn \htwc$}
\end{picture}
\vspace{-23pt}	 
\caption{\setb  The full symmetry $\ese$ of  $\lvfs$  broken through the necessary 
identification of an external 4-dimensional spacetime manifold $M_4$. 
The projected components $\bv_4 \inn \TM_4 \:(\equiv \bh \inn \htwc)$ are now embedded 
in $\mcY$ similarly as they had been in $\mcX$ in equation~\ref{hinvts} (see also 
equations~\ref{hthoxy} and \ref{xoct3}). 
 The subcomponents $\theta_{\lag,I,J,K}$ are the four right-handed Weyl spinors in the 
$\mcY$ components
 corresponding to the four left-handed Weyl spinors $\theta_{l,i,j,k}$ of $\mcX$ 
listed in equation~\ref{thelijk}.
 The scalar dark matter (DM) candidates and the
 `vector-Higgs' $\bv_4$, included in the list of matter fields, exhibit `Beyond the 
Standard Model' features. }
\label{mforese}
\end{figure}

It can also be noted that there is very little redundancy in identifying these 
Standard Model structures in figure~\ref{mforese}. In fact the only residual pieces of 
$\bv_{56} \equiv x \in F(\htho)$ are the four Lorentz scalar and $\suth_c \times 
\uo_Q$ invariant components $\{\alpha, \beta, n, N \}$, which hence provide a set of 
candidates to account for the `dark sector' in cosmology, augmenting the $n\in \rrr$ 
component alone of figure~\ref{mtogmaph}(b) and table~\ref{esibr}, as will be 
discussed further towards the end of section~\ref{ee6}.

  The Higgs sector of the Standard Model is formulated in terms of suitably 
parametrised postulated Lagrangian terms introduced to describe the phenomena of 
electroweak symmetry breaking without explaining their physical origin. For the 
present theory the necessary identification of an external spacetime $M_4$ from 
subcomponents of the full form of time, as pictured for the form $\lvfs$ in 
figure~\ref{mforese}, necessarily breaks the full symmetry of time and hence in 
principle accounts for the origin of the symmetry breaking phenomena
  of the properties of matter observed \textit{in} spacetime.  Central to this 
symmetry breaking structure is the projected 4-vector $\bv_4 \in \TM_4$ itself which 
is associated with the Standard Model scalar Higgs state as described in (\cite{Unifi} 
section~8.3). There are three principle arguments for this association:
\begin{itemize}
 \item The choice of a particular `direction' for $\bv_4(x) \in \TM_4$ 
   for any $x\inn M_4$, as pictured in  
figure~\ref{mforese}, when considered globally over the extended spacetime is 
analogous to the spontaneous symmetry breaking associated with the choice of direction 
of magnetisation through the alignment of atomic spins in a ferromagnet below the 
critical temperature (as discussed after $\cite{Unifi}$ figure~13.3). Further,
 the frame-dependent choice of the four real components of $\bv_4 \in \TM_4$ is 
 analogous to the gauge-dependent choice of four real components for the vacuum value 
of the Higgs complex doublet field in a `Mexican hat' potential.  

 \item Variation in the magnitude $\vert \bv_4(x) \vert$ on the tangent space of $M_4$ 
in the projection out of the full form $\lvn$ is directly associated with a conformal 
warping of the spacetime geometry (as described following $\cite{Unifi}$ figure~13.1). 
This in turn directly generates variations in energy-momentum via the Einstein field 
equation, interpreted as $-\kappa T^{\mu\nu} := G^{\mu\nu}$, motivating the link 
between $\vert \bv_4 \vert$ and the scalar Higgs as the `origin of mass'.  
 (See also equation~\ref{gfromavt} and the discussion towards the end of 
section~\ref{ee6}). 

\item As noted in subsection~\ref{ee42} at the level of the $\esi$ symmetry of $\lvt$
 a `mock electroweak theory' can be constructed in which  natural
  $\sutw \times \uo \subset \esi$ subgroups can be identified which are broken to 
$\uo_Q$ through impingement upon the `vector-Higgs' $\bv_4 \inn \TM_4$ components
  ($\cite{Unifi}$ subsections~8.3.1 and 8.3.2). This impingement is proposed to 
generate the masses for the heavy gauge bosons $W^{\pm}$ and $Z^0$ in the complete 
theory, with masses for the leptons and quarks deriving from terms in the expansion of 
the full form $\lvn$
 ($\cite{Unifi}$ subsection~8.3.3). 

\end{itemize}

 In non-standard models in which the Higgs is not taken to be a fundamental scalar the 
Higgs might be considered as, or replaced by, a composite of spin-$\frac{1}{2}$ states
 (see for example \cite{Suss2}).
In principle such an approach might be connected with the present theory since spinors 
can also be composed to form objects transforming as a vector, such as for example
 $\bv_4 \in \TM_4$. Here such a spinor decomposition of the $\bv_4 \inn \TM_4$ part of 
$\mcY \inn \htho$ in figure~\ref{mforese} could be mirrored by a spinor decomposition 
of the corresponding 4-vector $\{p,m,a_1,a_8\}$ components in $\mcX \inn \htho$, which 
is required to describe the left-handed neutrino state. More broadly a spinor 
decomposition of the full set of $X \inn \htwo$ and $Y \inn \htwo$ components is 
required to account also for the Lorentz transformation properties of the left and 
right-handed $u$-quark states respectively. 

  This suggests that in the full theory, with a higher-dimensional form \mbox{$\lvn$} 
beyond $n=56$, more spinors will be introduced in place of some of the vector and 
scalar states in figure~\ref{mforese}. Such a structure might then be consistent with 
a combination of the Standard Model and composite Higgs approaches, in a physical 
theory providing a full explanation for the origin of symmetry breaking as described 
in this section. 
 
   In summary, via a sequence of mathematically natural augmentations of the general 
form of time through cubic determinants to the quartic norm $L(\bv_{56}) = q(x) =1$ of 
equation~\ref{lvfsq}, with an $\eseg$ symmetry broken over the $M_4$ base manifold,
  a series of esoteric properties of the Standard Model have been
  directly or partially
   uncovered, as summarised in figure~\ref{mforese}, bearing a close resemblance to a 
generation of Standard Model quarks and leptons with elements of an electroweak 
symmetry breaking structure also identified. 

  The rank-7 Lie algebra $\ese$ is in fact large enough to incorporate the full rank-6 
subgroup $\sltc \times \SM \subset \ese$, in principle including an $\sutw_L \times 
\uo_Y$ electroweak symmetry. However, empirically such an $\sutw_L$ symmetry
 is required to act upon left-handed doublets $\binom{\nu}{e}_{\! L}$ and 
$\binom{u}{d}_{\! L}$, within which each state is a Weyl spinor, while it is still the 
case at this stage in figure~\ref{mforese} that  the `$\nu$'-lepton and `$u$'-quark 
states are identified by their internal $\suth_c \times \uo_Q$ transformation 
properties and the corresponding components are not spinors under the external 
$\sltc$, as noted above and similarly as described in the previous subsection 
following table~\ref{esibr}.
  Hence these electroweak features remain to be identified, and are expected to be 
closely related to the breaking of an internal $\sufo \subset \esi \subset  \ese$ 
to an $\suth_c \times \uo_Q \subset \sufo$ as discussed at the end of the previous 
subsection. 

   In addition, perhaps the most obvious absence in figure~\ref{mforese} is the 
Standard Model feature of a full three generations of quarks and leptons. Since 
empirically the $\sutw_L$  action also mixes the three generations and the standard 
scalar Higgs transforms as an $\sutw_L$ doublet, all of the further significant 
symmetry structures  and states  of the Standard Model remaining to be fully 
identified over those summarised in figure~\ref{mforese}  are mutually correlated,  
and hence in principle might collectively be sought and identified in a further 
extension of the theory. This motivates consideration of a further augmentation to a 
yet higher-dimensional form of time, with a symmetry beyond both $\esi$ and $\ese$, as 
we describe in the following section.


\section{$\ee$ Symmetry and Time}
\label{ee5}

\subsection{Completing the Standard Model}
\label{ee51}

  The progression of multi-dimensional forms of time, beyond $\lvf$ of 
equation~\ref{lvfr} with Lorentz symmetry and as described in the previous section, is 
summarised here in table~\ref{vftovfs} along with the main Standard Model properties 
identified in the symmetry breaking over the base manifold $M_4$.
\begin{table}[htbp]
\centering
\begin{tabular}{|l|ll|l|}
 \hline
       & space    & symmetry   &       Standard Model properties   \\
 \hline  
 $\lvf$ &  $\bv_4 \inn \htwc$  & $\sltc$   &   $M_4$ vacuum (`vector-Higgs' only)
      \vspace{-5pt} \\
 	    &  4-dimensional & quadratic form$\!\!$  &    figure~\ref{fourd}  \\ \hline
 $\lvni$ &  $\bv_9 \inn \hthc$ & $\slthc$   &   internal $\uo_Q$ on a Weyl spinor
       \vspace{-5pt}  \\
        &  9-dimensional &  cubic form  &   equation~\ref{slthbr}       \\ \hline      
 $L(\bv_{27}) \! = \! 1$ 
    & $\bv_{26} \equiv \mcX \inn \htho$ &  $\mbox{E}_{6(-26)}$   &  
     $\suth_c \times \uo_Q$ on 4 Weyl spinors$\!$ 
	  \vspace{-5pt}   \\ 
	    & 27-dimensional & cubic form &   table~\ref{esibr}     \\ \hline
 $L(\bv_{56}) \! = \! 1$
    & $\bv_{56} \! \equiv \! x \! \inn \! F(\htho)$ & $\mbox{E}_{7(-25)}$    &   
    Dirac spinors and $L$,$R$ asymmetry 
	     \vspace{-5pt}  \\ 
	    & 56-dimensional & quartic form   &   figure~\ref{mforese}       \\
   \hline
  \end{tabular}
  \caption{\setb Series of higher-dimensional forms of time and their symmetries 
together with the principle features of the Standard Model accumulated, as described 
in detail in section~\ref{ee4}.}
\label{vftovfs}
\end{table} 

  We can then ask how a further stage in the progression of forms of time beyond 
table~\ref{vftovfs} might be identified, and the extent to which it may account for 
the further properties of the Standard Model required as described at the end of the 
previous section. Inspection of table~\ref{vftovfs} suggests an augmentation of the 
progression of spaces with $\htwc \to \hthc \to \htho \to F(\htho) \to \mcT$, where 
$\mcT$ represents the $n$-dimensional space for the full form of time upon which a 
quintic or higher-order polynomial norm denoted $L(\bv_n) = Q(\btt) = 1$, with 
$\bv_n \equiv \btt \in \mcT$, might be defined.
  In seeking a natural mathematical extension with 
  $\sltc \to \slthc \to \esi \to \ese \to \hat{G}$ from the symmetries of 
table~\ref{vftovfs} there is a strong hint that $Q(\btt) = 1$, as the full form of 
time, may be invariant under $\hat{G} = \ee$ as the full symmetry of time,
 uniquely terminating this series in $\ee$ as the largest exceptional Lie group.
 Ideally having first identified such an $\ee$ symmetry of $\lvn$ 
  we might \text{then} consider whether the physical implications of the additional 
structure match the further Standard Model properties sought, completing the picture 
of known elementary particle states and potentially making predictions for new 
phenomena.

  As described in subsection~\ref{ee23} one possible augmentation of the $\eseg$ 
symmetry of the quartic form $q(x)$ of equation~\ref{fquartic}, with $x \in F(\htho)$, 
is in fact known, namely the `quasiconformal' non-linear realisation of $\eeg$ as the 
symmetry preserving the `light cone separation' $d(e,f)=0$ of equation~\ref{defcone}, 
with $e,f \inn eF(\htho)$ elements of the 57-dimensional extended Freudenthal triple 
system. The quartic symplectic distance $d(e,f)$ between any two points incorporates 
the norm of an arbitrary element $e =(x,\tau) \in eF(\htho)$ which can be written, via 
 equation~\ref{qusydi} on setting $f=(y,\kappa)=0$, as:
\begin{equation}
 \label{nedeo}
   N(e) \; = \; d(e,0) \; = \; q(x) - \tau^2 
\end{equation}
  
   The $\eseg \subset \eeg$ subgroup acts as a linear 56-dimensional representation on 
 $x \in F(\htho)$, leaving $q(x)$ invariant, and upon $\tau \in \rrr$ as a singlet. 
Hence the light cone $N(e) = 0$ can be written as:
\begin{equation}
  \label{qxtaul}
     q(x) = \tau^2 \qquad \equiv \qquad L(\bv_{56}) = 1 
\end{equation}
 upon a trivial normalisation, with an $\eseg$ symmetry. Hence equation~\ref{lvfsq}, 
that is the final entry of table~\ref{vftovfs}, can be identified as a substructure of 
the $\eeg$ symmetry of $d(e,f) = 0$. However this latter structure is clearly 
\textit{not} of the form $\lvn$ of equation~\ref{lvo}, and
 the question remains regarding  whether $\ee$  can describe the symmetry of the 
hypothetical homogeneous polynomial form $Q(\btt) = 1$, again incorporating $q(x)$ in 
some manner but now interpreted as an augmentation to a higher-dimensional form of 
time, similarly as $q(x)$ itself incorporates the cubic form $\det(\mcX)$ in 
equation~\ref{lvfsq}.

  In particular, the most natural extension might be expected to involve the smallest 
non-trivial representation of $\ee$, that is of 248 dimensions, and hence the full 
form of time can be provisionally written as $L(\bv_{248}) = Q(\btt) = 1$, with 
 $\bv_{248} \equiv \btt \in \mcT$. While such a form, as an extension of the 
mathematical structures listed in table~\ref{vftovfs}, has not been explicitly 
identified in the literature we can consider in general terms whether the further 
augmentation $\esi \to \ese \to \ee$ acting on a space of $27 \to 56 \to 248$ 
dimensions might in principle accommodate the further Standard Model properties 
required as outlined at the end of the previous section.

 As we noted there for the matter fields identified over the base space $M_4$ in 
figure~\ref{mforese} the most evident discrepancy at the level of the broken $\eseg$ 
action on the space $F(\htho)$ is the need to account for \textit{three} generations 
of Standard Model quarks and leptons. With this in mind it is suggestive to note that 
while the smallest non-trivial representation of $\ese$ branches to representations of 
the subgroup $\esi$ as described in equation~\ref{esetoesi}, the $\mathbf{248}$ 
representation of $\ee$ exhibits the branching pattern under a subgroup containing 
$\esi$ as: 
\begin{equation}
  \label{eetoesi}
  \ee \supset \esi \times \suth : \quad
    \mathbf{248} \, \to \, 
	(\mathbf{27},\mathbf{3}) + (\mathbf{\overline{27}},\mathbf{\overline{3}})
	+ (\mathbf{78},\mathbf{1}) + (\mathbf{1},\mathbf{8}) 
\end{equation}	  
  Hence at the level of the standard representation structure there is an apparent  
threefold nature to the embedding of the smallest non-trivial representations of 
$\esi$, that is the $\mathbf{27}$ and $\mathbf{\overline{27}}$, within the 
$\mathbf{248}$ of $\ee$. While the embedding of $\esi$ in $\ese$, with a 
representation structure of equation~\ref{esetoesi}, led to the identification of both 
the \textit{left} and \textit{right}-handed states of Dirac spinors, through the 
explicit construction summarised in figure~\ref{mforese}, in general terms 
equation~\ref{eetoesi} suggests that the augmentation from $\esi$ to $\ee$ might in 
principle also incorporate \textit{three generations} of such states.

  The question then concerns how such a threefold structure might be explicitly 
realised in a homogeneous polynomial form of time $\lvtfe$ in the context of the 
present theory. In particular we can consider how such a structure might be built up 
from the $\sltc \times \suth_c \times \uo_Q \subset \esig \subset \eseg$ action on 
$\mcX \in \htho$ and $\mcY \in \htho$, as the  $\mathbf{27}$ and 
$\mathbf{\overline{27}}$ components of $x \in F(\htho)$ of equation~\ref{ftscomp}, as 
described in subsections~\ref{ee42} and \ref{ee43}, to a full Standard Model subgroup:
\begin{equation}
 \label{lorsme}
   \sltc \times \suth_c \times \sutw_L \times \uo_Y \subset \eeg
\end{equation}   
    acting on a 248-dimensional space $\mcT$ incorporating three generations. If such 
a space $\mcT$ can be constructed, built up from $F(\htho)$ by following these hints 
from the Standard Model, the unbroken $\ee$ symmetry action on the homogeneous form 
$Q(\btt)$, with $\btt\in \mcT$, could then be further studied as a mathematical 
structure of interest in its own right.

 As noted in subsection~\ref{ee42} it is known from both Dynkin analysis for $\esi$ 
and explicit study of the $\esig$ action on the space $\htho$ that the full Lorentz 
and Standard Model subgroup
  \mbox{$\sltc \times \suth \times \sutw \times \uo$} cannot be accommodated inside 
this exceptional Lie group alone. 
However it was also noted that there are
  $\sutw \times \uo \subset \esig$ subgroups acting on the components of $\htho$ that 
exhibit some of the properties sought for standard electroweak theory, with the 
association between $\bv_4 \in \TM_4$ and the Higgs further motivated in 
subsection~\ref{ee43}, as described in detail in (\cite{Unifi} section~8.3). Hence, 
since the $\esig$ level can partially account for electroweak properties,  a realistic 
$\sutw_L \times \uo_Y \subset \eeg$
  might be expected to be composed of generators which straddle the $\esig \subset 
\eeg$ subgroup, neither completely contained as a subgroup of $\esig$ nor fully 
independent of it.
 Similarly it is also known that the decomposition of $\mcX \in \htho$ and $\mcY \in 
\htho$ under the subgroup $\sltc \subset \esig$ does not accommodate Weyl spinor 
components for the candidate $\nu$-lepton and $u$-quark states, as summarised in 
table~\ref{esibr} and further discussed for figure~\ref{mforese} for the $\eseg$ 
extension.

  For the above reasons the hypothetical full physical symmetry breaking pattern 
  $\sltc \times \suth_c \times \sutw_L \times \uo_Y \subset \ee$ of 
equation~\ref{lorsme} is generally  \textit{not} expected to be neatly aligned with an 
$\ee$ decomposition of the form of 
 equation~\ref{eetoesi}  containing the 
  $\sltc \times \suth_c \times \uo_Q \subset \esi$ structure already identified.
   Hence unlike the progression of forms listed in table~\ref{vftovfs} through to the 
case $\esi \to \ese$, for which the extension from $\bv_{27}$ to $\bv_{56}$ matched 
the component structure of equation~\ref{esetoesi}, the further augmentation to an 
$\ee$ symmetry of $\lvtfe$ is expected to exhibit a different breaking pattern. That 
is the final and largest `Russian doll' of the sequence may be somewhat `skewed' with 
respect to the others, with an $\ee$ decomposition and its representations not 
directly aligned with the intermediate substructures of table~\ref{vftovfs}.
 The question then concerns the possible construction of the higher-dimensional form
  itself.

  Further, as described in subsection~\ref{ee42}, standard Dynkin analysis of 
representations based on complex Lie algebras only serves as a guide here, since the 
octonion algebra is central to the description of these symmetry structures. Hence 
here we consider how we might proceed towards the construction of the desired $\ee$ 
symmetry by studying the mathematical patterns that emerge in the explicit algebraic 
augmentations of the forms of time that have already been identified.  

 In table~\ref{vftovfs} the 9-dimensional cubic form $\lvni$ with $\bv_9 \inn \hthc$ 
is neatly contained within the 27-dimensional cubic form $\lvt$ as can be seen by 
restricting the elements $a,b,c \inn \ooo$ of $\mcX \inn \htho$ in 
equation~\ref{hthoxy} to subspaces with
 $a,b,c \inn \ccc$ in equation~\ref{detpmn}. A similar restriction can be applied to 
the components $X \inn \htwo$ and $\theta \inn \ooo^2$ in equation~\ref{vcubic}, 
through which equation~\ref{stglvni} below can be obtained. Equation~\ref{vcubic} 
itself also demonstrates how in place of $\lvni$ an intermediate 10-dimensional form 
of time 
 $L(\bv_{10}) = \det(X) = 1$ can be considered, with $\bv_{10} \equiv X \inn \htwo$ 
embedded within the larger spaces as shown in equation~\ref{xoct3} and 
figure~\ref{mforese}. This second route is taken for example in (\cite{Novel} 
table~5), with the choice of these two possibilities 
  discussed for (\cite{KKone} equation~95) and before equation~\ref{lvts} here.

  Taking into account this choice of progression we have the following series of 
augmentations of homogeneous polynomial forms of time based on the extension of spaces 
and symmetries:

\begin{itemize}
\item  $\sltc$ on  $L(\bv_4) = \det(\bh) = 1 \quad \to
						   \quad \slthc$ on $L(\bv_9) = \det(\bv_9) = 1$ 
	
\item[] $\bv_4 \equiv \bh \in \htwc \to \bv_9 \in \hthc \;$
	   via $2 \times 2 \to 3 \times 3$ determinant (equation~\ref{hinvnine})	
	  \begin{equation}
	  \label{stglvni}
	  \det (\bh) \; \to \; \det(\bv_9)
                = n\det(\bh) - 2 \bh \!\cdot\! (\psi \psi^{\dag}) 
	  \end{equation}	
	  	
\vspace{5pt}										 
\item[{\bf or}]  $\sltc$ on  $L(\bv_4) = \det(\bh) = 1 \quad \to
						   \quad \sltwoo$ on $L(\bv_{10}) = \det(X) = 1$ 
	
\item[] $\bv_4 \equiv \bh \in \htwc \to \bv_{10} \equiv X \in \htwo \;$
	   via 	$\ccc \to \ooo$ division algebra  (\cite{Unifi} section~6.3)
	  \begin{equation}
	  \label{stglvte}
	    \det (\bh) \; \to \;  \det(X)
                = \det(\bh) - \vert a(6) \vert^2  
	  \end{equation}	
 
\vspace{5pt} 
\item  $\sltwoo$ on  $L(\bv_{10}) \! = \! \det(X) \! = \! 1 \quad \to
					 \quad \esig \! \equiv \! \sltho$
					  on $L(\bv_{27}) \! = \! \det(\mcX) \! = \! 1$ 
	
\item[] $\bv_{10} \equiv X \in \htwo \to \bv_{27} \equiv \mcX \in \htho \;$
	   via $2 \times 2 \to 3 \times 3$ determinant (equation~\ref{vcubic})	
	  \begin{equation}
	  \label{stglvts}
	   \det (X)\; \to \;  \det(\mcX)
                = n\det(X) - 2X \!\cdot\! (\theta \theta^{\dag})
	  \end{equation}	
	  
\vspace{5pt}										 
\item  $\esig$ on  $L(\bv_{27}) = \det(\mcX) = 1 \quad \to
						   \quad \eseg$ on $L(\bv_{56}) = q(x) = 1$ 
	
\item[] $\bv_{27} \equiv \mcX \in \htho \to \bv_{56} \equiv x \in F(\htho) \;$
	   via cubic $\to$ quartic form
	                 (equation~\ref{fquartic}) 
	  \begin{equation}
	  \label{stglvfs}
	   \det (\mcX) \, \to \,  q(x)
                = -2\lbrack \alpha\beta - (\mcX,\mcY)\rbrack^2 \, - \,
       8\lbrack\alpha \det(\mcX) + \beta\det(\mcY) - (\mcX^{\sharp},
	                                          \mcY^{\sharp})\rbrack
											  \qquad 
	  \end{equation}

\end{itemize}                 
 
   We note in particular that in progressing from 10 to 27 dimensions with $\htwo \to 
\htho$  the expression for $L(\bv_{27}) = \det(\mcX) =1$ with $\esig$ symmetry 
contains a single term $n \det(X)$ in equation~\ref{stglvts}, within which both $n$ 
and $\det(X)$ are invariant under $\sltwoo \in \esig$. On the other hand in the 
progression from 27 to 56 dimensions in the following stage the expression for 
$L(\bv_{56}) = q(x) =1$ with $\eseg$ symmetry contains \textit{two}  terms of a 
similar form, namely $\alpha \det(\mcX)$ and $\beta\det(\mcY)$ in 
equation~\ref{stglvfs}, with each of the four factors being invariant under the 
subgroup $\esig \subset \eseg$. We then conjecture that for a further stage in this 
progression there may be \textit{three} terms of the form $\lambda \:\! q(x)$, with 
$\lambda \inn \rrr$ and each factor invariant under an $\eseg \subset \eeg$ subgroup 
of the full symmetry $\eeg$ of the proposed full form of time, and hence we consider 
the extension:

\begin{itemize}
\item  $\eseg$ on  $L(\bv_{56}) = q(x) = 1 \quad \to
						   \quad \eeg$ on $L(\bv_{248}) = Q(\btt) = 1$ 
	
\item[] $\bv_{56} \equiv x \in F(\htwo) \to \bv_{248} \equiv \btt \in \mcT \;$
	   to hypothetical  $\ge\,$quintic form $Q(\btt)$
	  \begin{equation}
	  \label{stglvtfe}
	   q(x) \; \to  \; Q(\btt) \sim
		f\big((\lambda \:\! q(x) + \eta \:\! q(y) + \sigma \:\! q(z))
		    \,,\:\! (x,y,z)\big)	   
											  \qquad 
	  \end{equation}	  	
\end{itemize}  
 where $f$ is a function of
   $x,y,z \in F(\htho)$ and $\lambda, \eta, \sigma \in \rrr$. The three parameters 
  $\lambda, \eta, \sigma$ can be chosen such that they are proportionately 
  scaled by the  dilation 
    $\Delta \in \ee$, described for equation~\ref{gradxf}, in the same way as 
$\tau^{-2}$ from an element $e = (x,\tau) \inn eF(\htho)$ in equation~\ref{qusydi} for 
example. While in the latter case the dilation scaling for $\tau \inn \rrr$ is such 
that equation~\ref{defcone} is invariant, in the former case terms of the form
 $\lambda \:\! q(x)$   will then be invariant under the action of $\Delta \in \ee$. 
This structure is analogous to, and motivated by, the invariance of $n \det(X)$ under 
the dilation subgroup of $\esi$, as described for equation~\ref{vcubic}, and the 
invariance of $\alpha \det(\mcX)$ under the $\ese$ dilation (\cite{Unifi} 
equation~9.30).
    It is also suggested that a `trilinear product' $(x,y,z)$ may feature in the 
expression for $Q(\btt)$ in equation~\ref{stglvtfe}, by analogy with the bilinear 
products of the form $(\mcX, \mcY)$ in the expression for $q(x)$ in 
equation~\ref{stglvfs}. There may also be further terms involving $x,y,z$ in 
combination with $\lambda, \eta, \sigma$ 
   as well as possible additional parameters.

  The progression from $L(\bv_{10})=1$ to $\lvt$ introduced the further component 
$\theta \in \ooo^2$ in equation~\ref{stglvts}, as explained for equations~\ref{vcubic} 
and \ref{xoct3}, which decomposes as the set of Weyl spinors of equation~\ref{thelijk} 
under the external $\sltc \subset \esig$ symmetry as reviewed in 
subsection~\ref{ee42}. In turn the necessary incorporation of both $\mcX$ and $\mcY 
\in \htho$ in the extension from $\lvt$ to $\lvfs$ in equation~\ref{stglvfs} leads to 
the identification of both left and 
right-handed spinors, as described in subsection~\ref{ee43}. On noting this pattern 
the conjecture here is that an expression involving $x,y,z\in F(\htho)$ as proposed in 
equation~\ref{stglvtfe} for the full form $\lvtfe$ will account for three generations 
of Standard Model spinor states as an explicit expression for the suggestion made 
following equation~\ref{eetoesi}.

  The property of normed division algebras in equation~\ref{abnorm} and of matrix 
determinants in equation~\ref{abdet}, described in subsection~\ref{ee22}, motivated a 
consideration of these structures for symmetries of the forms of time as introduced in 
section~\ref{ee3}. However the cubic form $L(\bv_{27}) = \det(\mcX) = 1$, while 
employing both of these structures in equation~\ref{detpmn}, can also be written in 
terms of trace functions as equation~\ref{dettra}, while the quartic form 
  $L(\bv_{56}) = q(x) = 1$ of equation~\ref{lvfsq} cannot be expressed as a simple 
determinant function. On the other hand both of these temporal forms and their 
symmetries involve the octonion algebra in an essential manner. Similarly here for 
equation~\ref{stglvtfe}  the overall expression for $L(\bv_{248}) = Q(\btt) = 1$ is 
not expected to be a determinant function
while the determinant of various subcomponents may feature
 and octonion composition is expected to play a significant role in describing this 
structure and its symmetry. 
 
   The provisional structure of equation~\ref{stglvtfe} contains $3 \times (56 + 1) = 
171$ real variables, well short of the 248 dimensions for the smallest non-trivial 
representation of $\ee$. However, in generalising to the higher-dimensional form of 
time another key feature of the Standard Model required to be identified is a set of 
components, correlated with $X\in \htwo$ at the $\esig$ stage and also
 $Y\in \htwo$ at the $\eseg$ stage, as 
 described for table~\ref{esibr} and figure~\ref{mforese} respectively, which will 
 transform as a set of Weyl spinors under the external $\sltc \subset \eeg$ to account 
fully for the $\nu$-lepton and $u$-quark states. 
 While $X$ and $Y$ are embedded directly in the higher-dimensional structures of 
equations~\ref{stglvts} and \ref{stglvfs}, via equations~\ref{hthoxy} and \ref{xoct3}, 
for the further extension the need to open up $X = \binom{p \;\: \bar{a}}{a \; m},
            Y = \binom{P \;\: 
			\mathbf{\bar{\mathnormal{A}}}
			}{A \; M} \in \htwo$, or more specifically 
  the $a,A \in \ooo$ degrees of freedom, to act as spinors
  will necessarily increase the total number of components involved.

   Focussing on $X \inn \htwo$ this is
  studied in detail in (\cite{Unifi} section 9.1) where it is also described how in 
principle this can be achieved while maintaining the appropriate $\suth_c \times 
\uo_Q$ transformation properties of the components of $a\in \ooo$ listed in 
table~\ref{esibr} and figure~\ref{mforese} which already match those of the 
$\nu$-lepton and $u$-quark states. This will necessarily augment the 171 real 
components of equation~\ref{stglvtfe} with the
 aim of converging upon a homogeneous polynomial form $L(\bv_n)=Q(\btt) = 1$, 
potentially with $n=248$, and the ambition of incorporating the complete set of 
Standard Model states under the breaking of the full $\eeg$ symmetry over $M_4$.

   As a generalisation from $q(x) = 1$ this new structure developed from 
equation~\ref{stglvtfe} should be expected to not only incorporate the necessary 
spinor states for the $\nu$-lepton and $u$-quarks from the $X,Y \inn \htwo$ 
components, but to see these associated with the existing $e$-lepton and $d$-quark 
spinor states respectively from the corresponding $\theta \inn \ooo^2$ components in 
figure~\ref{mforese} as doublets of an $\sutw_L \in \eeg$ symmetry, all accommodated 
within the components of the homogeneous polynomial form $L(\bv_{248}) = Q(\btt) = 1$ 
of quintic or higher order. 

   As noted above in general in extending from $q(x) = 1$ with $x\in F(\htho)$ and 
with an $\eseg$ symmetry to the full hypothetical form $Q(\btt) = 1$ with $\eeg$ 
symmetry this 
form may involve three elements $x,y,z\in F(\htho)$, or a closely related structure.
 The final term in the provisional expression of equation~\ref{stglvtfe} alludes to
  a possible role for the triple product $T(x,y,z)$ as defined in 
equations~\ref{tqdef} and \ref{texpl} and/or the Jordan triple product
  $\{\mcX,\mcY,\mcZ\}$ of equation~\ref{jtphtho}. At the latter level of $\mcX, \mcY, 
\mcZ \in \htho$, and with $Q(\btt) \in \rrr$ being a scalar quantity, the cubic form 
$(\mcX, \mcY, \mcZ)$  of equation~\ref{cubicj} or \ref{cubiclin}, as a direct 
generalisation of $L(\bv_{27}) = \det(\mcX) = 1$ from equation~\ref{detxxx} or 
\ref{detpmn}, might also feature in a significant way. 

 As a tentative step towards deducing what the structure of $Q(\btt)$ might look like 
we take three elements $\mcX, \mcY, \mcZ \in \htho$, with the first two having the 
components of equation~\ref{hthoxy} while $\mcZ$ is assigned corresponding components 
$p',m',n' \in \rrr$ and $a',b',c' \in \ooo$. Then we find via substitution into 
equation~\ref{cubicj}, and cross-checking with (\cite{Ohwa} equation 214), that the 
cubic form $(\mcX,\mcY,\mcZ) \in \rrr$ can be written out explicitly as the following 
cubic polynomial function:
\begin{equation}
  \begin{array}{ll}
     (\mcX,\mcY,\mcZ)  = & \fh {\big(}
	 pMn' + p'mN + Pm'n + pm'N + Pmn' + p'Mn  {\big)}  \\
    &	- {\big(} p \langle B,b' \rangle + P \langle b',b \rangle +
	                 p' \langle b,B \rangle + m \langle C,c' \rangle +
					 M \langle c',c \rangle + m' \langle c,C \rangle   \\
	& \qquad \qquad  \qquad \qquad	 \qquad \qquad	\qquad \qquad					  
					+ n \langle A,a' \rangle + N \langle a',a \rangle +
					 n' \langle a,A \rangle {\big)}   \\
    & + \mbox{Re}{\big(} bCa' + cAb' + aBc' + bc'A + ca'B + ab'C
                   		{\big)} 
  \end{array}  \label{xyzlong}
\end{equation}	 								 
	 where the octonion inner product $\langle B,b' \rangle$ is defined in
	  equation~\ref{octinner}.
   This expression can be considered as a generalisation  either from 
equation~\ref{incomp} for the bilinear form $(\mcX,\mcY)$  or from 
equation~\ref{detpmn} for the cubic norm $\det(\mcX) = \frac{1}{3}(\mcX,\mcX,\mcX)$.
 One possible means of introducing Weyl spinors into the 
 $X = \binom{p \;\: \bar{a}}{a \; m} \in \htwo$ subcomponents of $\mcX \in \htho$ is 
to express $X$ in terms of a new 2-component object $\tthX = \binom{\bar{r}}{s} \in 
\ooo^2$ as the product:
\begin{equation}
  \label{xththfull}
   X \, = \, \left( \begin{array}{cc} p & \bar{a} \\ a & m \end{array} \right) 
     \, = \, \tthX\tthXd
	 \, = \, \left( \begin{array}{c} \bar{r} \\ s \end{array}  \right)
             \left( \begin{array}{cc} r & \bar{s} \end{array}  \right)
	 \, = \, \left( \begin{array}{cc} \bar{r}r & \bar{r}\bar{s} \\
	               sr & s\bar{s} \end{array} \right)
\end{equation}

  As described in (\cite{Unifi} section~9.1, see equation~9.3) $\tthX$ then itself 
decomposes into a set of four Weyl spinors under the external $\sltc \subset 
 \sltwoo \subset \esig \subset \eseg$ symmetry. This also shows how `further 
dimensions' are needed to incorporate the new spinors, with the 10 real components of 
$X \in \htwo$ here replaced by the 16 of
 $\tthX \in \ooo^2$. Applying a similar augmentation to the corresponding 
subcomponents \mbox{$Y = \binom{P \;\: \mathbf{\bar{\mathnormal{A}}}}{A \; M},
  Z = \binom{p' \;\: \bar{a}'}{a' \; m'} \in \htwo$} of
 $\mcY, \mcZ \in \htho$ we have collectively the replaced components:
\begin{equation}
   \begin{array}{ccc}
     p \to (\bar{r}r) \quad & \quad 
	      P \to (\mathbf{\bar{\mathnormal{R}}}R)
		         \quad & \quad  p' \to (\bar{r}'r')  \\
     m \to (s\bar{s}) \quad & \quad
	       M \to (S \mathbf{\bar{\mathnormal{S}}})
		          \quad & \quad m' \to (s'\bar{s}')  \\
	 a \to (sr) \quad & \quad A \to (SR) \quad & \quad a' \to (s'r') 
	\end{array}  	\label{pmarep}     	          
\end{equation} 
where the first column can be read off directly from equation~\ref{xththfull}.

  On applying all of the substitutions of equation~\ref{pmarep} in 
equation~\ref{xyzlong} it can easily be seen that while the second and fourth rows now 
contain only $4^{\mathrm{th}}$-order terms, in the first and third row we have 
$5^{\mathrm {th}}$-order terms, and hence we no longer have a \textit{homogeneous} 
polynomial expression. However on setting $\mcX = \mcY = \mcZ$ for 
equation~\ref{xyzlong} the $5^{\mathrm {th}}$-order terms cancel out.
 These observations are  similar to the case of substituting several spinors, $\tthX,  
\phi_{\! \mbox{\tiny{$X$}}} \ldots \in \ooo^2$,  via 
 $X = \tthX \tthXd + \phi_{\! \mbox{\tiny{$X$}}} \phi_{\! \mbox{\tiny{$X$}}}^{\dag}+ 
\ldots$ as an extension of equation~\ref{xththfull} into the expression for 
$\det(\mcX)$ in equation~\ref{detpmn} or \ref{stglvts}, which then also contains 
$4^{\mathrm {th}}$-order and $5^{\mathrm {th}}$-order terms, with the latter 
cancelling out for  $X = \tthX \tthXd$ alone, as described for (\cite{Unifi} 
equations~9.5--9.8).

  In principle the aim is then to incorporate structures similar to 
   equation~\ref{xththfull} into a generalisation from $q(x), q(y), q(z)$ and 
$(x,y,z)$  
    in equation~\ref{stglvtfe}, with $x,y,z \in F(\htho)$, in a manner such that the 
augmented terms  collect together in $Q(\btt)$ as a homogeneous polynomial form, with 
any inhomogeneous terms automatically cancelling out in the full expression, 
consistent with the underlying basis of equation~\ref{lvo}. This full form of time 
$Q(\btt)$ may closely involve the octonion triality symmetry in obtaining Weyl spinors 
under the external $\sltc \subset \ee$ while maintaining the appropriate charges under 
the internal subgroup $\suth_c \times \uo_Q \subset \esi \subset \ese$ of 
table~\ref{esibr} and figure~\ref{mforese} to describe $\nu$-lepton and $u$-quark 
states  (as also suggested for \cite{Unifi} equations~9.9--9.12). This intrinsic 
employment of octonions in the symmetry structure again signals that Dynkin analysis 
can only serve as an approximate guide, as we consider further in the following 
subsection.

  It may be that \textit{all} components of $\bv_{248} \equiv \btt \in \mcT$ transform 
directly as spinors or scalars under the external $\sltc$ of the broken $\ee$ action 
on \mbox{$L(\bv_{248}) = Q(\btt) = 1$}. In this case sets of components apparently 
transforming as both left and right-handed neutrino states might be identified. 
However the necessity of identifying an external spacetime arena $M_4$ with the 
components $\bv_4 \in \TM_4$ projected out of the $\bv_{248}$ components, as depicted 
for the simpler model in figure~\ref{mtogmaph} as well as in figure~\ref{mforese}, 
imply that some of these spinor components, for example the 
  `right-handed neutrino' slots, will necessarily be combined in a composite external 
4-vector $\bv_4$ (as discussed for \cite{Unifi} equation~9.52). This composition might 
be similar to that described in equation~\ref{xththfull}, except with $\bv_4 \equiv 
\bh \in \htwc$ on the left-hand side and in principle with several spinors 
$\theta,\phi,\psi\inn \ccc^2$ involved in forming 
 $\bv_4\equiv \bh = \theta\theta^{\dag}+\phi\phi^{\dag}+\psi\psi^{\dag} \inn \htwc$.
  With the $\ee$ symmetry breaking projection of $\bv_4 \in \TM_4$ closely related 
with the phenomena of the Standard Model Higgs mechanism this structure is analogous 
to composite Higgs models, as noted in the discussion towards the end of 
subsection~\ref{ee43}.

As also noted in subsection~\ref{ee43}
  in the Standard Model itself the Higgs mechanism is introduced as an ad hoc 
phenomenological   model to describe and parametrise, but not \textit{explain}, the 
empirical observations of electroweak symmetry breaking. In the present theory the 
origin of the symmetry breaking is in the necessary identification of 4-dimensional 
spacetime itself, as described for equation~\ref{lvnitr} and figures~\ref{mtogmaph} 
and \ref{mforese}, with the components \mbox{$\bv_4 \in \TM_4$} projected out of the 
full form of time, considered here as the hypothetical form $L(\bv_{248}) = Q(\btt) 
=1$ with an $\ee$ symmetry. More broadly, with the Standard Model as a whole 
considered as a phenomenological parametrisation of empirical observations, the aim in 
principle is to account for all of these phenomena, with the symmetry breaking 
structure of the theory key to this ambition.

 In summary, we have considered the progression from the vacuum case of $\lvf$ over 
$M_4$ in figure~\ref{fourd} through a series of higher-dimensional forms of time 
projected over the same base manifold resulting in matter fields resembling a series 
of properties of the Standard Model. These include the form $\lvni$ with a broken 
$\slthc$ symmetry for which an `electromagnetic' gauge field in interaction with a 
Weyl spinor was identified in figure~\ref{mtogmaph}, 
 succeeded by the fractional charges and colour triplets of table~\ref{esibr} for 
$\lvt$ under the broken $\esi$ symmetry and the left-right asymmetry in 
figure~\ref{mforese} for the components of $\lvfs$ under the broken $\ese$ symmetry.
  In addition to structures resembling one generation of Standard Model leptons and 
quarks the symmetry breaking mechanism itself exhibits features analogous to the Higgs 
sector. Given the simplicity of the theory, in deriving from the one dimension of time 
only, the degree of explanatory power already uncovered up to the level of this $\ese$ 
symmetry of time is noteworthy.

 In turn while a full  $\ee$ action is in principle large enough to incorporate three 
generations and the full phenomena of electroweak symmetry breaking in the projection 
over $M_4$ it is non-trivial to see in detail how this might arise since we currently 
lack a specific homogeneous polynomial form $L(\bv_{248}) = Q(\btt) =1$ as a natural 
extension from table~\ref{vftovfs}. 
On the other hand the potential existence of such an object, as provisionally 
described for equation~\ref{stglvtfe}, together with 
the fact that it is \textit{not} straightforward to see in detail how it might carry 
further explanatory power in terms of Standard Model structures, implies that the 
theory is \textit{testable} in this non-trivial theoretical sense. In the following 
subsection we consider how the pursuit of this higher-dimensional form of time might 
further relate to some of the $\ee$ studies reviewed in section~\ref{ee2}.

\subsection{Further Connections with $\ee$ Studies}
\label{ee52}

  The highest-dimensional form of time that we have considered explicitly,
   $L(\bv_{56}) = q(x) = 1$ with $x \in F(\htho)$ and an $\eseg$ symmetry as 
introduced in subsection~\ref{ee23}, extends to an $\eeg$ realisation on the 
57-dimensional space of the extended Freudenthal triple system as a symmetry 
preserving the light cone separation $d(e,f)=0$ with $e,f \in eF(\htho)$, as also 
described in subsection~\ref{ee23}. 
 While the structure of equation~\ref{defcone} arose from considering the spaces 
$\htho$ and $F(\htho)$ as `generalised spacetimes' \cite{Gunay3,Gunay2} here we are 
primarily interested in the `general form of time'  \cite{Unifi,Novel,KKone} as 
motivated here in section~\ref{ee3} and derived for equation~\ref{lvo}.
 Here 4-dimensional `spacetime' itself is identified through a particular `form of 
time', as described for equation~\ref{fourtran} and figure~\ref{fourd}, with the 
necessary projection of the corresponding components $\bv_4 \equiv \bh \in \htwc$ out 
of the full form of time, pictured in figure~\ref{mforese} for $\lvfs$, breaking the 
full symmetry of time and leading to the empirical properties of the matter fields 
identified over $M_4$ as presented through section~\ref{ee4}.

 In the previous subsection we considered the possible construction of a homogeneous 
polynomial form $\lvtfe$ over a 248-dimensional space with an $\ee$ symmetry as a 
further progression from the forms of time listed in table~\ref{vftovfs}, in principle 
involving three elements of $F(\htho)$ as suggested in equation~\ref{stglvtfe}, with 
the aim of incorporating the complete set of Standard Model states and symmetry 
structures, including a full three generations of leptons and quarks. We can also ask 
whether this currently hypothetical $\ee$ action on $L(\bv_{248}) = Q(\btt) = 1$ might 
be related to or built upon the $\ee$ realisation on the 57-dimensional space 
$eF(\htho)$ by in some sense opening up and rearranging the components within the  
expression $d(e,f) = 0$ in a manner compatible with a homogeneous form $\lvn$, 
potentially with $n=248$. 

  This is analogous to the fact that $\eseg$ can itself be considered as a conformal 
group for the space $\htho$ that leaves the cubic light cone $\det(\mcX-\mcY) = 0$ 
invariant, as described for equation~\ref{htholc}, while also being the symmetry group 
of the homogeneous quartic form $q(x)$ of equation~\ref{fquartic}.
  As noted towards the end of subsection~\ref{ee23} employment of the split octonions 
leads to a conformal realisation of the real form E$_{7(7)}$ as set of non-linear 
actions on the space h$_3\ooo_s$, as described for (\cite{Gunay3} equation~47). In 
addition to leaving a 27-dimensional cubic light cone invariant (\cite{Gunay3} 
equation~45), the same group E$_{7(7)}$ acts via a linear 56-dimensional 
representation on a quartic invariant (\cite{Gunay3} equation~18). 
 An explicit identification of a relationship between these two kinds of symmetry 
actions as a realisation in $\rrr^{27}$ and a representation in $\rrr^{56}$ for the 
analogous case of $\htho$ for the non-split octonions and the real form $\eseg$    
  could provide a clue for how the 57-dimensional realisation of $\eeg$ acting upon 
$d(e,f) = 0$ might be unfolded into  a possible $\eeg$ action leaving invariant a 
homogeneous polynomial form $Q(\btt)=1$ as a 248-dimensional representation.
 Further, the observation that  $L(\bv_{56}) = q(x) = 1$  with $\eseg$ symmetry is 
embedded within the $\eeg$ symmetry of $d(e,f) = 0$, as described for  
equations~\ref{nedeo} and \ref{qxtaul}, may assist in the identification of this 
augmented form of time.

  The quartic distance $d(e,f)$, defined in equation~\ref{qusydi}, itself incorporates 
terms involving $q(x\!-\!y)$ and the form $\{x,y\}$ of equation~\ref{biasqf}, with 
$x,y \in F(\htho)$ collectively formed of 114 real components, and hence might already 
be associated with \textit{two} generations of fermions in the context of 
figure~\ref{mforese}. 
 From equation~\ref{tqdef} both the antisymmetric bilinear form $\{x,y\}$ and the 
symmetric quartic form $q(x,y,z,w)$ (the linearisation of the quartic norm $q(x)$ in 
equation~\ref{qxyzw}) are closely related to the symmetric triple product $T(x,y,z)$, 
and hence they might also relate to the provisional `trilinear product' denoted 
$(x,y,z)$ in equation~\ref{stglvtfe}, presumed to be associated with a full 
\textit{three} generations of quarks and leptons.

  Terms in the full form of time  $L(\bv_{248}) = Q(\btt) =1$, provisionally 
introduced in  equation~\ref{stglvtfe},  
 might also then in principle contain factors such as 
  $\{x,y\}$, as well as $q(x)$ and $\lambda$, each of which is invariant
   under an
   $\eseg \subset \eeg$ subgroup. Again, this is analogous to the $\esig \subset 
\eseg$ invariance of the $\alpha, \beta, \det(\mcX)$ and $\det(\mcY)$ factors in the 
expression for $L(\bv_{56}) = q(x) =1$ of equation~\ref{stglvfs}.
 In all cases these structures ultimately need to incorporate a modification of $q(x)$ 
and $\det(\mcX)$ to open up the $X\in \htwo$ or $a\in \ooo$ subcomponents, for example 
similarly as described in equation~\ref{xththfull}, to incorporate further $\sltc \in 
\eeg$ spinor states while being compatible  with an overall homogeneous polynomial 
form for the full $L(\bv_{248}) = Q(\btt) =1$ expression.

  As noted in the discussion following equation~\ref{eetoesi} the need to identify the 
above spinor decompositions implies that any embedding of the full Standard Model 
structure is not expected to directly match the $\ee \supset \esi \times \suth$ 
branching pattern of that equation. It has also been seen from 
equations~\ref{xyzlong}--\ref{pmarep} that augmenting the cubic form 
$(\mcX,\mcY,\mcZ)$ of equations~\ref{cubicj} and \ref{cubiclin} to incorporate more 
spinor components leads to quartic and quintic terms. On the other hand since the 
terms such as 
$\lambda \:\! q(x)$ in equation~\ref{stglvtfe}  are already of quintic order, 
incorporating spinors in such an expression is expected to lead to terms of at least 
sixth or seventh order. This raises the question of whether the full form 
$L(\bv_{248}) = Q(\btt) = 1$ might in fact converge upon an eighth order polynomial, 
that is an octic $\ee$ invariant \cite{CedP,Tala,GarG,BeRu} as reviewed in 
subsection~\ref{ee21} and which might then itself provide a significant guide.

  Indeed, as was described in subsection~\ref{ee21}, in seeking a 248-dimensional form 
$\lvtfe$  the lowest-order invariant homogeneous polynomial incorporating an 
irreducible representation of $\ee$ beyond the quadratic Killing form is such an 
eighth order expression. More generally the $\ee$ tensor invariants are defined in 
terms of the Casimir operators in the 248-dimensional $\ee$ algebra.
A similar association is not possible between elements of the 133-dimensional $\ese$ 
Lie algebra and the 56-dimensional $\eseg$ quartic invariant $q(x)$, however  a 
correspondence can still be made between $q(x)$ and an invariant fourth order tensor. 
The relation between such a quartic invariant tensor and the Freudenthal triple system 
is indicated for E$_{7(7)}$ acting on $F(\htho_s)$, the case alluded to above, in 
(\cite{Gunay3} section~2.3). For the case of $\eseg$ the invariant tensor is 
explicitly constructed in \cite{Gunay2} on employing an
   $\mbox{SO}(2,10) \subset \eseg$ basis. (As noted after equations~\ref{commhtwc}
    and \ref{htholc} the non-compact group $\mbox{SO}(2,10)$ is the conformal group 
for $\htwo$, while $\eseg$ is the conformal group for $\htho$). 
  With elements $X^{\mu,a}$ and $\psi^{\alpha}$ transforming as an SO$(2,10)$ vector 
and spinor respectively the quartic invariant $\mcI_4$ includes for example the term 
(see \cite{Gunay2} equation~82 for details of the notation):
\begin{equation}
  \label{iquart}
   \mcI_4 \;=\; \ldots \; + \; 2\epsilon_{ab}X^{\mu,a}X^{\nu,b} \psi^{\alpha}
      (C\Gamma_{\mu\nu})_{\alpha\beta}\psi^{\beta}\; +\; \ldots
\end{equation} 

   The octic $\ee$ invariant $X_8$ is written out explicitly in (\cite{CedP} 
equation~2.3) in an $\mbox{SO}(16) \subset \ee$ basis, where SO(16) is a maximal 
subgroup of the compact real form of $\ee$.
 With generators $T^{ab}$ in the adjoint and $\phi^{\alpha}$ in the spinor 
representation of SO(16) there are many terms in this case including as an example:
\begin{equation}
  \label{xoctic}
  X_8 \; =  \; \ldots \; 
-\frac{3}{64}T^{ab}T^{cd}T^{ef}T^{gh}(\phi_{\alpha}\phi^{\alpha})
    (\phi^{\beta}(\Gamma_{abcdefgh})_{\beta\gamma}\phi^{\gamma}) \; +  \;\ldots
\end{equation}
   where again the details and notation are described in the reference.  
  The above  octic invariant $X_8$ is constructed for the compact real form 
E$_{8(-248)}$, and hence we require a `twisted form' of this expression to identify an 
octic invariant for the real form $\eeg$.
This might be constructed via an $\mbox{SO}(4,12) \subset \eeg$ basis, where the 
non-compact subgroup SO$(4,12)$ is the quasiconformal group associated with 
10-dimensional spacetime $\htwo$ (\cite{Gunay2} equation~7), while SO$(2,12)$, as 
employed for equation~\ref{iquart} above, is the conformal group of the same space. 
(While $\eeg$ and $\eseg$ are the quasiconformal and conformal groups associated with 
the `generalised spacetime' $\htho$). The Lie algebra itself for the compact real form  
E$_{8(-248)}$ is presented with respect to the maximal compact subgroup SO(16)
 in (\cite{CedP} equation~2.1), with a corresponding decomposition for $\eeg$ in the 
subgroup SO$(4,12)$ basis described for example in (\cite{Lisi2} equation~4.3).

   While providing no detailed analysis here the suggestion is to consider whether the 
construction of an octic invariant $X_8$, similar to that of equation~\ref{xoctic}, 
for the real form $\eeg$ can be seen as an augmentation from the quartic $\eseg$ 
invariant $\mcI_4$ of equation~\ref{iquart}, in principle via a subgroup substructure
 $\mbox{SO}(2,10) \subset \mbox{SO}(4,12)$ within the embedding $\eseg \subset \eeg$.
With the tensor invariant $\mcI_4$ correlated with the 56-dimensional space $F(\htho)$ 
underlying the quartic invariant $q(x)$ under the $\eseg$ symmetry 
 in principle it may be possible to identify a parallel relation for the augmented 
invariant $X_8$ in terms of a 248-dimensional octic polynomial expression
 explicitly relating to the   
 familiar elements of $\det(X)$, $\det(\mcX)$ and $q(x)$ in the progression of 
equations~\ref{stglvte}--\ref{stglvfs} in leading to an $\eeg$ invariant $\lvtfe$ as 
provisionally proposed in equation~\ref{stglvtfe}. 
With $X \inn \htwo$, $\mcX \inn \htho$ and $x\inn F(\htho)$ care is needed in this 
translation from the tensor invariants to polynomial forms in $\rrr^{56}$ and 
$\rrr^{248}$ owing to the algebraic properties of the octonions.
  In the latter case for the $\eeg$ symmetry this would
 ideally involve the $X\in \htwo$ parts already expressed in terms of an $\sltc 
\subset \eeg$ spinor decomposition such as equation~\ref{xththfull}, and in a manner 
in principle related to the property of octonion triality as noted in the previous 
subsection and discussed further below.

   The initial aim would then be to identify an explicit structure for an octic $\eeg$ 
invariant over $\rrr^{248}$ as a candidate for the homogeneous polynomial form 
$L(\bv_{248}) = Q(\btt) =1$ as a further progression from the forms listed in 
table~\ref{vftovfs}. Subsequently the embedding of the symmetry breaking structure 
$\sltc \times \suth_c
  \times \uo_Q \subset \esig \subset \eseg \subset \eeg$ might be used as an initial 
basis upon which to seek further structures of the Standard Model, subsuming the 
findings of figure~\ref{mforese},  by essentially reading off the symmetry properties 
of all the physical matter fields deriving from the breaking of $\eeg$ over the base 
space $M_4$.   

   From the above discussion the full form $L(\bv_n) = Q(\btt) = 1$ that we seek is 
expected to lie somewhere between an unfolding of the $\eeg$ realisation on the 
quartic light cone form $d(e,f) = 0$ of equations~\ref{qusydi} and \ref{defcone}, with 
$e,f \in eF(\htho)$ and possibly $n\neq 248$,
 and an octic $\eeg$ invariant, with $n=248$, based on a twisted form of the invariant 
$X_8$ described for equation~\ref{xoctic}. An intermediate homogeneous polynomial 
form, for example of quintic order, would also correspond to a \textit{realisation} 
rather than a \textit{representation} of $\eeg$, since the invariants of the 
248-dimensional adjoint  representation are expected to be homogeneous polynomials of 
order $2, 8, 12, 14, 18, 20, 24$ or $30$ as described in subsection~\ref{ee21}.   
   
   A group \textit{realisation} can be defined as a map from elements of the group 
into an algebraic structure with isomorphic composition properties, which in general 
involves non-linear actions (as for several conformal group generator elements of
 $\mbox{su}(2,2) \equiv \mbox{so}(2,4)$
 in equation~\ref{usuact}) and in principle can act on a space of arbitrary dimension. 
The $\eeg$ Lie algebra realisation on the 57-dimensional space of elements 
$e=(x,\tau)\in eF(\htho)$ is expressed through actions such as those of 
equation~\ref{kkact}, which are highly non-linear.

    On the other hand a group \textit{representation} is a map from the group elements 
into a set of matrices, expressing linear transformations on a space of a specific 
dimension. The smallest non-trivial irreducible representation of $\ee$ is the 
248-dimensional adjoint representation, which can hence be  expressed by a  subset of 
the general linear matrices $\mbox{GL}(248,\rrr)$ or $\mbox{GL}(248,\ccc)$ acting on 
the 248-dimensional vector space of the $\ee$ Lie algebra, as a real or complex space, 
itself.  

  However, even for the hypothetical case of a non-linear realisation of $\eeg$ on for 
example a quintic homogeneous form $L(\bv_n) = Q(\btt) = 1$, with $n \neq 248$, when 
projected over $M_4$ the \textit{broken} symmetry subgroup could still be expressed by 
a reducible linear representation on the components of $\bv_n$. 
  Indeed for the $\eeg$ realisation on the 57-dimensional space of elements $e = 
(x,\tau) \in eF(\htho)$ the $\eseg \subset \eeg$ subgroup acts via a linear 
representation on the $x \in F(\htho)$ components (\cite{Unifi} equations~9.29--9.32) 
and the $\tau \in \rrr$ component (as a singlet), with this substructure employed in 
equation~\ref{qxtaul} for example. In principle not only 
 $\sltc \times \suth_c \times \uo_Q \subset \eseg$ of figure~\ref{mforese} but a full  
 subgroup $\sltc \times \suth_c \times \sutw_L \times \uo_Y \subset \eeg$
   could act via a linear representation on the multiplets of matter fields over 
$M_4$, with  interaction terms and a charge structure read off for direct comparison 
with the Standard Model,
 and with non-linear $\eeg$ actions not surviving the symmetry breaking. 
  The rearrangement and augmentation of the light cone $d(e,f) = 0$ into the form of a 
norm $L(\bv_n) = Q(\btt) =1$  might itself not exhibit the full $\eeg$ symmetry of the 
former structure while still leaving room to identify the full Standard Model  in the 
further  symmetry breaking in the necessary identification of the external spacetime 
$M_4$. 

  For the case of employing the smallest non-trivial $\ee$ representation the matter 
fields will be described by the broken symmetry acting on a 248-dimensional space. 
Since this smallest non-trivial representation is the adjoint representation of $\ee$,  
the external Lorentz and internal gauge symmetry properties of the Standard Model 
identified in this way might be expected to correlate with the structure of the 
248-dimensional $\ee$ Lie algebra itself. The possibility of this latter construction 
is directly and explicitly examined in \cite{Lisi}, as reviewed in 
subsection~\ref{ee21}, in which all states of the Standard Model together with 
gravitational field parameters are associated with elements of the root system of the 
$\ee$ Lie algebra. While a resemblance is identified in \cite{Lisi} between these 
structures of the  $\ee$ Lie algebra and the structure of the Standard Model of 
particle physics, as noted in subsection~\ref{ee21} the three generations of 
`fermions' are only obtained with respect to mutual SO(8) triality maps and this 
intrinsic feature cannot be avoided while remaining strictly within this 
framework~\cite{DiGa}.
In \cite{Lisi} the required triality maps relate  factors of the external Lorentz 
$\sltc \equiv \sutw \times \sutw$ 
 (for the `complexified' case as noted for equation~\ref{sssuesi})
 and an internal $\sutw_L \times \sutw_R$ symmetry via the subgroup:
\begin{equation}
 \label{sssse}
   \sutw \times \sutw \times \sutw_L \times \sutw_R \subset \mbox{SO}(8)
                               \subset \ee
\end{equation} 
   in the `graviweak' sector of the theory, with one of the $\sutw_R$ generators 
associated with hypercharge $Y$. Here the first $\sutw \times \sutw$ factor is 
analogous to the corresponding external symmetry factor in equation~\ref{sssuesi} or 
\ref{sssesi} while the internal $\sutw_L \times \sutw_R$ is analogous to that in 
equation~\ref{sspasa}; combining gravitational and electroweak structures in 
equation~\ref{sssse}.
  
  By contrast, in the conceptual picture of the present work particle states 
corresponding to Standard Model fermions and the Higgs are associated with components 
of the representation space $\bv_{248} \equiv \btt \in \mcT$ of the form $L(\bv_{248}) 
= Q(\btt) =1$, assuming the full $(n=248)$-dimensional case, upon which the subgroup 
symmetry  of equation~\ref{lorsme} acts.
 Hence with the Lorentz symmetry and internal gauge symmetry generated by elements of 
the real form $\eeg$ only the gauge bosons of the Standard Model, 
 rather than all states, are identified explicitly within the $\ee$ Lie algebra 
itself. Further, as a development from the $\esig$ action on $\htho$ of 
subsection~\ref{ee42} and the $\eseg$ action on $F(\htho)$ of subsection~\ref{ee43}, 
the octonion algebra $\ooo$ is expected to play a central role in the construction of  
the
 $\eeg$ symmetry of $L(\bv_{248}) =Q(\btt) = 1$.

  With the non-associative composition of the octonions  able to describe a high 
degree of symmetry, as noted in subsections~\ref{ee22} and \ref{ee42}, and employed in 
the explicit $\eeg$ action on the components of $\bv_{248}$, standard arguments for 
associative symmetry groups via the root system of their complex Lie algebras and 
Dynkin analysis may not directly apply here. The possibility of such a discrepancy was 
demonstrated in subsection~\ref{ee42} in which only Weyl spinors of a single 
handedness were identified in table~\ref{esibr} for the explicit breaking of the 
$\esig$ symmetry on the components of $\bv_{27} \equiv \mcX \in \htho$, while both 
left and right-handed states are seen for the closely related Dynkin analysis in 
table~\ref{dynkbr}.   

  In particular it was noted in the discussion following equation~\ref{pmarep} in the 
previous subsection and after equation~\ref{xoctic} above that the properties of 
octonion triality, associated with an $\mbox{SO}(8) \subset \esig$ subgroup as 
described in (\cite{Unifi} section~9.1), may be intimately involved in the 
identification of $\sltc$ Weyl spinors in a decomposition of the $X,Y \in \htwo$ 
components to fully account for the $\nu$-lepton and $u$-quark states, building on the 
structures identified in figure~\ref{mforese}.  
  The same, or a closely related, triality might also be intimately  involved in the 
identification of a full three generations of Standard Model fermions in the 
components of $\lvtfe$ under a broken $\sltc \times \suth_c \times \sutw_L \times 
\uo_Y \subset \eeg$ symmetry, effectively untangling the corresponding three 
`triality-related' generations identified in the $\ee$ Lie algebra root lattice 
in~\cite{Lisi} relating to the SO(8) subgroup of equation~\ref{sssse}.
  (The relation between SO(8) triality and the octonion algebra is explored for 
example in \cite{Man1} while
  the SO(8) triality structure employed in (\cite{Gunay5} equations~87--90), with
   $\mbox{SO}(8) \subset \mbox{SO}(2,12) \subset \eseg \subset \eeg$ might be 
significant in relation to the decompositions described following 
equations~\ref{iquart} and \ref{xoctic} here).

  In the present theory the above structures closely relate to the need to identify an 
electroweak symmetry $\sutw_L \times \uo_Y \subset \eeg$, acting on the components of 
$\lvtfe$, as would be required to consider triality relations between the $\sutw$ 
factors in equation~\ref{sssse}.
 Since both in \cite{Lisi,Lisi2} for $\ee$ and here in figure~\ref{mforese} at the 
$\eseg$ level one generation of the $e$-lepton and $d$-quark states is identified, 
this common feature might provide a handle to help identify the first generation 
$\nu$-lepton and $u$-quark spinor states from the $X,Y \in \htwo$ components here, 
potentially correlating with those  already identified in \cite{Lisi,Lisi2}. These 
states can be sought through their connection in 
$\binom{\nu}{e}_{\! L}$ and $\binom{u}{d}_{\! L}$ doublets under an internal $\sutw_L$ 
symmetry, and may act as a stage on the way to identifying a full three generations of 
leptons and quarks. 

  For the full subgroup symmetry of equation~\ref{lorsme}  the manner in which the 
action of the electroweak factor $\sutw_L \times \uo_Y \subset \eeg$ explicitly 
augments the 
 $\uo_Q \subset \esig \subset \eseg$ component of table~\ref{esibr} and 
figure~\ref{mforese} will closely relate to the `skew' in the largest `Russian doll' 
of $\eeg$ acting upon $\lvtfe$ with respect to the subsumed structures of 
table~\ref{vftovfs}, as proposed in the previous subsection.
 The structures of table~\ref{vftovfs} would seem to lead more directly to an $\ee$ 
composition incorporating a set of three `Jordan pairs', each of which might here be 
denoted 
 $(\htho,\overline{\mbox{h}}_3\ooo)$, as described for (\cite{MaTr} figure~1 and the 
top line of equation~1.1). This structure reflects the $\ee \supset \esi \times \suth$ 
subgroup branching of equation~\ref{eetoesi} (\cite{MaTr} equation~2.26) and could be 
interpreted here as describing three generations of the `quarks' and `leptons' listed 
in figure~\ref{mforese}. Similarly as the spaces $\htho$ and
 \raisebox{0.2ex}{`}$\:\!\overline{\mbox{h}}_3\ooo\:\!$\raisebox{0.2ex}{'}
  need to be opened up to correctly describe the full spinor structure, so the 
breaking of $\ee$ is not expected to be directly aligned with this $\esi \times \suth$ 
branching.

  Further, while the $\esi \times \suth \subset \ee$ subgroup branching of 
equation~\ref{eetoesi} is apparent in (\cite{Lisi} figure~4) there 
 the Standard Model embedding is more directly aligned within an $\ff \times \gt 
\subset \ee$ subgroup (as more apparent in \cite{Lisi} figure~3). That is, in 
\cite{Lisi} the external gravitational and internal electroweak sector derive from a 
graviweak $\mbox{SO}(8) \subset \ff$, incorporating the subgroup in 
equation~\ref{sssse}, with the internal colour symmetry accommodated via $\suth_c 
\subset \gt$. While the latter symmetry can be correlated with the $\suth_c \subset 
\gt \subset \esig$ of table~\ref{esibr}
 it has  been concluded here that a departure from that full $\esig$ representation 
structure in an $\esig \subset \eseg \subset \eeg$ embedding seems to be needed in 
order to identify the $\nu$-lepton and $u$-quark spinor states in a decomposition of  
$X,Y \in \htwo$.
  Hence on the way to equation~\ref{lorsme} an intermediate decomposition:
\begin{equation}
  \label{ffgtee}
   \mbox{SO}(8)\, (\subset \ff)\, \times\, \suth_c\, (\subset \gt) \;
    \subset\; \ff \:\!\times \:\!\gt \;\subset \;\ee
\end{equation}
  might be sought, with the first factor SO(8) further containing the `graviweak' 
subgroup of equation~\ref{sssse}. Out of this full symmetry the electromagnetic gauge 
symmetry may be obtained as $\uo_Q \subset \sutw_L \times \uo_Y \subset \mbox{SO}(8) 
\subset \ff \subset \ee$, incorporating electroweak symmetry breaking, rather than 
directly through the chain $\uo_Q \subset \sufo \subset \esi \subset \ese$ via 
 equations~\ref{sssesi} and \ref{esisufo} as for the intermediate forms of time of
  table~\ref{esibr} and figure~\ref{mforese}.
   Since $\ff \times \gt \subset \ee$ is a maximal subgroup the branching pattern of 
equation~\ref{ffgtee} cannot be directly aligned with an intermediate 
 $\esi \subset \ee$ or $\ese \subset \ee$ subgroup.

 In conclusion, the significant new feature introduced here is that  particle states 
are to be identified from a combination of both the components of the full form of 
time $\lvtfe$, a homogeneous polynomial form potentially of octic order, together  
with broken subgroups of an $\ee$ symmetry, constructed through octonion composition, 
rather than directly in terms of the 248 elements of the $\ee$  root system  or from 
any analysis of the complex $\ee$ Lie algebra itself. In this manner the aim is to 
identify the full structure of the Standard Model, ironing out the discrepant features 
found in \cite{Lisi,Lisi2} while avoiding the prohibitive theorems of \cite{DiGa} 
through this different perspective on the role of $\ee$.

  On the mathematical side a possible connection with the `magic square' of Lie 
algebras, briefly reviewed in subsection~\ref{ee21}, may shed further light on the 
role of the octonions and the uniqueness of $\ee$ as employed in this structure.
 The chain of real forms of exceptional Lie algebras
  $\mbox{F}_{4(-52)} \to \esig \to \eseg \to \eeg$ appears in the final column of the 
  $4 \times 4$ magic square M$(\kkk,\kkk')$ involving  a split division algebra $\kkk$ 
and $3\times 3$ Hermitian matrices over a non-split division algebra $\kkk'$ as 
described in (\cite{BaSu}, \cite{DrayMW} table~1). In \cite{DrayMW} the $\eseg$ Lie 
algebra entry in this `half-split' magic square is linked to the 
 $\eseg$ group action on the space $F(\htho)$, as the conformal group associated with 
the $\esig$ action on $\htho$ (see also the comment following \cite{Gunay2} 
equation~6).

  Similarly the conformal group $\mbox{SU}(2,2) \equiv \mbox{SO}(2,4)$ (the former 
being the double cover of the latter) associated with the Lorentz group in 
4-dimensional spacetime $\htwc$, as introduced here in subsection~\ref{ee23}, is 
linked with the entry in the $4 \times 4$ `half-split' magic square involving $2\times 
2$ matrices over $\hhh_s \otimes \ccc$, where $\hhh_s$ denotes the split quaternions, 
as described in (\cite{KinDr} table~1). 
 This table (see also \cite{DrayMW} table~2) also contains the groups SO$(2,10)$ and 
SO$(4,12)$, as the $\hhh_s \otimes \ooo$ and $\ooo_s \otimes \ooo$ entries 
respectively, which were discussed above for the substructure decomposition of the 
invariant tensors of equations~\ref{iquart} and \ref{xoctic} respectively. 
  In the literature there are hints that a generalisation of such magic square 
constructions for $2\times 2$ matrices might ultimately lead to a geometric 
interpretation for the exceptional Lie group $\ee$ associated with the final entry of 
the magic square for $3 \times 3$ matrices (see for example~\cite{KinDr} section~7, 
\cite{Kin2} section~2.3 and chapter~4). 

  A relation between real forms of Lie algebras obtained from the magic square and the 
corresponding real forms identified as conformal or quasiconformal symmetries of 
generalised spacetimes was also noted towards the end of subsection~\ref{ee23}.
 As one possible connection with the identification of physical structures, including  
Standard Model states, discussed in this subsection we note that in 
 (\cite{Gunay6} equation~7) the subgroup decomposition $\ff \times \gt \subset \ee$ of 
equation~\ref{ffgtee} above  is described for $\ee$ as the largest group of the magic 
square.
  These structures associated with the magic square might then also relate to the 
potential uniqueness of $\eeg$ for application as the full symmetry of time for the 
present theory.

  With the emphasis on the conceptual picture described in section~\ref{ee3},
  as developed through the explicit forms of time of section~\ref{ee4} as summarised 
in table~\ref{vftovfs} and leading to the prediction of a central role for the 
exceptional Lie group $\eeg$ in the previous subsection,
   while drawing together the various mathematical threads discussed in this 
subsection, the ambition is to converge upon the full form of time $L(\bv_{248}) = 
Q(\btt) = 1$ invariant under an $\eeg$ symmetry and assess the broader properties of 
the full physical theory. 
 The underlying basis of the theory, constructed from a single dimension of time, is 
arguably the simplest possible starting point for any unification scheme. 
  The construction of higher-dimensional manifestations of the general form of time 
$\lvn$ has been guided by this underlying simplicity, which ultimately
  translates into a high degree of symmetry for the proposed full form $L(\bv_{248}) = 
Q(\btt) = 1$,  exhibiting the structure of $\eeg$ as a real form of the largest 
exceptional Lie group.

   Through this exploration of the structure implicit within the one-dimensional flow 
of time the physical properties of matter in spacetime might be fully and directly 
deduced as described in this paper. As noted at the end of the previous subsection
  the identification of the full form of time $L(\bv_{248}) = Q(\btt) = 1$ as a 
natural mathematical structure accounting for these physical properties  then presents 
a non-trivial puzzle. Further possible means towards finding a solution have been 
suggested in this subsection. In the following section we consider these developments 
in the context of the broader ambitions of the full theory, and in relation to further 
aspects of   the existing literature reviewed in section~\ref{ee2}, aiming towards the 
goal of collectively tying all of these threads together into a complete theory.


\section{Comments on the Overall Physical Theory}
\label{ee6}

   In this section we describe how the main topic of this paper, concerning $\ee$ as a 
candidate for the full symmetry of time, relates to the other areas in which the 
theory has developed as described in \cite{Unifi}, with the four main branches of the 
theory summarised in (\cite{Unifi} figure~15.1). In particular this regards the 
connections with Kaluza-Klein theory, quantum field theory (QFT) and cosmology, as 
well as with the Standard Model of particle physics.

   The symmetry breaking structure for a higher-dimensional form of time over the 
projected external 4-dimensional spacetime, as introduced here for the $\slthc$ case 
in subsection~\ref{ee41} through equation~\ref{lvnitr} and figure~\ref{mtogmaph}, 
leads directly to the framework of a  principle fibre bundle $P\equiv M_4 \times G$ as 
employed by a range of Kaluza-Klein theories (including \cite{Cho, Katan}). This 
connection is  examined in detail in
 (\cite{Unifi} chapters 2--5, \cite{KKone}), with the geometric structure of 
figure~\ref{mtogmaph}(b) in this paper analysed more thoroughly for (\cite{KKone} 
figure~3(b)). This leads to a relation between the external curvature, described by 
the components of the Einstein tensor $G^{\mu\nu}$, and the internal curvature, with 
components $F^{\alpha\:\!\mu\nu}$, of the form (\cite{KKone} equation~93):
\begin{equation}
 \label{gchift}
  G^{\mu\nu} \: = \: 
  2\chi(  - F^{\alpha\:\!\mu}_{\ph{\alpha\mu} \rho}F_{\alpha}^{\ph{\alpha}\rho\nu}
	                -\frac{1}{4} g^{\mu\nu} \, F^{\alpha}_{\ph{\alpha}\rho 
\sigma}F_{\alpha}^{\ph{\alpha}\rho\sigma})   \: =: \:  -\kappa T^{\mu\nu}   
\end{equation}
  which also \textit{defines} the energy-momentum tensor $T^{\mu\nu}$ for this 
spacetime geometry solution. Here $\mu,\nu,\rho,\sigma$ are spacetime indices, 
$\alpha$ is a Lie algebra index and the real parameters $\chi$ and $\kappa$ can both 
be considered as normalisation constants for the empirical units ultimately employed
 for gravitation, gauge fields and energy-momentum.
 
     As described in (\cite{KKone} subsections~2.3 and 5.3) the breaking of the full 
symmetry group, denoted $\hat{G}$, of the full form of time $\lvh$ 
(where $\bvh = \bv_n$ for the largest $n$ considered)
over the base manifold $M_4$ is absolute. That is the surviving symmetry for the 
\textit{physical} theory can be written as (\cite{KKone} equation~23):
\begin{equation} 
  \label{dirprod}
   \mbox{Lorentz} \times G \, \subset \, \hG
\end{equation}
 where the Lorentz group for 4-dimensional spacetime may be replaced by its double 
cover $\sltc$ and $G$ is the residual internal symmetry group. For example in 
equation~\ref{slthbr} we have $\hat{G} = \slthc$ and $G= \uo$. 
 With the external Lorentz and internal $G$ symmetries both originating from the same 
unification group $\hat{G}$
such an \textit{absolute} symmetry breaking is required in order to avoid a 
trivial $S$-matrix for the quantised theory as an implication of the Coleman-Mandula 
theorem~\cite{ColMan}, as noted following equation~\ref{slthbr} in 
subsection~\ref{ee41} and discussed in (\cite{KKone} subsection~5.3).

 An avoidance of the prohibitive restrictions of the Coleman-Mandula theorem without 
invoking supersymmetry and in a manner relating to the identification of the external 
spacetime structure itself can also be found in other theories. For example the 
graviweak models described in
\cite{NePe} place gravitational and weak interactions on an equal footing, with the 
full symmetry of theory manifest in a `topological phase' without a metric in 
spacetime. Only in the `broken phase' can a residual symmetry of a global Lorentz and 
a local internal symmetry be identified, compatible with the Coleman-Mandula theorem
 (\cite{NePe} section~7).
 
 For the present theory the identification of spacetime $M_4$ itself out of the 
symmetries of a 4-dimensional substructure of the form $\lvh$ breaks the full symmetry 
$\hat{G}$, again compatible with the restrictions of the Coleman-Mandula theorem in 
the QFT limit. 
  Here we have considered equation~\ref{dirprod}
   explicitly for the cases of $\hat{G} = \slthc$, $\hat{G}= \esig$ and $\hat{G} 
=\eseg$ in the three subsections of section~\ref{ee4} and have been led to propose 
$\hat{G} = \eeg$ as the full symmetry for the full form of time $L(\bv_{248}) = 
Q(\btt) = 1$ in section~\ref{ee5}. In all cases this theory is manifestly 
background-free in that we begin simply with the one dimension of time only, expressed 
through the general \textit{mathematical} form of equation~\ref{lvo} with symmetry 
$\hat{G}$, and only \textit{then} break  this symmetry through the \textit{necessary} 
identification of an external 4-dimensional spacetime, building upon the basic vacuum 
 structure of equation~\ref{fourtran} and figure~\ref{fourd}, through which the 
\textit{physical} theory is constructed. 
 
 For the minimal physical model based on $\hat{G} = \slthc$ acting on $\lvni$
  in figure~\ref{mtogmaph}
  both the spinor field $\psi(x)$ and the gauge field $A(x)$,
   associated with the internal $\uo$ symmetry, and their mutual interaction arise 
naturally through the geometric structures described in subsection~\ref{ee41}.
In principle the postulates of QFT could simply be \textit{applied} to quantise the 
gauge and spinor matter fields arising over $M_4$, as deduced not only
 for equation~\ref{slthbr} but also for example in table~\ref{esibr} and 
figure~\ref{mforese}  for the higher-dimensional full forms of time. On the other hand 
the ambition of the present theory has been to avoid the need for `postulates' and 
instead at every stage to make the case for \textit{deriving} the structure of the 
 complete unified theory upwards from the original basic conceptual picture, as has 
been the case for equations~\ref{gchift} and \ref{dirprod} in leading to connections 
with Kaluza-Klein theory and with the Standard Model.
This leads to a proposal that all quantum phenomena arise through a degeneracy of 
solutions for identifying the external spacetime $M_4$ out of the components of the 
multi-dimensional form of time and its symmetries. That is the external geometry of 
equation~\ref{gchift} can be generalised for these multiple possible solutions as 
expressed by:
\begin{equation}
 \label{gfromav}
  G^{\mu\nu} = f(A,\bvh)
\end{equation}
   where $f(A,\bvh)$ denotes a rank-2 tensor function of the gauge fields $A(x)$, both 
Abelian and non-Abelian, and components of the multi-dimensional flow of time $\bvh$, 
including the $\psi(x)$ spinor components (see also the discussion of \cite{Unifi}
 equation~5.32). 
 Again we emphasise that, rather than \textit{beginning} with 4-dimensional spacetime 
containing a set of fields and \textit{then} applying restrictions associated with a 
one-dimensional constraining structure, we begin purely with the one-dimensional flow 
of time through which physical relations such as equation~\ref{gfromav} are derived 
which inherently incorporate constraints owing to the simplicity of this underlying 
 foundation.
    
	The manner in which the many possible solutions for equation~\ref{gfromav} can in 
principle be related to calculations in QFT is described in (\cite{Unifi} chapters~10 
and 11),
 by comparison with the mathematical methods of canonical quantisation.
 The permitted field exchanges implicitly integrated into equation~\ref{gfromav}
   within the constraints of the theory are summarised and discussed for
     (\cite{Unifi} equation~11.29), taking the place of Lagrangian terms in a standard 
QFT. A provisional link between the determination of an event probability based on the 
degeneracy of equation~\ref{gfromav} and the calculation of a process cross-section is 
made via the optical theorem of QFT as described for 
(\cite{Unifi} equation~11.46). 

   As noted in (\cite{KKone} subsection~5.3) it is proposed that in the quantum field 
theory limit the spin-statistics theorem will also apply here, accounting for the 
bosonic nature of the gauge fields and the fermionic nature of the spinor fields, and 
hence also of the lepton and quark particle states subsequently identified,
 for example in figure~\ref{mforese}. 
 Indeed the phenomena of `particle' entities themselves, as observed in the laboratory 
to exhibit bosonic or fermionic properties, are also to be accounted for through this 
`quantisation' of the fields (\cite{Unifi} sections~10.1, 11.3 and 15.2).  
 Hence here supersymmetry is neither required  to sidestep the Coleman-Mandula 
theorem, as noted above, nor to combine bosons and fermions in a consistent unified 
field theory, and there is no corresponding extension to the Standard Model expected.

 More generally, rather than making a direct comparison with any of a range of 
existing quantisation schemes (such as the canonical or path integral approach) the 
degeneracy of spacetime solutions can be considered as the source of quantisation in 
its own right, converging upon the structural form of a standard QFT, and in turn the 
various theorems listed above, only in the flat spacetime limit, and upon quantum 
mechanics in the non-relativistic limit.

  However with curved spacetime solutions $G^{\mu\nu}(x) \neq 0$ generally implied for
   equation~\ref{gfromav} gravity is fully embraced in this picture right down to the 
level of high energy physics (HEP) particle interactions. Such laboratory phenomena 
can be considered as microscopic solutions for general relativity, on a scale at which 
the intrinsic indeterminacy in the composition of the matter fields underlying a local 
spacetime geometry in equation~\ref{gfromav} becomes evident in the quantum 
uncertainty exhibited in such experiments (see also the discussion of \cite{Unifi} 
figure~15.2).  
In this picture the gravitational field itself is \textit{not} quantised, and on the 
contrary it is gravity in the form of the external spacetime geometry that provides 
the means through which all other fields \textit{are} quantised, through the 
degeneracy on the right-hand side of equation~\ref{gfromav}. Hence, in principle, the 
need to apply the postulates of quantum theory can be avoided, while the aim remains 
of reproducing the successes of QFT calculations in describing HEP phenomena, and of 
accounting for quantum effects more generally, in a suitable limiting approximation. 

   Deviations from a flat spacetime background are well beyond the reach of direct 
observation on the scale of HEP experiments, and gravity is generally completely 
neglected as an assumption underlying the corresponding QFT calculations. Further, as 
a pragmatic tool for relating observables the machinery of such QFT calculations says 
very little about what is actually physically \textit{happening} at the interaction 
point in a HEP experiment. On placing the postulates of quantum theory at the heart of 
the calculation the evolution of the physical system is generally described in terms 
of a superposition of states in an abstract Hilbert space. In recording an observation 
at some stage the `measurement problem' is inevitably encountered, involving the grey 
area through which this superposition apparently `collapses' into a single state 
registered by the macroscopic apparatus. The thought experiment that places 
Schr\"{o}dinger's hypothetical cat in superposition of an 
 $\vert$alive$\rangle$ and a $\vert$dead$\rangle$ state emphasises the conceptual 
difficulty in where to draw the line between the `quantum' and `classical' worlds.    

 Here on the other hand we can \textit{begin} with the world of classical 
probabilities on the large scale, based on the relative frequencies of possible 
outcomes for events taking place \textit{in} spacetime, such as the result of tossing 
a coin. As we then zoom in to a microscopic physical state we retain this classical 
notion of probability, in the sense of being determined by the relative frequency of 
possible outcomes which, at the level of equation~\ref{gfromav}, ultimately correspond 
to the degeneracy of solutions \textit{for} the local spacetime geometry itself. This 
would apply for example to the decay of an atomic nucleus that might lie at the centre 
of a
 `Schr\"{o}dinger's cat' type of experiment. Hence from this perspective there is 
essentially no discontinuity in the method of probability calculation in terms of the
 relative frequencies of possible outcomes. 

   With the gravitational field itself not quantised
   the spacetime geometry  is considered to be completely smooth, with 
 the Einstein field equation of general relativity fully preserved here by simply
  \textit{defining} the  energy-momentum tensor identically with 
equation~\ref{gfromav}, that is: 
 \begin{equation}
 \label{gfromavt}
  G^{\mu\nu} \: = \: f(A,\bvh)   \: =: \:  -\kappa T^{\mu\nu}   
\end{equation} 
  which also generalises equation~\ref{gchift}.
  While $G^{\mu\nu}(x)$ describes the external spacetime geometry, $T^{\mu\nu}(x)$ can 
be interpreted as describing the underlying matter field composition, including its 
quantum properties.
   The spacetime smoothness is valid down to arbitrarily small scales consistent with 
the initial conceptual picture motivated in the opening of section~\ref{ee3} in the 
discussion of equation~\ref{propint} regarding vanishingly small inertial frames, as 
also noted in the discussion following equation~\ref{slthbr}.
 In combining general relativity and quantum theory in this unified framework 
essentially it the latter that is subsumed within a generalisation of the former.

  While writing three volumes on the foundations, applications and successes of 
  quantum field theory, in the context of the Standard Model of particle physics, 
Weinberg notes that (\cite{Wein} volume I, chapter 12 opening):
\begin{quotation}
    It is generally believed today that the realistic theories that we use to 
 describe physics at accessible energies are what are known as `effective field
  theories'. \ldots these are low-energy approximations to a more fundamental theory 
that may not be a field theory at all. 
\end{quotation}
  
  It is such a fundamental theory that we are aiming for here with the one-dimensional 
progression in time alone acting as the progenitor both of 
   matter fields and spacetime itself. The physical properties of these structures 
derive from the symmetries and components of the multi-dimensional forms of time 
studied in section~\ref{ee4}. These developments now culminate in the proposal that  
 the calculational structures of QFT also arise in the limiting approximation of a 
flat  4-dimensional spacetime base manifold $M_4$, which is built through 
equation~\ref{gfromav}  out of the  symmetry breaking structures of the full $\eeg$ 
action on the full form of time $L(\bv_{248}) = Q(\btt) = 1$, as proposed in 
section~\ref{ee5}, through a degeneracy of multiple possible solutions. 

This building of spacetime itself in a manner intimately related to the structure of 
$\ee$ may share some of the ambition described in (\cite{MaTr} section~3), although 
the original motivation and the general conceptual picture is very different.
 In (\cite{MaTr} section~3) a \textit{discrete} spacetime structure is proposed to 
emerge through elementary interactions determined by the $\ee$ Lie algebra structure,
 while here we are considering a \textit{continuous} spacetime with a smooth geometry, 
on the left-hand side of equation~\ref{gfromav}, arising through the basic arithmetic 
composition and symmetries of the continuous flow of a single dimension of time 
directly expressed in the proposed multi-dimensional form $L(\bv_{248}) = Q(\btt) = 1$ 
with a full $\eeg$ symmetry. 
  In the context of the quantisation of the theory we can further
 assess here the degree to which we may be uniquely led to $\ee$, and in particular 
the non-compact real form $\eeg$, as the full symmetry of time.

  The chain of vector spaces $\htwc \to \hthc \to \htho \to F(\htho)$ for the 
multi-dimensional forms of time and their corresponding symmetries was considered in 
section~\ref{ee4}, as summarised in table~\ref{vftovfs}, and led to the predicted full 
symmetry $\eeg$ acting on the yet higher-dimensional space $\mcT$ for the next stage.
 However, the alternative augmentation of spaces $\htwc \to \hthc \to \mbox{h}_4\ccc 
\to \ldots \mbox{h}_m\ccc$ with $m\times m$ Hermitian complex matrices
of real dimension $m^2$ up to an arbitrary positive integer $m$ might also have been 
considered in place of the path followed after equation~\ref{lvfr}. The corresponding 
full symmetry
  $\hat{G} = \mbox{SL}(m,\ccc)$ for the $m^2$-dimensional homogeneous polynomial form 
of time:
\begin{equation}
  \label{lvmm}
 L(\bv_{m^2}) \; = \; \det(\bv_{m^2}) \; = \; 1 
   \qquad \mbox{with} \qquad   \bv_{m^2} \in \mbox{h}_m\ccc
\end{equation} 
  would imply a breaking pattern over the base manifold $M_4$ corresponding to 
equation~\ref{dirprod} of:
\begin{equation}
 \label{slmcbr}
   \sltc \times \mbox{SL}(m-2,\ccc) \times \uo \; \subset \; \mbox{SL}(m,\ccc)
\end{equation}
 for $m\ge 4$, with the internal symmetry $G = \mbox{SL}(m-2,\ccc) \times \uo$ 
including a non-compact group factor. 
  The application of matrix structures to describe symmetries of time was initially 
motivated following equation~\ref{abdet}, although it was noted in the discussion 
following equation~\ref{stglvtfe} that higher-dimensional forms of time do not need to 
take the form of a matrix determinant. Nevertheless the structure
  of equation~\ref{lvmm} for $m=2,3,4\ldots$ might be considered a simple and natural 
progression towards higher-dimensional forms of time.

   In quantum field theory the Lagrangian for a model with an internal gauge field 
generally includes a kinetic energy term of the form:   
\begin{equation}
 \label{lagkff}
   \lag  = - \frac{1}{4}K_{\alpha\beta} F^{\alpha}_{\ph{\alpha}\mu\nu}
                    F^{\alpha \:\!\mu\nu}
\end{equation}   
   where  $K_{\alpha\beta}$ is the Killing metric for the Lie algebra of the gauge 
group and $F^{\alpha}_{\ph{\alpha}\mu\nu}$ is the gauge curvature introduced in 
equation~\ref{gchift}. 
  The requirement of positive kinetic energy is ensured for a negative definite 
Killing form which in turn implies that the gauge group is required to be a product of 
compact simple groups and $\uo$ factors (see also for example~\cite{Wein} volume II, 
section 15.2). Hence
  a non-compact internal symmetry group cannot feature in a consistent quantum field 
theory.

 For the present theory, in place of any Lagrangian terms such as 
equation~\ref{lagkff} constraints such as equation~\ref{gchift} are employed,
 as discussed following equation~\ref{gfromav}, and
 with the energy-momentum tensor \textit{defined}
  through the Einstein field equation $-\kappa T^{\mu\nu} := G^{\mu\nu}$ in the 
general case as described for equation~\ref{gfromavt}.
  The aim of converging upon a consistent quantum field theory in the flat spacetime 
limiting approximation, and with terms similar to that in equation~\ref{lagkff} 
appearing in equation~\ref{gchift}, suggests that here also  the internal symmetry $G$ 
in equation~\ref{dirprod} will be required to be a compact group for the overall 
theory to be consistent, hence disfavouring the higher-dimensional forms described for 
equations~\ref{lvmm} and \ref{slmcbr}.

  This requirement relates to both the manner in which equation~\ref{gchift} itself is 
derived, as proposed via a perturbed action integral  over the base manifold $M_4$ in 
(\cite{KKone} equation~91), and the specific structure of the breaking of the full 
symmetry $\hat{G}$ of the full form of time $\lvh$, to equation~\ref{dirprod}, when 
projected over the same base space $M_4$.
In the context of the present theory alone  compactness of the internal gauge groups 
may be a requirement for obtaining a consistent solution for the spacetime geometry 
through equation~\ref{gfromav}, a result which would then translate into a similar 
constraint in the QFT limit expressed in terms of energy-momentum via 
equation~\ref{gfromavt}, in principle equivalent to the condition of positive kinetic 
energy.

  Here we are necessarily employing a non-compact unification group $\hat{G}$ in 
equation~\ref{dirprod} in order to incorporate the non-compact Lorentz group $\soot$ 
 and hence some care is needed to see how the compactness of the internal group 
component $G$ might be
 guaranteed. One observation is that for the chain of spaces for the general form of 
time $\htwo \to \htho \to F(\htho) \to \mcT$, summarised for 
equations~\ref{stglvte}--\ref{stglvtfe}, the corresponding full symmetry groups:
\begin{equation} 
 \label{gseries}
  \hat{G} \; = \; \sootn \to \esig \to \eseg \to \eeg 
\end{equation}
are the `most compact' of the non-compact real forms of the several options available 
for each complex Lie group in the chain $\mbox{SO}(10) \to \esi \to \ese \to \ee$. 
That is, each of the real groups in equation~\ref{gseries} has the most negative 
signature of Killing form possible other than the compact real form of the group 
itself. This observation, together with the fact that the internal symmetry groups 
identified so far for $\esig$ and $\eseg$ \textit{are} compact (namely  $G = \suth_c 
\times \uo_Q$ in table~\ref{esibr} and figure~\ref{mforese}), might in part explain 
the uniqueness of this series of augmentations to the general form of time, with the 
symmetries of equation~\ref{gseries}, for constructing a complete and consistent 
physical theory.

  As noted in subsections~\ref{ee21} and \ref{ee52} the $\ee$ unification framework as 
strictly proposed in \cite{Lisi} has been shown to be ultimately untenable owing to 
the lack of a sufficiently \textit{non-compact} real form $\ee$ capable of 
accommodating three generations of Standard Model spinor states \cite{DiGa}. Here in a 
sense we have the opposite issue, and it is a question whether the real form $\eeg$ is 
sufficiently
 \textit{compact} to ensure a compact  internal symmetry group $G$ in 
equation~\ref{dirprod}. It is proposed then that the \textit{absolute} symmetry 
breaking described for that equation can be compatible not only with the 
Coleman-Mandula theorem, as discussed above, but also with the requirement of compact 
internal symmetry groups in the QFT limit of the theory.
The exact nature of the $\eeg$ symmetry breaking will itself depend on the structure 
of the currently hypothetical full form of time $L(\bv_{248}) = Q(\btt) = 1$ as the 
subcomponents $\bv_4 \inn \TM_4$ are projected out over the base manifold $M_4$ for 
this higher-dimensional generalisation from figures~\ref{mtogmaph} and \ref{mforese}.

  More generally, consistency with QFT properties should also include for example the 
optical theorem and spin-statistics theorem as also alluded to earlier in this 
section. On the other hand,
since we are \textit{not}
 applying postulates of quantum theory from the outset it also remains a possibility 
that non-compact factors of an internal symmetry $G$ may simply be 
 suppressed or filtered out as not giving possible solutions for the external geometry 
under equation~\ref{gfromav}, leaving only the compact factors featuring in the 
physics. This also raises broader questions concerning the interrelations between the 
various branches of the theory as discussed more generally in (\cite{Unifi} 
section~15.1).

The requirement of a compact internal symmetry subgroup $G \subset \eeg$ for 
compatibility with the QFT limit can be contrasted with the
  heterotic branch of string theory, briefly reviewed in subsection~\ref{ee21},  that 
is guided  inevitably towards the compact group $\ee \times \ee$  as an
 internal symmetry in order to obtain a consistent QFT free from anomalies.
 A fundamental difference in the role of the $\ee$ symmetry is that 
for the present theory the full symmetry $\hat{G} = \eeg$ in equation~\ref{dirprod} 
contains both the external Lorentz symmetry and an internal symmetry $G$, which is 
large enough to encapsulate the Standard Model gauge groups with little to spare, 
 while for the  case of string theory  the compact $\ee \times \ee$   purely internal 
gauge symmetry  can very comfortably incorporate the Standard Model as will be 
discussed further below.

 For many theories a set of quantisation postulates are assumed and applied from the 
outset. In string theory the dynamical degrees of freedom of strings and membranes in 
a higher-dimensional spacetime are quantised, with the gravitational field  itself 
incorporated into this scheme of a proposed consistent theory of `quantum gravity'.
   A very different approach to `quantisation' is taken in the present theory, as 
described in this section with reference to equation~\ref{gfromavt}, with many 
features of `classical
 general relativity' fully preserved.
 It might however also be considered whether the `anomaly-free' properties of $\ee$ 
might have application for the present theory in the QFT limit, or rather for a 
subgroup of $\ee$ since only a broken fragment survives as the internal gauge 
symmetry. 
 This is part of the broader question concerning the emergence of a consistent QFT 
from the theory presented here, as considered more generally in this section.

  Possible connections between the present framework and string theory might be 
established through shared mathematical structures. For example the exceptional Jordan 
algebra, with the $\esi$ cubic invariant of equation~\ref{cubicj} as reviewed here in 
subsection~\ref{ee22} and central to the present theory, has also played a significant 
role in string theory (see for example \cite{CoHo}), with further potential links via 
the mathematical structures described in subsection~\ref{ee23} to supergravity as 
noted for \cite{Gunay3,Gunay2,Gunay4,Gunay5} at the end of that subsection.
  
  Another common feature is the notion of `extra-dimensional' structures. The 
significant conceptual difference is that while string theory is formulated from the 
outset in for example a 10 or 11-dimensional spacetime in this paper we begin with the 
much simpler structure of one dimension of time only, as motivated in 
section~\ref{ee3}. Form one dimension we are led directly to the multi-dimensional 
form of time via the simple mathematical identities leading to equation~\ref{lvo}.
 The full symmetry $\hat{G}$ and the full form of time $\lvh$ it acts upon 
collectively give rise to the components of matter fields over a 4-dimensional 
spacetime $M_4$, which itself is constructed out of substructures of $\lvh$ thus 
breaking the full symmetry.

  For string theory the question remains regarding the identification of a  
4-dimensional spacetime vacuum solution exhibiting properties corresponding to the 
Standard Model (see for example~\cite{Braun}) out of a vast landscape of 
possibilities, as noted in subsection~\ref{ee21}. The structures of string theory are 
typically much larger than needed to accommodate the Standard Model alone. For example 
as a grand unification scheme the $\ee \times \ee$ gauge group embedded within the 
structure of heterotic string theory,
 on employing the subgroup decomposition of equation~\ref{eetoesi} to one factor of 
$\ee$  (\cite{Polch} equations~16.3.5 and 16.3.6),
 is large enough to describe 36 generations of quarks and leptons (\cite{Polch} 
section~16.3).

 This is in stark contrast to the present theory where we have converged upon the 
predicted $\eeg$ symmetry of $L(\bv_{248}) = Q(\btt) =1$, which in principle is just 
large enough to account for the required three generations, after already directly 
identifying a series of significant esoteric properties of the Standard Model through 
the natural augmentations of the forms of time described in section~\ref{ee4} and 
summarised in table~\ref{vftovfs}.  In particular there is very little redundancy in 
the explicit case of the breaking of the $\hat{G} = \eseg$ symmetry of $L(\bv_{56}) = 
q(x) =1$ over the base space $M_4$ as depicted in figure~\ref{mforese}, which already 
largely accounts for a single generation with very little in the way of a `hidden 
sector'. The full three generations are proposed to enter via the components of $x,y,z 
\in F(\htho)$ in an  expression for the full temporal form $Q(\btt)$, based upon 
equation~\ref{stglvtfe}, again necessarily with a fairly tight fit, but now with a 
full $\eeg$ symmetry.

  In fact at the $\eseg$ level depicted in figure~\ref{mforese} the only components of 
the 56-dimensional form of time \textit{not} closely associated with an aspect of the 
Standard Model are the four Lorentz scalars $\alpha, \beta, n, N \in \rrr$. As 
$\suth_c \times \uo_Q \subset \eseg$ invariants these fields provide possible 
candidates for the cosmological dark matter or dark energy as alluded to in 
subsection~\ref{ee43} and described in 
 (\cite{Unifi} section~13.1). These components, which potentially make a significant 
impact upon the large scale structure of the universe, are housed in a complementary 
sector of the components of the space $F(\htho)$ compared with the Standard Model 
states that are seen in the laboratory, as can be seen from equations~\ref{hthoxy}, 
\ref{xoct3} and \ref{ftscomp} and figure~\ref{mforese}. On the other hand the full 
$\eeg$ symmetry action on the components of the currently hypothetical form 
$L(\bv_{248}) = Q(\btt) = 1$ may be needed to uncover both the full Standard Model and 
the details of any new physics that may be significant either    in the laboratory or 
on the cosmological scale.
 
   We also note that in addition to the compact internal gauge groups already 
identified, in general non-compact `dilation' transformations might also play a 
significant role. For example for the case of the full symmetry $\hat{G} = \slthc$ 
broken over the base space $M_4$, as depicted in figure~\ref{mtogmaph}, the breaking 
pattern of equation~\ref{slthbr} can be augmented to the subgroup:
\begin{equation}
 \label{slthdn}
   \sltc \times \uo \times \mbox{D}(1) \; \subset \; \slthc
\end{equation}
  where $\mbox{D}(1)$ is a dilation action, as described for (\cite{KKone} equation 
25). (An additional dilation `$\times\, \mbox{D}(1)$' can also be appended to the 
subgroup on the left-hand side of equation~\ref{slmcbr} for any $m\ge 4$).
 Other possible dilations are described in the opening of (\cite{Unifi} section~13.2). 
These  include a similar dilation to that in equation~\ref{slthdn} at the $\esig 
\equiv \sltho$ level as discussed for equation~\ref{vcubic} (\cite{Unifi} 
equations~8.35, 13.5 and 13.6), corresponding also to the second removed `Dynkin dot' 
in figure~\ref{dynesi}(b).
 A further dilation is included at the $\eseg$ stage as noted for 
equations~\ref{gradx} and \ref{fquartic} (\cite{Unifi} equation~9.30), and potentially 
also at the $\eeg$ stage as denoted by $\Delta$ after equation~\ref{gradxf} and 
discussed after equation~\ref{stglvtfe}.

  While being symmetries of the full form of time $\lvh$, \textit{none} of these 
dilation transformations represent a physical symmetry of the external spacetime $M_4$ 
after the symmetry breaking, rather each of them  changes the `scale' of the
 4-vector $\bv_4 \in \TM_4$ projected out of the full set of $\bvh$ components. 
 Hence the subgroup decomposition of equation~\ref{slthdn}, for this full $\lvni$  
example, reduces to the subgroup $\sltc \times \uo  \subset \slthc$ as the physical 
symmetry of 4-dimensional spacetime as analysed for equation~\ref{slthbr}.

  However a possible role for the dilation transformations in the very early universe 
is considered in (\cite{Unifi} section~13.2) in which an initially unstable value of 
the magnitude $\vert \bv_4 \vert \simeq 0$ for the projection of $\bv_4 \inn \TM_4$ 
out of the full form $\lvh$
 for cosmic time $t \simeq 0$ rapidly increases and converges upon the present day 
stable value with $\vert \bv_4 \vert = h_0 > 0$ via a phase transition in the `Big 
Bang', as sketched in (\cite{Unifi} figure 13.3). At the $\eseg$ level
 these dilations can also drive any of the scalar invariants $\alpha, \beta, n, N$ to 
extreme values for $t \to 0$, allowing  very different properties for the `dark 
sector' in the very early universe. For the full theory
 the Big Bang phase transition then 
   marks the very early stage at which both the dark sector stabilises and  the 
familiar properties of the Standard Model of particle physics first emerge. 

  In addition to the potentially rapid growth in $\vert \bv_4 \vert$ in time marking a 
significant evolution of the spacetime geometry in the very early universe, small 
variations in this scalar field about $\vert \bv_4(x) \vert = h_0$ in space and time 
since the phase transition have been associated with the Standard Model Higgs field, 
as noted in the second bullet point following figure~\ref{mforese}. In this context it 
is these small
 variations in the magnitude of $\bv_4(x) \inn \TM_4$ projected out of $\lvh$ that 
directly impact upon the local spacetime geometry $G^{\mu\nu}(x)$ of 
equation~\ref{gfromavt},  through which the energy-momentum $T^{\mu\nu}(x)$ is 
defined, and hence $\delta \vert \bv_4(x) \vert$ acts as an apparent source of mass.

   In turn variations in other components of $\bvh$, such as any of
      $\{\alpha,\beta,n,N\}$, are correlated with variations of $\vert \bv_4 \vert$ 
under the constraint $\lvh$, and hence any of the components of $\bvh$ can be a 
constituent of a `massive' field in spacetime $M_4$, as alluded to in the third bullet 
point after figure~\ref{mforese}. Such a field may be `dark' if the underlying 
components are invariant under the internal gauge symmetry, as for 
$\{\alpha,\beta,n,N\}$, and hence potentially only apparent on the cosmological scale, 
or `visible' if the components interact with the gauge fields, as is the case for the 
Dirac spinor states for example in figure~\ref{mforese}, and hence in principle 
observable in the laboratory.
 Gauge fields $A(x)$ themselves, such as the electromagnetic field, directly carry 
energy-momentum via $f(A,\bvh)$ in equation~\ref{gfromavt} as expressed through the
 Kaluza-Klein relation of equation~\ref{gchift}, and can also become `massive' through 
impingement on the $\bv_4 \inn \TM_4$ components in the symmetry breaking structure, 
as is proposed for $W^{\pm}$ and $Z^0$ gauge bosons in the electroweak sector, 
(\cite{Unifi} section~8.3). The potential of this theory for applications in cosmology 
more generally is described in (\cite{Unifi} chapters~12 and 13).

  The Standard Model has been the main focus of this paper, which has summarised and 
built upon (\cite{Unifi} chapters~6--9, \cite{Novel}).
Through a combination of arguments, developed through sections~\ref{ee4} and \ref{ee5} 
and continued in this section,  we are guided to the non-compact real form  $\hat{G} = 
\eeg$ as playing at least a highly significant role as the symmetry of time.
 Further questions regarding the degree of uniqueness of the theory, including the 
nature of the external geometry of the base space $M_4$ itself, are discussed in 
\cite{Unifi} section~13.3).
 A range of potential empirical predictions for the laboratory that may follow from 
the developments of the theory proposed in this paper will be assessed in the 
following section. In the meantime,
 the connections made between the symmetry breaking patterns for higher-dimensional 
forms of time and the empirical properties of particle states already observed in 
modern day high energy physics laboratories, 
   as summarised for example in table~\ref{esibr} and figure~\ref{mforese}, together 
with the simplicity of the underlying conceptual picture, the broader developments of 
the theory discussed in this section and the potential for further progress make the 
overall case for the basic conception of the theory.


\section{Conclusions}
\label{ee7}

  In this paper we have described how a physical theory can be derived from the 
continuum of time expressed as a multi-dimensional form, from the structure of which 
both an external 4-dimensional spacetime and the matter fields within it can be 
identified. This has led to the proposal of $\eeg$ as the main candidate for the full 
symmetry of the highest-dimensional form of time supported by an analysis of the 
Standard Model structures that have already been uncovered for intermediate forms of 
time and consideration of the  remaining particle multiplet properties required.

  The conceptual basis and more general development of the theory has been presented 
in \cite{Unifi, Novel, KKone}, while here we have emphasised the connections with the 
exceptional Lie group $\ee$ and related literature including
 \cite{Polch,Braun,Lisi,DiGa,MaTr,Gari,CedP,Tala,GarG,BeRu,
   Baez1,BaSu,Gunay3,Gunay2,Rios} as briefly reviewed  in section~\ref{ee2}.
  Unlike several other unification schemes the theory here is not initially motivated 
by a notion of mathematical `beauty' or `uniqueness' but rather follows the clear and 
simple conceptual picture of constructing a full theory through an analysis of the 
structures and symmetries  of the basic arithmetic composition of the one-dimensional 
flow of time itself.

   The structure of an extended 4-dimensional spacetime and forms of matter  
collectively derived as a manifestation of this one dimension are then carried 
implicitly within and  \textit{simultaneously with} the flow of time, without the need 
to introduce an independent `material substratum' -- posited to model empirical 
observations.
 This approach can be contrasted with a range of theories and models that aim to 
account for the properties of matter by initially postulating \textit{extra spatial
 dimensions} over and above 4-dimensional spacetime. Here we take a much simpler 
starting point and begin with \textit{fewer dimensions}, namely the one dimension of 
time alone.

  The possibility of constructing a full physical theory from the one dimension of 
time has been motivated in section~\ref{ee3}. There it was  noted that the general 
multi-dimensional form of time, as derived for equations~\ref{propgen} and \ref{lvo}, 
contains within it as a particular case the quadratic form of 4-dimensional spacetime 
of equations~\ref{propeta} and \ref{lvfr}. Through the symmetries of $\lvf$  the basic 
geometric structure of an inertial frame can be identified, as described for 
equation~\ref{fourtran} and figure~\ref{fourd}.  
 Subsequently in the three subsections of
 section~\ref{ee4}  `matter fields' were obtained  from the residual components in the 
projection of the higher-dimensional forms of time of 
 equations~\ref{lvni}, \ref{lvtspmn} and \ref{lvfsq} over the external spacetime 
$M_4$, with properties determined by the corresponding breaking of the $\slthc$, 
$\esig$ and $\eseg$ symmetry respectively.

  In this manner a series of Standard Model properties have been identified, including 
spinor fields with fractional charges in colour triplets and a left-right asymmetry, 
as summarised in figure~\ref{mforese}.
  These structures resemble
   one generation of Standard Model quarks and leptons, firmly rooting the theory in 
the empirical world through this correspondence of physical properties.
    This empirical success parallels and vindicates the argument that the full 
mathematical structure of the theory is firmly anchored in the observable world in 
deriving from the elementary mathematical structure of the real line $\rrr$ 
representing the continuous universal flow of time. That is, as had been proposed in  
section~\ref{ee3}, the effectiveness of the theory is not `unreasonable'.

   In pursuing this theory further the ambition of uncovering the symmetry 
transformation properties of the complete set of Standard Model states, taking into 
account the known structures and relations of the exceptional Lie groups, led to the 
proposal of $\eeg$ as the ultimate unification symmetry in subsection~\ref{ee51}, with 
the capacity in principle to incorporate three generations of quarks and leptons.
  The proposed provisional form of 
 $L(\bv_{248}) = Q(\btt) = 1$ in equation~\ref{stglvtfe} is motivated through a 
combination of the mathematical patterns observed for 
equations~\ref{stglvni}--\ref{stglvfs} as well as the ambitions for the full physical 
theory. However beneath these considerations the theory rests on the firm and simple 
foundation of the universal nature of the one-dimensional flow of time, through which 
the structures of the physical world are directly determined.

  This underlying simplicity, which led to the `general form of time' in 
equation~\ref{lvo}, is perhaps analogous to the guide of `simplicity and 
elegance'~\cite{Ham1} employed by 
 Hamilton in deducing the arithmetic rules for his `algebra of time' in the 1830s, as
  reviewed in section~\ref{ee3}. 
  As also noted there the philosophical influences behind Hamilton's algebraic work, 
regarding our perception of the world through the necessary forms of space and time, 
also provide some of the basis for the theory presented in this paper. 
  In more recent decades a notion of `mathematical beauty' has to some degree 
motivated many developments in theoretical physics, including the perceived elegance 
of the high degree of symmetry possessed by the structure of the Lie group $\ee$. In 
addition to this aesthetic appeal
   the existence of $\ee$ itself might be considered unique mathematically, as 
discussed in section~\ref{ee2}. However here these aspects of $\ee$ are not the 
original motivation for considering this symmetry group. 
  
    Indeed the original conception of the theory, as introduced in section~\ref{ee3} 
in leading to equation~\ref{lvo}, had no immediate connection with the structure of 
$\ee$ or any of the exceptional Lie groups. The development of a more explicit 
construction of the theory, as summarised towards the end of section~\ref{ee3}, led to 
consideration of $\esig$ and $\eseg$ as symmetries of time, structures which in turn 
have been consolidated through the connections with the Standard Model described in 
section~\ref{ee4}. This progression has inevitably directed our attention towards 
$\eeg$ as the potential symmetry of the full form of time as explained in 
subsection~\ref{ee51}.  
 This proposal is supported by several other studies of structures relating to $\ee$, 
both in the physics and the mathematics literature, as described in 
subsection~\ref{ee52} and further in section~\ref{ee6} where the broader developments 
of the present theory have also been  reviewed. 
   The ambition is then to tie together the various strands in the literature 
regarding $\ee$ by establishing a unique role for this largest exceptional Lie group 
in the theory presented in this paper, hence relating $\ee$ directly to the 
fundamental structure of the physical world.
  
   While $\esi$ and $\ese$ symmetries of homogeneous polynomial forms are already well 
known, equations~\ref{lvts} and \ref{lvfs} respectively, the apparent absence of an 
explicit augmentation to an $\ee$ symmetry on a possible form of time $\lvtfe$ means 
that the suggestion  that $\ee$ itself, as the largest of this series of exceptional 
Lie groups, might play a significant role as a symmetry of time is in part based on 
mathematical aesthetics. 
 Further, regarding  the construction of the currently hypothetical full form of time 
$L(\bv_{248}) = Q(\btt) =1$, 
 this project might be guided to some degree by mathematical structures that could be 
considered `beautiful' or `unique' or `natural'  on observing patterns in the 
progression from equation~\ref{stglvni}
 to \ref{stglvfs},
 elements of which have helped to shape the provisional form of 
equation~\ref{stglvtfe} as noted above.

  If, following $\esig$ and $\eseg$, a further augmentation to an $\eeg$ symmetry of a 
homogeneous from $\lvtfe$ was already known it would simply be a case of extended the 
analysis of subsections~\ref{ee42} and \ref{ee43} for this higher-dimensional form of 
time and reading off the symmetry breaking pattern while looking for further 
correspondence with known Standard Model structure. On the other hand if such an 
$\eeg$ symmetry cannot be identified prior to this analysis, in practice hints from 
the Standard Model itself, including the need to open up further spinor states and 
identify three generations in the extension from figure~\ref{mforese}, might be 
employed to help construct the mathematical form $\lvtfe$.

  As noted at the end of subsection~\ref{ee51}, this unification scheme is then   
testable in the theoretical sense through the \textit{prediction} of this $\eeg$ 
symmetry of a homogeneous polynomial form as a further augmentation from the 
structures of table~\ref{vftovfs} capable of incorporating a full explanation of the 
Standard Model symmetry properties, and with little  redundancy. From the current 
progression of symmetries of time through 
 $\sltc \to \slthc \to \esig \to \eseg$ the aim of completing the Standard Model 
structure in a further stage with an $\eeg$ symmetry of a homogeneous form 
$L(\bv_{248}) = Q(\btt) =1$, subsuming the previous stages and utilising known 
mathematical properties of $\ee$ such as  the quasiconformal symmetry of the quartic 
light cone of equation~\ref{defcone}  
 or  the symmetry of an octic invariant as described for equation~\ref{xoctic},   
 at this moment presents us with a non-trivial puzzle, as noted at the end of 
subsection~\ref{ee52}.
 
  We have sought ways in which this problem might be approached and have considered 
what a solution might look like, but it is by no means obvious that it should exist at 
all. As noted in sections~\ref{ee4} and \ref{ee5} the octonion algebra is expected to 
play a key role in the explicit construction of this $\ee$ symmetry and hence care is 
needed in any direct interpretation of more general studies involving for example  
Dynkin analysis of complex Lie algebras. However the discussion of 
subsection~\ref{ee52} involving the relevant mathematical structures suggests that the 
problem of finding a solution for the explicit action of an $\eeg$ symmetry on a 
 concrete expression for $L(\bv_{248}) = Q(\btt) =1$ should be tractable \textit{if} a 
solution exists at all.
 Hence the search for a rigorous solution satisfying the appropriate mathematical and 
empirical criteria provides a robust \textit{theoretical} test. 

Further, since significant Standard Model properties have already been established up 
to the $\eseg$ stage summarised in figure~\ref{mforese}, any new features beyond the 
Standard Model identified at the $\eeg$ level could lead directly and unambiguously to 
\textit{empirical} predictions  
 that might be tested in the high energy physics laboratory or through cosmological 
observations. In particular, three potential areas for `new physics' to arise can be 
identified as follows:

\begin{itemize}
\item The simplest observation is that the rank-8 Lie group $\ee$ is large enough to 
contain the rank-8 subgroup (\cite{Unifi} equation~9.51):
\vspace{-6pt}
\begin{equation}
   \mbox{Lorentz} \times \suth \times
      \sutw \times \uo \times \sutw \times \uo  \subset \ee
\vspace{-7pt}
\end{equation}
  Hence in addition to the external Lorentz symmetry and the Standard Model
   gauge symmetry $\SML$ (see equation~\ref{lorsme}) in principle $\ee$ can contain a 
further internal
   $\sutw \times \uo$, implying the possibility of further gauge interactions. 
 The additional $\sutw$ might be identified as an `$\sutw_R$' factor, analogous to 
that in equation~\ref{sspasa} or \ref{sssse},
 while the additional $\uo$ factor could be expressed by the
  dilation  $\Delta$ of equation~\ref{gradxf}  for $\eeg$ as the non-compact real form 
of interest here, with a possible role as discussed following equation~\ref{slthdn}.

\item With variations in the 4-vector magnitude $\vert \bv_4 \vert$ associated with 
the Higgs field, as discussed after figure~\ref{mforese}, non-standard physics for the 
electroweak symmetry breaking sector of the Standard Model might be expected. As noted 
towards the end of subsection~\ref{ee51} under the external $\mbox{Lorentz} \subset 
\eeg$ symmetry the components of
 $L(\bv_{248}) = Q(\btt) = 1$ may transform as spinors and scalars only. In this case 
it is the components transforming as `right-handed neutrinos' that are proposed to be 
subsumed into the necessary composite 4-vector projection $\bv_4 \inn \TM_4$ onto the 
external spacetime out of $\lvtfe$. Hence in principle this sector of the theory may 
also have implications for neutrino physics.

\item While under the $\eseg$ action on $L(\bv_{56}) = q(x) = 1$ there are only four 
components $\{\alpha,\beta,n,N\}$ of $\bv_{56} \equiv x \inn F(\htho)$ that are not 
directly associated with Standard Model structures in figure~\ref{mforese}, in 
identifying a full three generations of leptons and quarks from the hypothetical 
$\eeg$ action on $L(\bv_{248}) = Q(\btt) = 1$, with $\bv_{248} \equiv \btt \inn \mcT$, 
there could be as many as $\sim \!\! 50$
 components of the 248-dimensional space $\mcT$ remaining. In principle these may give 
rise to new particle phenomena detectable in HEP experiments or have observable 
consequences on a galactic or cosmological scale as an extended `dark sector'. 

\end{itemize}

   The specific mathematical structure of an $\eeg$ symmetry acting on the full form 
of time $\lvtfe$ and its symmetry breaking pattern over the base manifold $M_4$ will 
be needed to analyse these potential empirical predictions in detail. A more complete 
theory, in particular incorporating quantum theory as proposed in section~\ref{ee6}, 
might also be required for precise calculations of the properties of possible new 
particle states, as well as of the known ones. However one general feature is that 
there is only limited room for new physics beyond the Standard Model, with for example 
no set of `mirror states' or supersymmetric partners associated with the familiar 
Standard Model particles. In this sense the theory at this stage is already 
`falsifiable'  since the discovery of, for example, supersymmetry states would not be 
compatible with this theory without a significant, and internally poorly motivated, 
modification.  

   In the meantime the main arguments for the merit of this theory echo the ambitions 
outlined in the three bullet points at the end of section~\ref{ee1}.
   The theory  is firmly grounded in the basic notion of the flow of time that infuses 
all of our experiments and observations in the world. 
   Given the interest dating over many decades in a wide range of theories based on 
extra spatial dimensions, the observation that it is possible to construct a full 
physical theory simply from the one dimension of time alone, and the explanatory power 
that has already emerged in particular regarding connections with the Standard Model, 
underpins the plausibility of this approach. 
  It is possible to extract significantly more out of this theory than is put in 
through the simple underlying assumptions.
In addition, the development of the theory points towards a unique role for $\ee$ as 
the full symmetry of the full
multi-dimensional form of time, as also emphasised in this paper, highlighting the 
prospects for further progress and the potential predictive power of the theory.


{\setlength{\baselineskip}{0.91\baselineskip}

\par}


\par}

\end{document}